\documentclass[12pt,amsmath,amssymb,showkeys,prb]{revtex4}
\usepackage{xcolor}
\usepackage{setspace}
\usepackage{graphicx}
\usepackage{amssymb}
\usepackage{amsmath}
\usepackage{bbold}
\usepackage{mathrsfs}
\usepackage{calligra}
\usepackage{bm}

\newcommand{\citen}[1]{%
  \begingroup
    \romannumeral-`\x 
    \setcitestyle{numbers}%
    \cite{#1}%
  \endgroup
}
\newcommand{\e}{\mathrm e}

\newcommand{\scd}{{\rm SCD}}

\usepackage{mathrsfs}

\DeclareSymbolFont{matha}{OML}{txmi}{m}{it}
\DeclareMathSymbol{\varw}{\mathord}{matha}{119}   
\newcommand{\kB}{k_{\rm B}} 
\newcommand{\lb}{l_{\rm B}} 

\renewcommand{\i}{{\rm i}}  

\begin{document}

$\null$ \hfill {\bf July 25, 2024}
\vskip 1.0in

\begin{center}

{\Large\bf 
Differential Effects of Sequence-Local versus Nonlocal}\\

\vskip 0.3cm

{\Large\bf 
Charge Patterns on Phase Separation and}\\

\vskip 0.3cm

{\Large\bf Conformational Dimensions of Polyampholytes as}\\

\vskip 0.3cm

{\Large\bf Model Intrinsically Disordered Proteins}\\

\vskip .5in
{\bf Tanmoy P{\footnotesize{\bf{AL}}}},$^{1,\dagger}$
{\bf Jonas W{\footnotesize{\bf{ESS\'EN}}}},$^{1,\dagger}$
{\bf Suman D{\footnotesize{\bf{AS}}}},$^{1,2,\dagger}$
and
{\bf Hue Sun C{\footnotesize{\bf{HAN}}}}$^{1,*}$

$\null$

$^1$Department of Biochemistry,
University of Toronto, Toronto, Ontario M5S 1A8, Canada\\
$^2$Department of Chemistry, Gandhi Institute of Technology and
Management, Visakhapatnam, Andhra Pradesh 530045, India\\

$\null$\\
$\null$\\
$\null$\\

{\tt To appear in "J. Phys. Chem. Lett."}
\end{center}
$\null$\\

\noindent
$^\dagger$Contributed equally.

\vskip 1.3cm

\noindent
$^*$Correspondence information:\\
{\phantom{$^\dagger$}}
Hue Sun C{\footnotesize{HAN}}.$\quad$
E-mail: {\tt huesun.chan@utoronto.ca}\\
{\phantom{$^\dagger$}}
Tel: (416)978-2697; Fax: (416)978-8548\\
{\phantom{$^\dagger$}}
Department of Biochemistry, University of Toronto,
Medical Sciences Building -- 5th Fl.,\\
{\phantom{$^\dagger$}}
1 King's College Circle, Toronto, Ontario M5S 1A8, Canada.\\

\vfill\eject

\noindent
{\large\bf Abstract}\\

\noindent
Conformational properties of intrinsically disordered proteins (IDPs) are
governed by a sequence-ensemble relationship. To differentiate the impact of
sequence-local versus sequence-nonlocal features of an IDP's charge pattern on
its conformational dimensions and its phase-separation propensity, the charge
``blockiness'' $\kappa$ and the nonlocality-weighted sequence charge 
decoration (SCD)
parameters are compared for their correlations with isolated-chain radii of
gyration ($R_{\rm g}$s) and upper critical solution temperatures (UCSTs) of
polyampholytes modeled by random phase approximation, field-theoretic
simulation, and coarse-grained molecular dynamics. SCD is superior to $\kappa$ 
in predicting $R_{\rm g}$ because SCD accounts for effects 
of contact order, i.e., nonlocality, on dimensions of isolated chains. 
In contrast, $\kappa$ and SCD are
comparably good, though nonideal, predictors of UCST because frequencies of
interchain contacts in the multiple-chain condensed phase are less 
sensitive to sequence
positions than frequencies of intrachain contacts of an isolated chain, as
reflected by $\kappa$ correlating better with condensed-phase interaction 
energy than SCD.  
\\

\vfill\eject

\noindent
Extensive recent research has elucidated myriad 
properties of biomolecular condensates
and their essential roles in diverse biological 
functions (reviewed, e.g., in refs.~\citen{cliff2017,rosen2021}).
These membraneless compartments are underpinned to
a significant degree by liquid-liquid phase separation (LLPS) 
of intrinsically disordered proteins (IDPs), though more complex thermodynamic 
processes such as gelation/percolation and dynamic mechanisms---involving 
not only IDPs but folded protein domains as well as nucleic acids---also
contribute prominently.\cite{PappuChemRev,Mingjie2023,HXZhouRev2024}
The propensity for an IDP to undergo LLPS is dependent upon its sequence of
amino acids.\cite{Nott15,linPRL,TanjaRohitNatChem2022}
Many IDPs are enriched in charged and aromatic residues. Accordingly, 
beside $\pi$-related interactions,\cite{robert,Chen2015,bauer2022,koby2023}
electrostatics is an important driving force for many aspects of IDP 
properties, as exemplified by its notable roles in IDP conformational 
dimensions,\cite{rohit2013,kings2015} the stability of biomolecular 
condensates,\cite{Nott15,knowles2022} and 
a condensate's capability to selectively recruit IDPs with 
different sequence charge patterns.\cite{Sabari2023} 
\\

\noindent
An IDP can exist and function as essentially isolated 
chain molecules and/or collectively in a multiple-chain condensate, or 
certain intermediate oligomeric configurations in between. With this in mind,
analyzing the relationship between the behaviors of 
isolated and condensed IDPs is instrumental not only for inferring 
condensed-phase properties from those of computationally and experimentally 
simpler isolated-chain systems. More fundamentally, it offers
insights into how sequence-encoded IDP properties are modulated 
by IDP concentration.
For homopolymers with short-spatial-range interactions, it has been
known since the 1950s via the Flory-Huggins (FH) theory that the radius 
of gyration $R_{\rm g}$ of an isolated polymer anticorrelates with 
the polymer's LLPS propensity. This is because 
$R_{\rm g}$ decreases monotonically with the FH $\chi(T)$ 
interaction parameter\cite{Flory1951} at a given absolute temperature $T$
and the LLPS critical 
temperature\cite{Flory1952} $T_{\rm cr}^{\rm FH}$ 
increases with $\chi(T)$ when the interactions are purely enthalpic, i.e.,
$\chi(T)\sim 1/T$ and thus
$T_{\rm cr}^{\rm FH}=T\chi(T)/\chi_{\rm cr}$ where 
$\chi_{\rm cr} =(1+1/\sqrt{N_{\rm p}})^{2}/2$ and $N_{\rm p}$ is polymer chain
length,\cite{KnowlesJPCL} as is verified by recent simulations and more 
sophisticated polymer theories.\cite{Panag1998,WangWang2014}
Inspired by recent interest in sequence-specific IDP LLPS, similar 
relationships for heteropolymers are 
predicted between LLPS propensity and isolated-chain $R_{\rm g}$
(ref.~\citen{lin2017}), coil-globule transition,\cite{BestPNAS2018} 
and two-chain association in a dilute 
solution.\cite{Alan} Likewise, dilute-phase $R_{\rm g}$ and demixing 
temperature was seen to correlate experimentally for
variants of the P domain of core stress-granule marker polyA-binding 
protein undergoing heat-induced LLPS.\cite{Riback2017}
\\

\noindent
The predicted dilute/condensed-phase correlations for heteropolymers
are significant but not perfect,\cite{lin2017,BestPNAS2018} 
as underscored by a recent
extensive simulation study of the human proteome.\cite{KrestenbioRxiv2024} 
The imperfection is instructive about the physical chemistry of 
concentration-dependent IDP interactions.
We focus here on electrostatics as a first step.
Building on the substantial 
recent works on polyampholyte configurations and
LLPS,\cite{rohit2013,kings2015,lin2017,joanJPCL2019,joanPNAS,Wessen2021,Pal2021}
we attend to a hitherto less explored consequence of sequence specificity,
namely how the impact of sequence-local patterns involving 
charges proximate to one another differs from that of 
sequence-nonlocal patterns encompassing charges far apart along the chain.
The differing effects of local versus nonlocal interactions has long
been recognized 
in globular proteins. These include folding stability\cite{Dill1990} and the 
effects of intrachain contact order\cite{ChanDill1990} on conformational 
ensembles as well as folding 
kinetics.\cite{Dilletal1993,Plaxco1998,Chan98,AnnuRevPhysChem2011}
But much about the corresponding effects for IDPs remains to be elucidated.
To make progress, here we contrast the predictive powers of two common 
sequence charge parameters, namely the ``blockiness'' measure $\kappa$ 
(ref.~\citen{rohit2013}) and ``sequence charge decoration'' 
(SCD, ref.~\citen{kings2015}). 
For a $N_{\rm p}$-bead polyampholyte with charge sequence 
$\sigma_\alpha$ ($\alpha=1,2,\dots,N_{\rm p}$),
$\kappa$ captures primarily sequence-local aspects of the charge pattern
as its terms account for charge asymmetry 
only in blocks of $g=$ $5$ or $6$ consecutive beads relative to the 
overall charge asymmetry of the polyampholyte in the form of
$\kappa\sim$
$\sum_{\beta=1}^{N_{\rm p}-g+1}[(\sum_{\alpha=\beta}^{\beta+g-1}\sigma_\alpha)^2/(\sum_{\alpha=\beta}^{\beta+g-1}\vert\sigma_\alpha\vert)$ 
$-$ 
$(\sum_{\alpha=1}^{N_{\rm p}}\sigma_\alpha)^2/(\sum_{\alpha=1}^{N_{\rm p}}\vert\sigma_\alpha\vert)]^2$ 
(refs.~\citen{rohit2013,suman2}).
In contrast,
${\rm SCD} \equiv \sum_{\alpha=2}^{N_{\rm p}} \sum_{\beta=1}^{\alpha-1} 
\sigma_{\alpha}\sigma_{\beta}
\sqrt{\alpha - \beta}/N_{\rm p}$
takes into account nonlocal sequence charge pattern
as it accords higher weights for 
sequence-nonlocal pairs of charges.\cite{kings2015}
\\

\noindent
Following several seminal
studies,\cite{rohit2013,kings2015,lin2017,joanJPCL2019}
we consider $N_{\rm p}=50$ overall-neutral polyampholytes with 
$\sigma_\alpha=\pm 1$ (in units of the protonic charge) and 
$\sum_{\alpha=1}^{N_{\rm p}}\sigma_\alpha=0$. Substantiating a preliminary
analysis,\cite{suman2} we rigorously determine the joint distribution 
$P({\rm SCD},\kappa)$ among all such sequences by 
first estimating 
the populated region in the $({\rm SCD},\kappa)$-plane 
via a diversity-enhanced genetic algorithm\cite{CamargoMolinaMandal2018} 
and then obtaining $P({\rm SCD},\kappa)$ in the identified region using a
Wang-Landau approach\cite{WangLandau2001} (all SCD $<0$;\cite{Alan} see 
text and Fig.~S1a of Supporting Information for details). 
The resulting heat map for $P({\rm SCD},\kappa)$ 
in Fig.~1a exhibits a moderate correlation
between $-{\rm SCD}$ and $\kappa$ (Pearson 
correlation coefficient $r=0.684$).  
\\

\noindent
Among these 50mers, we select for further analysis
26 sequences with a broad 
coverage in Fig.~1a and are illustrative of
sequence variations of interest (Fig.~1b).
Beside the 12 ``sv'' sequences as examples of the
30 original sv sequences\cite{rohit2013} are the
4 ``as'' sequences as controls for their $-{\rm SCD}$--$\kappa$ 
anticorrelation opposing the overall positive 
correlation.\cite{suman2} To probe the differential effects of 
sequence-local versus nonlocal charge patterns, we construct
4 ``c$\kappa$'' sequences with diverse SCDs but essentially 
the same $\kappa$, and
4 ``cSCD'' sequences with diverse $\kappa$s but essentially 
the same SCD. We also consider sequences obs1 and ebs1
with odd and even numbers of charge blocks, respectively, to assess
effects of like versus opposite charges at the two chain ends.
These sequences and their $\kappa$ and $-$SCD values are listed
in Table~S1 and Fig.~S1b of Supporting Information and marked in Fig.~1a.
By focusing on this set of sequences with the same chain length, 
we address sequence charge 
patterns' impact on the thermodynamics of polyampholytes.
As such, further investigations of dynamic and other material 
properties,\cite{koby2020,MittalNatComm2024,Wingreen2024}
broader questions about sequence specificity for IDPs of different 
chain lengths,\cite{RuffMiMB2021,PappuJMB2021}
polyampholytes in high salt,\cite{kings2018}
and for sequences
containing short spatial range hydrophobic-like 
interactions\cite{moleculargrammar,Zheng2020,WessenDasPalChan2022,Mittal2023,MittalNatChem2024}
and/or with high net charges\cite{kings2020,Caprin1_arXiv}
are left to future studies.
\\

\noindent
Theories and computational models are available
to address sequence-specific LLPS of polyampholytes 
(see, e.g., refs.~\citen{Mahdi2000,Ermoshkin2003,panag2005,Fredrickson2006,singperry2020}, reviewed in ref.~\citen{Jacobs2023}).
Here we apply three complementary 
methods: analytical random phase
approximation (RPA),\cite{Mahdi2000,Ermoshkin2003}
field-theoretic simulation 
(FTS),\cite{joanPNAS,Parisi1983,Klauder1983,Fredrickson2006}
and coarse-grained molecular dynamics (MD)
with the ``slab'' sampling method for 
phase equilibria\cite{Panagiotopoulos_Scaling_2017}
(formulations described in Supporting Information), which
have afforded numerous physical 
insights.\cite{linPRL,joanJPCL2019,dignon18,Pal2021,WessenDasPalChan2022}
Based on the same path-integral polymer model,
FTS is more accurate than RPA in principle because it
does not require an approximation like RPA and can be 
extended to tackle nonelectrostatic interactions.\cite{WessenDasPalChan2022}
Nonetheless, FTS is limited by finite resolution, simulation box size,
and treatments of excluded volume.\cite{joanJPCL2019,Pal2021}
Compared to RPA and FTS, coarse-grained MD accounts better for structural 
and energetic features of 
IDPs\cite{Zheng2020,Mittal2023,SumanPNAS,Mpipi,Kresten2021}
but is computationally more costly.
To focus on electrostatics,
we adopt the ``hard-core repulsion'' MD 
model with no nonelectrostatic attraction.\cite{suman2}
The utility of combining the complementary advantages of RPA, FTS, and MD 
is illustrated by recent studies of the dielectric 
properties of condensates\cite{Wessen2021} and the effects 
of salt and ATP on condensed polyampholytic and polyelectrolytic 
biomolecules.\cite{Caprin1_arXiv}
\\

\noindent
We employ all three methods for LLPS.
$R_{\rm g}$s and pairwise bead-bead contacts---which are 
not amenable to RPA currently---are computed by MD and the following
FTS approach:
In a multiple-chain system, the root-mean-square radius of gyration 
of the $i$th polymer
${R_{\rm g}^{(i)}}^2 = 
\langle\sum_{\alpha=1}^{N_{\rm p}} \sum_{\beta=1}^{N_{\rm p}} 
(\bm{R}_{i,\alpha} - \bm{R}_{i,\beta})^2\rangle /{2N_{\rm p}^2}$ 
$=$
$\int d\bm{r} \int d\bm{r}' \langle \hat{\rho}_{\rm c}^{(i)}(\bm{r}) 
\hat{\rho}_{\rm c}^{(i)}(\bm{r}') \rangle (\bm{r}-\bm{r}')^2/{2N_{\rm p}^2}$,
where $\langle \cdot\cdot\cdot \rangle$ denotes Boltzmann averaging, 
$\bm{R}_{i,\alpha}$ is the position of the $\alpha$th bead along 
the chain, and $\hat{\rho}_{\rm c}^{(i)}(\bm{r}) \equiv
\sum_{\alpha=1}^{N_{\rm p}} \delta(\bm{r} - \bm{R}_{i,\alpha})$ 
is the position ($\bm{r}$)-dependent bead center density. Since
the correlation function
$G^{(i)}(|\bm{r}-\bm{r}'|)$ $=$
$\langle \hat{\rho}_{\rm c}^{(i)}(\bm{r})\hat{\rho}_{\rm c}^{(i)}(\bm{r}') 
\rangle$ 
that depends on the relative distance $|\bm{r}-\bm{r}'|$ 
is amenable to FTS,\cite{Pal2021}
\begin{equation}
\label{eq:Rg_defn_FTS}
{R_{\rm g}^{(i)}}^2 
= \frac{V}{2N_{\rm p}^2} \int d\bm{r} G^{(i)}(|\bm{r}|) |\bm{r}|^2 \; ,
\end{equation}
where system volume $V=\int d\bm{r}$, can now be computed by
FTS.
Similarly, with a generalized correlation 
$G_{\alpha,\beta}^{(i),(j)}(|\bm{r}|)$
between the $\alpha$th bead of the $i$th chain and the $\beta$th bead
of the $j$th chain,
\begin{equation}
\omega_{\alpha,\beta}^{i,j}= V \int_{0}^{2b} d\bm{r} \;
G_{\alpha,\beta}^{i,j}(|\bm{r}|)
\end{equation}
is seen as the frequency of contact between the two beads, i.e., when
their centers are within a small distance (chosen here as $2b$ where
$b$ is the reference bond length between sequentially consecutive beads). 
Thus, through appropriate choices of $i$ and $j$, intrachain contacts of
an isolated chain as well as intrachain and interchain contacts in the
condensed phase (Fig.~1c) can be computed by FTS via 
$G_{\alpha,\beta}^{(i),(j)}(|\bm{r}|)$.
A derivation of this formulation 
based on the general FTS approach\cite{Fredrickson2006,FredricksonGanesanDrolet2002,RigglemanRajeevFredrickson2012,MiMB2023}
is provided in the Supporting Information.
\\

\noindent
As examples, Fig.~1d,e show the RPA, FTS, and MD phase diagrams for
six sequences. Phase diagrams for all 26 sequences 
in Fig.~1b are provided in Fig.~S2 of 
the Supporting Information. To facilitate comparisons,
temperatures are given as reduced temperature 
$T^*\equiv b/\lb$ where $\lb$ is Bjerrum length. Because of the 
models' different effective energy scales arising
from various approximations and treatments of 
excluded volume, the critical temperatures, $T^*_{\rm cr}$s,
predicted by RPA, FTS, and MD can be substantially 
different for the same sequence 
(Fig.~S1b in Supporting Information). Nonetheless, the variation
in $T^*_{\rm cr}$ predicted by the models are well 
correlated (Fig.~1f, Pearson correlation
coefficients $r\gtrsim 0.95$), indicating that the models are capturing 
essentially the same sequence-dependent trend of LLPS propensity.
As to the relationship between $T^*_{\rm cr}$ and $T^*$-dependent
root-mean-square isolated-chain $R_{\rm g}$, a $T^*$ sufficiently high for
a large sequence-dependent variation in $R_{\rm g}$ is chosen for each 
of the models in Fig.~1g.
Consistent with an earlier study on sv sequences,\cite{lin2017}
$R_{\rm g}$s of isolated polyampholytes are well 
correlated with their $T^*_{\rm cr}$s in all three models (Fig.~1g).
Notably, however, there is an appreciable $R_{\rm g}$--$T^*_{\rm cr}$
scatter in the MD model involving the cSCD, c$\kappa$, obs1, and ebs1 
sequences which as a group is less conformative 
to the moderate $-$SCD--$\kappa$ correlation than the sv sequences (Fig.~1a).
\\


\noindent
The impact of sequence-local versus nonlocal charge pattern on $R_{\rm g}$ 
and $T^*_{\rm cr}$ is assessed by comparing the extent to which
they are (anti)correlated with $\kappa$ and SCD (Fig.~2).
$R_{\rm g}$ depends on $T^*$. Since the charge-pattern-dependent variation
in $R_{\rm g}$ among sequences with moderate to high $-$SCD values 
is small at low $T^*$ as they adopt conformations with 
similarly high compactness, two $T^*$s are chosen for each of the FTS 
and MD models in Fig.~2a,b,e,f with the higher $T^*$ producing ample
$R_{\rm g}$ variations across the entire ranges of $-$SCD and $\kappa$.
Corresponding $R_{\rm g}$ data for additional $T^*$s are provided in Fig.~S3 
of Supporting Information.
\\

\noindent
It is clear from Fig.~2a,b,e,f that $R_{\rm g}$ anticorrelates
significantly better with $-$SCD than $\kappa$.
For the sv sequences, $R_{\rm g}$ anticorrelates reasonably well with
both $-$SCD and $\kappa$. Indeed, the significant $\kappa$--$R_{\rm g}$
scatter seen in both FTS and MD (Fig.~2e,f) involves the as, cSCD, and
c$\kappa$ sequences we introduced. 
Despite the large variations in $\kappa$ among the as and cSCD 
sequences, their $R_{\rm g}$s are very similar.
For the c$\kappa$ sequences, despite their essentially identical $\kappa$, 
their $R_{\rm g}$s are very different.
By comparison, the excellent $-$SCD--$R_{\rm g}$ anticorrelation is maintained
when challenged by these sequences (Fig.~1a,b).
In this light, the good $\kappa$--$R_{\rm g}$ anticorrelation observed 
previously for the sv sequences\cite{rohit2013} is largely attributable 
to the good correlation between the $\kappa$ values of this particular 
set of sv sequences and their $-$SCDs.
\\

\noindent
In contrast to $\kappa$ and SCD's clearly different performance 
for $R_{\rm g}$ (which is the original target of both 
parameters\cite{rohit2013,kings2015}, neither of these parameters was
derived originally for multiple-chain properties), their correlations
with $T^*_{\rm cr}$ are more comparable (Fig.~2c,d,g,h).
Unlike the excellent $-$SCD--$R_{\rm g}$ anticorrelation,
both the $-$SCD--$T^*_{\rm cr}$ and $\kappa$--$T^*_{\rm cr}$
correlations are good but not excellent. 
While $-$SCD is better than $\kappa$ with $T^*_{\rm cr}$ 
for RPA and FTS (Fig.~2c,g), $\kappa$ is slightly better 
for MD (Fig.~2d,h). Notably, for all 
three models---RPA, FTS, and MD---the variation in $T^*_{\rm cr}$ is 
larger among the c$\kappa$ than among the cSCD sequences.
\\

\noindent
The origin of the performances of $-$SCD and $\kappa$ for $R_{\rm g}$ is
explored by first considering how $R_{\rm g}$ is related to
intrachain contacts for a homopolymer with favorable short-spatial-range
interactions (model described in Supporting Information). 
In this baseline model, $R_{\rm g}$s conditioned upon
pairwise $\alpha,\beta$ 
contacts (Fig.~3a, top map, upper triangle) indicate
that more nonlocal (higher-order) contacts lead to 
smaller $R_{\rm g}$. Sequence-local and nonlocal
interactions thus have different implications for $R_{\rm g}$. This basic 
observation offers a semi-quantitative rationalization for SCD's
better performance with regard to $R_{\rm g}$ because,
as entailed by the original approximate analytical theory,\cite{kings2015}
nonlocal electrostatic interactions (larger $|\alpha - \beta|$) 
are more heavily 
weighted in SCD than local interactions (smaller $|\alpha - \beta|$).
\\

\noindent
Additional information is provided by contact patterns (Fig.~3a--d).
The intrachain contact frequencies $P^{[{\rm i}]}_{\alpha,\beta}$ and 
$P^{[{\rm c}]}_{\alpha,\beta}$ are the average numbers of
contacts between the $\alpha$th and $\beta$th beads, respectively, of an 
isolated chain (at infinite dilution) and a chain in the condensed phase, 
whereas the interchain $C^{[{\rm c}]}_{\alpha,\beta}$ is the average number 
of contacts between the $\alpha$th bead of one chain and the $\beta$th 
bead of another chain. 
By definition,
$P^{[{\rm i}]}_{\alpha,\beta}=P^{[{\rm i}]}_{\beta,\alpha}$,
$P^{[{\rm c}]}_{\alpha,\beta}=P^{[{\rm c}]}_{\beta,\alpha}$, and
$C^{[{\rm c}]}_{\alpha,\beta}=C^{[{\rm c}]}_{\beta,\alpha}$.
For the baseline homopolymer,
the isolated-chain $P^{[{\rm i}]}_{\alpha,\beta}$'s 
pattern is typical of Gaussian or self avoiding walk
conformations, with substantially lower frequencies for higher order 
contacts\cite{ChanDill1990} (Fig.~3a, top map, lower triangle). 
The condensed-phase intrachain pattern 
(Fig.~3a, bottom map, lower triangle) shares a similar trend but 
with less variation in contact frequency
(cf. heat map scales
for $-\ln P^{[{\rm i}]}_{\alpha,\beta}$ 
and $-\ln P^{[{\rm c}]}_{\alpha,\beta}$ in Fig.~3a).
In contrast, the interchain 
$C^{[{\rm c}]}_{\alpha,\beta}$ is quite insensitive to $\alpha,\beta$ 
except it is slightly higher when either $\alpha$, $\beta$, or both, 
are at or near the chain ends (Fig.~3a, bottom map, upper triangle).
\\

\noindent
Polyampholyte contact data are illustrated here by an example sequence.
The MD (Fig.~3b) and FTS (Fig.~3c) patterns are quite well 
correlated, their differences likely arise from the differing 
treatments of excluded volume in the two approaches.\cite{joanJPCL2019,Pal2021}
Data for the other 25 
sequences in Fig.~1b are in Fig.~S4 of Supporting Information.
In Fig.~3b, the $R_{\rm g}$ map exhibits sequence-specific 
features as well as a contact-order dependence (top map, upper triangle)
indicating differential sequence-local versus nonlocal 
effects on $R_{\rm g}$. Similar to the homopolymer (Fig.~3a), 
the intrachain pattern of an isolated polyampholyte 
(Fig.~3b,c top map, lower triangle) is similar to that of a polyampholyte 
in the condensed phase (Fig.~3b,c bottom map, lower triangle). 
Unlike the homopolymer, the isolated-chain intrachain 
pattern (Fig.~3b,c top map, lower triangle) is similar also to the 
condensed-phase interchain pattern (Fig.~3b,c bottom map, upper triangle).
This feature, which is echoed by the comparison in Fig.~3e below, applies to
the other 25 sequences as well. 
\\

\noindent
Averages of $P^{[{\rm i}]}_{\alpha,\beta}$, 
$P^{[{\rm c}]}_{\alpha,\beta}$, or $C^{[{\rm c}]}_{\alpha,\beta}$ 
for a given $|\alpha - \beta|$ are illustrated here using
the homopolymer model and the example sequence (Fig.~3d,e). 
Similar salient features are exhibited by the other 25 sequences (Fig.~S5 of
Supporting Information). Condensed-phase interchain 
$C^{[{\rm c}]}_{\alpha,\beta}$ is least sensitive to 
$|\alpha - \beta|$ for the homopolymer (Fig.~3d) and for 
polyampholytes (Fig.~3e). Notably,
the polyampholyte interchain $C^{[{\rm c}]}_{\alpha,\beta}$ (orange curve)  
is closer to the isolated-chain intrachain $P^{[{\rm i}]}_{\alpha,\beta}$ 
(green curve) than to the condensed-phase intrachain 
$P^{[{\rm c}]}_{\alpha,\beta}$ (blue curve). 
The root-mean-square distance $R_{\alpha\beta}$ between beads $\alpha$
and $\beta$ is highly sensitive to sequence\cite{rohit2013} and 
temperature for an isolated polyampholyte
(Fig.~3f) as its $R_{\rm g}$ decreases at low $T^*$.
In contrast, Fig.~3g shows that $R_{\alpha\beta}$ depends only weakly 
on $T^*$. Depending on the sequence, $R_{\rm g}$ can increase or decrease
slightly with $T^*$ (Fig.~S6 of Supporting Information). 
As in homopolymer melts,\cite{Flory1949} condensed-phase polyampholytes
adopt open, essentially Gaussian-like conformations 
($R_{\alpha\beta}/\sqrt{|\alpha - \beta|}\sim$ constant except for
small $|\alpha - \beta|$ in Fig.~3g), a
phenomenon also seen in recent simulations of biomolecular 
condensates.\cite{MittalACS2023,PappuNatComm2023,Johnson2024,Mittal2024} 
\\

\noindent
To gain further insights, we compare sequence-specific contact patterns 
such as those in Fig.~3b by the following symmetrized form of 
the Kullback-Leibler divergence\cite{KL} between contact frequencies
$\{C^{(s)}_{\alpha\beta}\}$ and $\{C^{(s^\prime)}_{\alpha\beta}\}$
(contact maps) of a pair of sequences $s,s^\prime$:
\begin{equation}
D_{s,s^\prime}=
\sum_{\alpha=1}^{N_{\rm p}-d_{\rm m}}\sum_{\beta=\alpha + d_{\rm m}}^{N_{\rm p}}
\Bigl[ c^{(s)}_{\alpha\beta} - c^{(s^\prime)}_{\alpha\beta} \Bigr]
\ln \left [
c^{(s)}_{\alpha\beta}/c^{(s^\prime)}_{\alpha\beta}
\right ] \; ,
\end{equation}
where $d_{\rm m}$ serves to exclude local contacts---which are often 
highly populated---from overwhelming the contact pattern's quantitative 
characterization,
and $c^{(s)}_{\alpha\beta}\equiv C^{(s)}_{\alpha\beta}/\sum_{\alpha^\prime=1}^{N_{\rm p}-d_{\rm m}}\sum_{\beta^\prime=\alpha + d_{\rm m}}^{N_{\rm p}}
C^{(s)}_{\alpha^\prime\beta^\prime}$ are normalized frequencies.
The comparisons of $P^{[{\rm i}]}_{\alpha,\beta}$
with $P^{[{\rm c}]}_{\alpha,\beta}$ (Fig.~4a) and with
$C^{[{\rm c}]}_{\alpha,\beta}$ (Fig.~4b) entail significant scatter.
Nonetheless, the relatively low $D_{s,s^\prime}$ values for $s=s^\prime$
(black circles) in Fig.~4b is consistent with the impression 
from Fig.~3b,c that the patterns of isolated-chain intrachain 
and condensed-phase interchain contacts are similar for a given polyampholyte.
This trend prevails for several other $d_{\rm m}$ values
and another isolated-chain $T^*_{\rm iso.}$ (Fig.~S7 of Supporting Information),
underpinning correlations between sequence-dependent isolated-chain
and condensed-phase properties. 
\\

\noindent
The relationship between isolated-chain and condensed-phase properties
is further elucidated by examining the total number of contacts made by 
a polyampholyte. Whereas the number of intrachain contacts
of an isolated chain correlates poorly with that of a condensed-phase
chain (Fig.~4c) because of their significantly
different $R_{\rm g}$s (see above), an excellent correlation is seen
between the number of intrachain contacts of an isolated chain with
the number of interchain contacts in the condensed phase (Fig.~4d),
echoed by the excellent correlation between isolated-chain
and condensed-phase potential energies (Fig.~4e).
Similar trends are seen for other values of $T^*_{\rm iso.}$ 
(Fig.~S8 and Fig.~S9a of Supporting Information).
$T^*_{\rm cr}$ anticorrelates reasonably well with isolated-chain potential 
energy $E^{[{\rm i}]}$ (Fig.~4f), which expectedly correlates with
the number of interchain contacts (Fig.~S9b in Supporting Information).
Notably, $E^{[{\rm i}]}$, in turn, anticorrelates quite well with $\kappa$
(Fig.~4h) but not so well with $-$SCD (Fig.~4g), indicating that in some 
situations $\kappa$ can be a better predictor of LLPS propensity, suggesting 
that differential effects of sequence-local versus nonlocal interactions 
may be less prominent for LLPS than for isolated-chain $R_{\rm g}$. 
This understanding is underscored by the fitting coefficients 
$c_\kappa$ for $\kappa$ and $c_{\rm SCD}$ for SCD in Fig.~4i,j 
indicating that variation in 
$R_{\rm g}$ can be accounted for essentially entirely by SCD 
($c_\kappa\approx 0$) 
but variation in $T^*_{\rm cr}$ is rationalized approximately equally 
by $\kappa$ and SCD ($c_\kappa\approx c_{\rm SCD}$).
$\kappa$ is not a good general predictor for $R_{\rm g}$ though
$\kappa$ correlates well with $E^{[{\rm i}]}$ because
$R_{\rm g}$ is not determined solely by $E^{[{\rm i}]}$. 
For instance, two chains each constrained by one contact with 
the same energy but different contact orders 
can have very different $R_{\rm g}$s. 
\\

\noindent
To recapitulate, many IDPs can exist in dilute and condensed phases serving
different biological functions. Single-chain IDP $R_{\rm g}$s in dilute
solutions, readily accessible experimentally,\cite{SongSAXS2021}
have been used to benchmark MD potentials for 
sequence-dependent IDP properties.\cite{KrestenNat2024,Alex2024}
Isolated-chain contact maps,\cite{Zheng2023} related topological
constructs such as SCDM\cite{HuihuiKingsBJ2020} and energy 
maps,\cite{Devarajan2022} and their relations with condensed-phase 
interchain contact maps\cite{Kresten2021,bauer2022} have proven useful 
in recent computational analyses.
For instance, isolated-chain intrachain and condensed-phase 
interchain contacts are similar for heterochromatin protein 1 
paralogs\cite{Phan2023} and EWS sequences\cite{Johnson2024} but not 
the TDP-43 C-terminal domain.\cite{mohanty2023}
Here, our findings indicate a fundamental divergence in the differential 
impact of local versus nonlocal sequence patterns on isolated, 
single-chain and condensed-phase multiple-chain properties. 
The differential impact is prominent for isolated-chain conformational
dimensions due to chain connectivity.\cite{ChanDill1990} 
It is substantially less 
for LLPS propensity because
the multiple-chain nature of condensed-phase interactions 
dampens---though not entirely abolish---the effects of contour separations
between residues along a single connected sequence due to the immense
number of configurations in which residues from different chains may interact.
This is a fundamental factor in the dilute/condensed-phase relationship
of IDPs that needs to be taken into account when devising improved 
sequence pattern parameters for the characterization of physical
and functional IDP molecular features.\cite{Sabari2023,Moses2021}
Inasmuch as such parameters' aim is instant estimation of LLPS 
propensity, theoretical quantities 
that can be numerically computed efficiently---such as the $T^*_{\rm cr}$ 
predicted by RPA-related theories\cite{WessenDasPalChan2022}---may just 
serve the purpose practically even if they are not closed-form 
mathematical expressions. 
\\


\noindent
{\large\bf Supporting Information}\\
Methodological and formulational details, supporting table, and 
supporting figures
\\

\noindent
{\large\bf Acknowledgements.}
This work was supported by Canadian Institutes of Health Research (CIHR) grant
NJT-155930 and Natural Sciences and Engineering Research Council of
Canada (NSERC) grant RGPIN-2018-04351 to H.S.C.
We are grateful for the computational resources provided generously 
by the Digital Research Alliance of Canada.


\vfill\eject
%
%



\setcounter{figure}{0}
\renewcommand{\figurename}{{\bf Figure}}
\renewcommand{\thefigure}{{\bf \arabic{figure}}}

\centerline{\large\bf Figures (Main Text)}


\begin{figure*}[ht]
\vskip -0.3cm
{\includegraphics[width=0.95\columnwidth,angle=0]{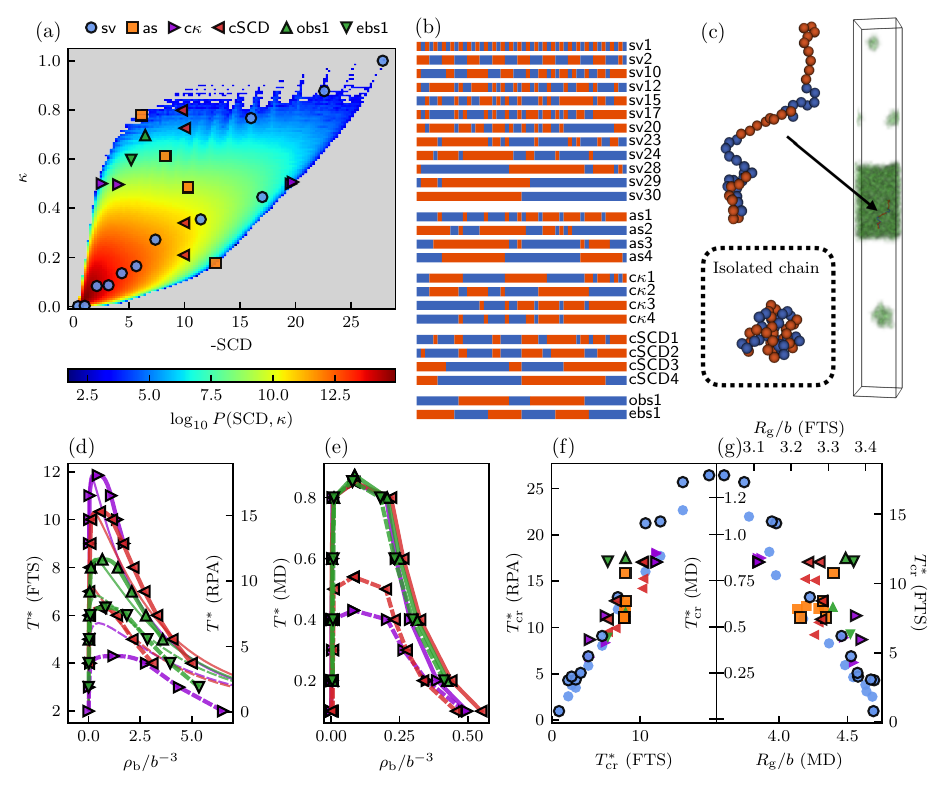}}
\vskip -0.4cm
\caption{Polyampholytes with representative variations in local and 
nonlocal sequence charge patterns.
(a) The SCD-$\kappa$ sequence-space population distribution 
of 50-bead polyampholytes is depicted by a heat map 
(bottom scale, region with zero population is in gray), wherein the
26 sequences studied (b) are marked by symbols (a, top).  
(b) The positive (blue) and negative (red) charge patterns of these sequences.
(c) Coarse-grained MD snapshots showing a conformation in the condensed 
phase (top) and an isolated conformation (dotted box). 
(d,e) Phase diagrams in RPA (thin continuous curves), FTS (symbols) (d),
and MD (symbols) (e) for sequences c$\kappa$1, c$\kappa$4, cSCD1, cSCD4, 
ebs1, and obs1.
Thick solid and dashed lines connecting symbols in (d) and (e) are merely 
guides for the eye.
(f) $T^*_{\rm cr}$s of all 26 sequences predicted by different theories.
The $T^*_{\rm cr}$--$T^*_{\rm cr}$ Pearson correlation coefficients for
RPA-FTS, RPA-MD, and FTS-MD are, respectively,
$r=0.997$, $0.945$, and $0.948$.
Data symbols in (f) and (g) involving MD are identified by black edges. 
(g) Correlation between $T^*_{\rm cr}$ and $R_{\rm g}$ 
in FTS (at $T^*=20$, $r=-0.954$) or in MD (at $T^*=10$, $r=-0.842$).
A complete list of $T^*_{\rm cr}$s and phase diagrams for all sequences in (b)
predicted by the RPA, FTS, and MD models are provided in Figs.~S1 and S2 of the 
Supporting Information.
}
\label{fig1}
\end{figure*}

\vfill\eject

\begin{figure*}[ht]
{\includegraphics[width=0.5\columnwidth,angle=0]{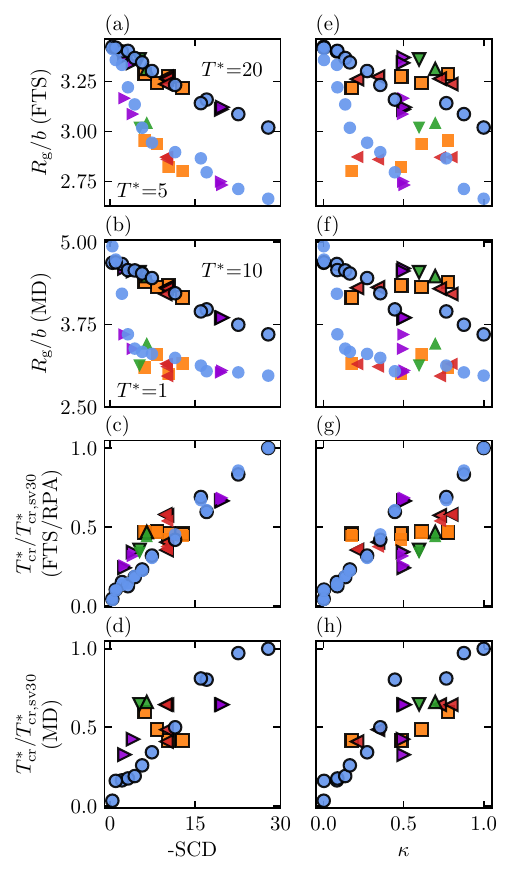}}
\caption{SCD and $\kappa$ as predictors for isolated-chain $R_{\rm g}$ 
and multiple-chain LLPS critical temperature $T^*_{\rm cr}$.
Symbols for polyampholytes are those in Fig.~1a.
(a,b,e,f) $R_{\rm g}$ computed by FTS (a,e) and MD (b,f) versus
$-$SCD (left) or $\kappa$ (right). Symbols with and without black 
edges are for different temperatures ($T^*$s).
$-$SCD--$R_{\rm g}$ correlation coefficients are $r=-0.983$ for FTS 
at $T^*=20$ (a) and $r=-0.991$ for MD at $T^*=10$ (b), the corresponding
$\kappa$--$R_{\rm g}$ values are $r=-0.665$ (e) and $-0.627$ (f).
(c,d,g,h) $T^*_{\rm cr}$ in FTS, RPA [depicted, respectively, 
in (c,g) by symbols 
with and without black edges] and MD (d,h) are plotted in units of the 
$T^*_{\rm cr}$ for sequence sv30 ($T^*_{\rm cr,sv30}$). The 
$-$SCD--$T^*_{\rm cr}$
$r$ values are: $0.943$ for RPA, $0.932$ for FTS (c), 
and $0.834$ for MD (d). The corresponding $\kappa$--$T^*_{\rm cr}$
values are $r=0.827$ for RPA, $0.841$ for FTS (g), and $0.886$ for MD (h).
}
\label{fig2}
\end{figure*}

\clearpage

\vfill\eject

\begin{figure*}[ht]
{\includegraphics[width=0.95\columnwidth,angle=0]{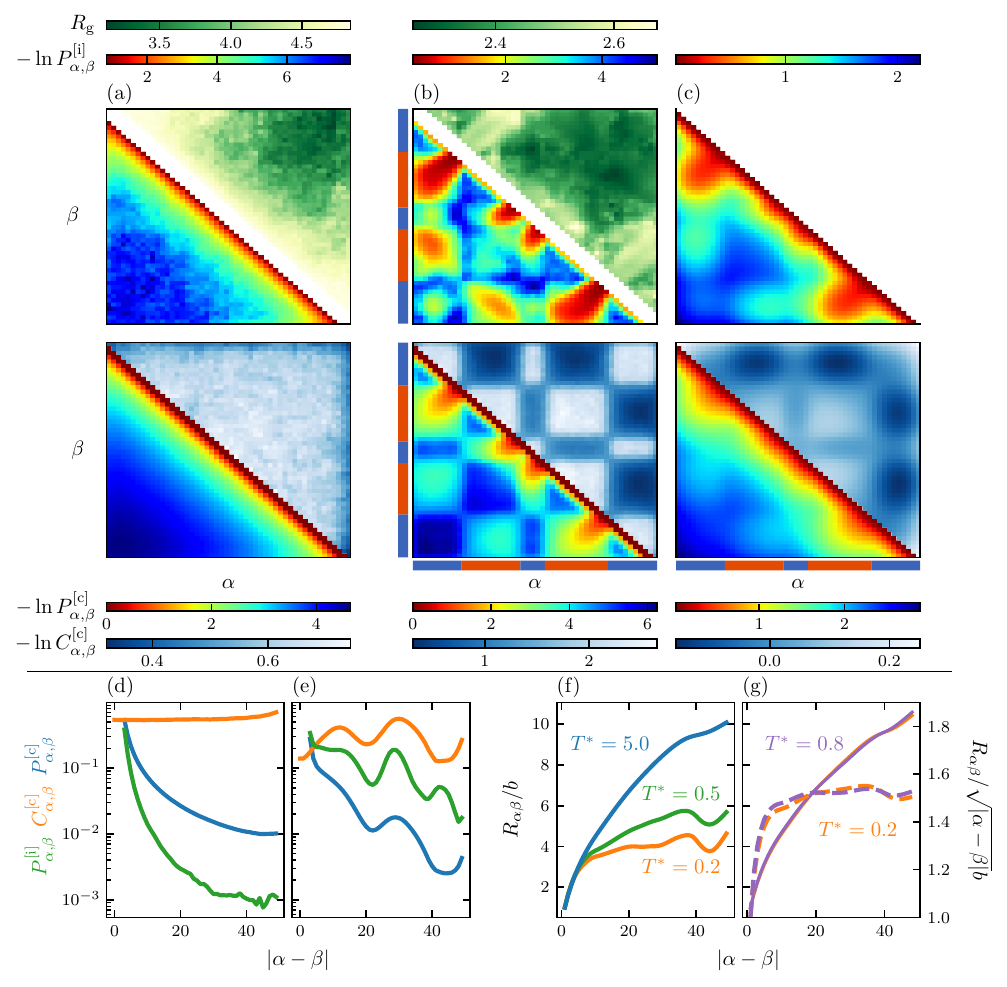}}
\caption{Contact patterns of polyampholytes as an isolated
chain versus in the phase-separated condensed phase.
(a--c) Top heat maps (color scales above) show isolated-chain contact 
frequencies (lower triangle) and $R_{\rm g}$ conditional upon
intrachain $i,j$ contacts (upper triangle); 
bottom maps (color scales below) show condensed-phase intrachain 
(lower triangle) and interchain contact frequencies (upper triangle);
$\alpha,\beta=1,2,\dots,50$ are bead labels along the chain.
Contact frequencies $P^{[{\rm i}]}_{\alpha,\beta}$, 
$P^{[{\rm c}]}_{\alpha,\beta}$, 
and $C^{[{\rm c}]}_{\alpha,\beta}$ are defined in the text.
Data are shown for (a) the baseline homopolymer
(MD, $T^*=2.0$) and (b) the sequence obs1 
(MD, $T^*=0.2$). The charge pattern of obs1 (as in Fig.~1b) 
is depicted along the axes in (b). 
(c) Corresponding FTS contact maps for obs1 at $T^*=3.0$. 
(d,e) Contact frequencies (color coded as indicated) in MD as functions of
contact order $|\alpha - \beta|$ for 
(d) the baseline homopolymer ($T^*=2.0$) and 
(e) obs1 ($T^*=0.2$).
(f,g) MD-simulated obs1 intrachain root-mean-square distance $R_{\alpha\beta}$
(solid curves, left vertical scale) at select temperatures for an isolated 
chain (f) and in the condensed phase (g); to facilitate analysis (see text),
$R_{\alpha\beta}/\sqrt{|\alpha - \beta|}$ plots 
(dashed curves, right vertical scale)
are also provided in (g).
}
\label{fig3}
\end{figure*}

\clearpage

\vfill\eject

\begin{figure*}[ht]
{\includegraphics[width=0.95\columnwidth,angle=0]{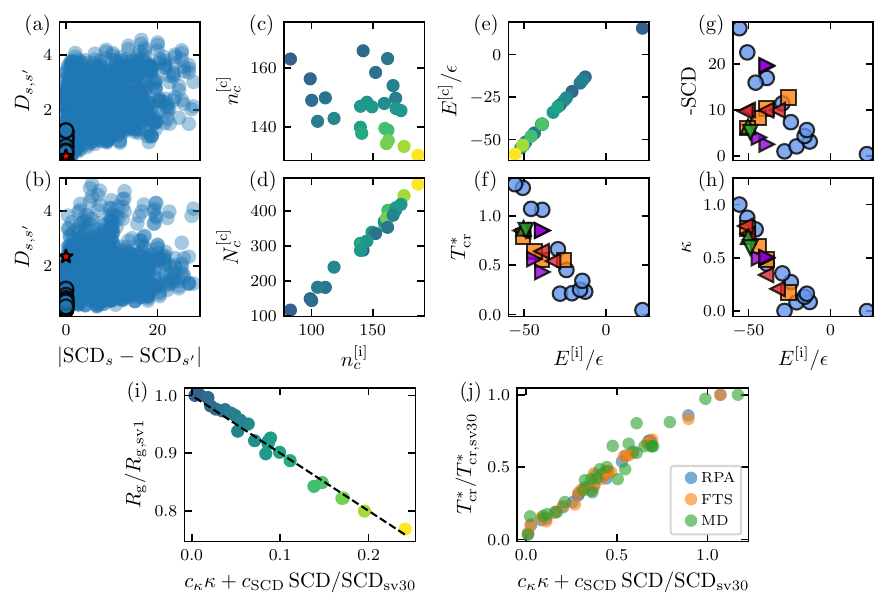}}
\caption{Impact of local versus nonlocal sequence charge 
pattern on polyampholyte contacts and interaction energies in
the MD model.
(a,b) $D_{s,s'}$ is symmetrized Kullback-Leibler divergence between the 
normalized $|\alpha - \beta|\geq 4$ 
contact frequencies of sequences $s$ and $s^\prime$
(see text).  The scatter plots show $D_{s,s'}$ 
versus $s$--$s^\prime$ SCD difference. Each datapoint (translucent
blue circle) provides the $D_{s,s'}$ between the intrachain contacts of 
an isolated chain ($s$) and (a) the intrachain or (b) interchain contacts 
of a condensed-phase chain ($s^\prime$) 
where $s,s^\prime=$ 25 of the sequences in Fig.~1b (all except sv1),
with $T^*_{\rm iso.}=0.2$, $0.1\le T^*_{\rm cond.}\le 0.3$, 
and $s=s^\prime$ datapoints 
as black circles. Red stars are $D_{s,s'}$ values for the 
baseline homopolymer at $T^*=2.0$.
(c,d) Scatter plots of number of intrachain contacts of isolated chains 
$n_c^{[{\rm i}]}$ with number of (c) intrachain contacts $n_c^{[{\rm c}]}$ 
($r=-0.396$) or (d) interchain contacts $N_c^{[{\rm c}]}$ in the condensed 
phase ($r=0.992$) for the 25 sequences analyzed in (a,b). 
(e--h) Scatter plots of isolated-chain potential energy $E^{[{\rm i}]}$
(in units of $\epsilon$, for all 26 sequences) with (e) condensed-phase 
potential energy $E^{[{\rm c}]}$ ($r=0.998$), (f) $T^*_{\rm cr}$ ($r=-0.808$), 
(g) SCD ($r=0.505$), or (h) $\kappa$ ($r=-0.855$).
Datapoints for different sequences in (c--e,i) are color-coded for SCD, 
with lighter color corresponding to larger $-$SCD (as in Figs.~S8 and S9). 
Symbols in (f--h) corresponds to those in Fig.~1a.
All quantities in (c--h) are computed for $T^*=0.2$ except a lower
$T^*=0.04$ sufficient for LLPS of sv1 is used to obtain $E^{[{\rm c}]}$ 
for sv1 in (e).
(i,j) Best-fit linear combinations of $\kappa$ and SCD as predictors
for (i) $R_{\rm g}$ (in units of $R_{\rm g}$ of sv1, $R_{\rm g,sv1}$, at
$T^*=10$) and (j) $T^*_{\rm cr}$ for all sequences in Fig.~1a. 
(i) The dashed line is $R_{\rm g}$ $=$ 
$1-(c_{\kappa}\kappa + c_{\rm SCD}{\rm SCD}/{\rm SCD}_{\rm sv30})$
with $c_\kappa=-0.0017$, $c_{\rm SCD}=0.243$ ($r=-0.991$),
and SCD$_{\rm sv30}$ is the SCD of sv30.
(j) The best-fit parameters are $a_\kappa=0.39$, $c_{\rm SCD}=0.68$ 
for RPA ($r=0.996$), $c_\kappa=0.42$, $c_{\rm SCD}=0.65$ for FTS ($r=0.994$), 
and $c_\kappa=0.63$, $c_{\rm SCD}=0.54$ for MD ($r=0.962$). 
}
\label{fig4}
\end{figure*}

\clearpage

\vfill\eject
 
\vfill\eject
$\null$ \hfill {\bf July 4, 2024}
\vskip 1.0in

\begin{center}

{\Huge\bf Supporting Information}\\

\vskip 0.3cm

{\Large\it for}

\vskip 0.3cm

{\Large\bf 
Differential Effects of Sequence-Local versus Nonlocal}\\

\vskip 0.3cm

{\Large\bf 
Charge Patterns on Phase Separation and}\\

\vskip 0.3cm

{\Large\bf Conformational Dimensions of Polyampholytes as}\\

\vskip 0.3cm

{\Large\bf Model Intrinsically Disordered Proteins}\\

\vskip .5in
{\bf Tanmoy P{\footnotesize{\bf{AL}}}},$^{1,\dagger}$
{\bf Jonas W{\footnotesize{\bf{ESS\'EN}}}},$^{1,\dagger}$
{\bf Suman D{\footnotesize{\bf{AS}}}},$^{1,2,\dagger}$
and
{\bf Hue Sun C{\footnotesize{\bf{HAN}}}}$^{1,*}$

$\null$

$^1$Department of Biochemistry,
University of Toronto, Toronto, Ontario M5S 1A8, Canada\\
$^2$Department of Chemistry, Gandhi Institute of Technology and
Management, Visakhapatnam, Andhra Pradesh 530045, India\\

%
\end{center}
$\null$\\

\noindent
$^\dagger$Contributed equally.

\vskip 1.3cm

\noindent
$^*$Correspondence information:\\
{\phantom{$^\dagger$}}
Hue Sun C{\footnotesize{HAN}}.$\quad$
E-mail: {\tt huesun.chan@utoronto.ca}\\
{\phantom{$^\dagger$}}
Tel: (416)978-2697; Fax: (416)978-8548\\
{\phantom{$^\dagger$}}
Department of Biochemistry, University of Toronto,
Medical Sciences Building -- 5th Fl.,\\
{\phantom{$^\dagger$}}
1 King's College Circle, Toronto, Ontario M5S 1A8, Canada.\\

\vfill\eject
\renewcommand{\theequation}{{\rm S}\arabic{equation}}
\setcounter{equation}{0}

\centerline{\Large\bf Methodological and Formulational Details}

$\null$

\noindent
{\large\bf Estimating the sequence-space distribution $P(\scd,\kappa)$} 

As outlined in the maintext, we investigate the relationship between 
the $\scd$ and $\kappa$ parameters and its ramifications on isolated-chain
and condensed-phase conformational properties by first
calculating their distribution $P(\scd,\kappa)$ for electrically
overall neutral 50mer polyampholytes (sequences with 50 beads). Following
convention in an earlier work that used lysine (K) for the positively 
charged ($+1$) and glutamic acid (E) for the negatively charged ($-1$) 
monomers/residues along the chain sequence,\cite{rohit2013} 
we refer to the sequences under consideration as K/E sequence in the 
following discussion, keeping in mind, however, that K and E here refer 
only to $+1$ and $-1$ polymer beads with no sidechain structure, as in
several previous simplified models.\cite{kings2015,lin2017,joanJPCL2019,suman2}
Values for $\kappa$ are computed in the present work using the 
$\kappa=(\kappa_5 +\kappa_6)/2$ expression defined
in ref.~\citen{suman2}.

The $P(\scd,\kappa)$ distribution is defined in such a way that
$P(\scd',\kappa') \, \Delta \scd \, \Delta \kappa$ is the number of possible
sequences with $(\scd,\kappa)$ values in a small region $\Delta \scd \, \Delta
\kappa $ (two-dimensional bin) 
in the vicinity of $(\scd',\kappa')$. We compute $P(\scd,\kappa)$
using the Wang-Landau (WL) algorithm,\cite{WangLandau2001} which is a highly
efficient flat-histogram method for estimating multi-peaked densities of
states. However, before setting up the WL algorithm, prior knowledge about the
the mathematically possible $(\kappa,\scd)$ combinations is required since the
WL algorithm relies on checking that all $(\kappa,\scd)$ bins are evenly
sampled.

To find the region in the $(\scd,\kappa)$-plane that can support 50mer
K/E sequences, we utilize a genetic algorithm (GA) that is capable of
scanning the space of sequences efficiently. Specifically, this GA takes an 
input target point $(\scd_{\rm target},\kappa_{\rm target})$ in sequence
space and attempts to generate a new sequence
$\{\sigma_{\alpha}\}$ ($\alpha=1,2,\dots,N$, where the chain length $N=50$) 
with $(\scd,\kappa)$ that maximizes the fitness function
\begin{equation} \label{eq:GA_fitness_function}
f(\{\sigma_{\alpha}\}) 
= -(\kappa - \kappa_{\rm target})^2 - 0.001(\scd - \scd_{\rm target})^2 \; .
\end{equation}
For a given target $(\scd_{\rm target},\kappa_{\rm target})$, the 
algorithm proceeds as follows:
\begin{enumerate}
\item Generate an initial ``population'' 
of $n_{\rm population}=100$ random electrically neutral K/E sequences
(each with 25 K and 25 E).
\item Each sequence in the population is used to generate 
$n_{\rm offspring}=50$ ``offspring'' sequences. Every offspring sequence 
is generated from its parent sequence using one of the following prescriptions:
\begin{itemize}
\item \textit{Single flip {\rm (}50\%{\rm )}}: A randomly selected pair of K/E
residues are interchanged.  
\item \textit{Cluster move {\rm (}45\%{\rm )}}: A randomly selected block 
of same-charge residues is moved one step (either left or right chosen 
at random), e.g., ...EEEEEKKKK... $\rightarrow$ ...KEEEEEKKK... . As such,
this is a special case of single flip. It serves to enhance sampling of blocky
sequences.
\item \textit{New sequence {\rm (}5\%{\rm )}}: An entirely new electrically 
neutral sequence with no relation to the parent sequence is randomly generated. 
\end{itemize}
The three prescriptions are chosen at random with probabilities provided
in parentheses above. We do not use crossover to generate offspring 
sequences (i.e.~where multiple parent sequences are combined to generate 
offspring) as initial exploratory runs did not reveal any major advantage 
when using crossover.
\item We next select $n_{\rm survivors}=30$ out the $n_{\rm population} \cdot
n_{\rm offspring}$ offspring sequences using a diversity enhanced survivor
selection algorithm. First, we evaluate the fitness $f_i$ of each offspring
sequence $\{\sigma_\alpha^{(i)}\}$, 
with $i=1,\dots,n_{\rm population} \cdot n_{\rm
offspring}$, according to Eq.~\eqref{eq:GA_fitness_function}, and select the
sequence corresponding to the highest (least negative) 
$f_i$ as the first survivor. The fitness
of the remaining sequences $\{\sigma_{\rm \alpha}^{(j\neq i)}\}$ 
are then modified
as $f_j \rightarrow f_j - \e^{-10 d_{ij}}$ where $d_{ij} =
\sum_{\alpha=1}^N\left(\sigma_{\alpha}^{(i)} - \sigma_{\alpha}^{(j)}
\right)^2/4N$. This step punishes sequences that are identical or
near-identical to the first selected survivor. The next survivor $i'$ is then
selected based on the updated fitnesses and the remaining sequences acquires
new punishments depending on their similarity with $i'$. This procedure is
repeated until $n_{\rm survivors}$ have been selected.  
\item The survivors now
constitute the population of the next generation and steps 2--3 are iterated
until either a sequence with $f_i \geq -10^{-5}$ has been found 
or the 10th generation of iterations is reached.
\end{enumerate}
The diversity enhanced survivor selection procedure prevents the algorithm from
being trapped in the vicinity of a local fitness maximum since sequences are
selected both according to having a high fitness and being dissimilar to other
sequences with high fitness. The method was introduced 
to scan for experimentally viable 
parameter regions in extensions of the Standard Model in particle 
physics,\cite{CamargoMolinaMandal2018} 
but the survivor selection procedure was not outlined in full detail in
ref.~\citen{CamargoMolinaMandal2018}.

We coarse-grain the $(\scd,\kappa)$-plane into $120 \times 120$ bins that cover
the rectangular region $\scd\in [\scd_{\rm sv30},0]$
(i.e., $-\scd\in [0,-\scd_{\rm sv30}]$), $\kappa \in [0,1]$,
where $\scd_{\rm sv30}$ is the SCD value of the diblock sequence sv30 
in ref.~\citen{rohit2013}. The
bin spanning the area  $[-\scd_i,-\scd_i+\Delta \scd]$ and $[\kappa_j, \kappa_j
+ \Delta \kappa]$ is indexed by $(i,j)$ (with $i,j =1,\dots,120$). The GA is
first run for targets $(\scd_{\rm target},\kappa_{\rm target})$ on the sites of
an evenly spaced $7 \times 7$ grid (marked by white dots in the left panel
of Fig.~S1a). Every new bin visited during the scan is
recorded and the associated sequence is stored. With the exception of the first
target iteration, the above step 1 of the GA is modified such that the initial
population is generated as offspring of the previously stored sequence with
nearest $\scd$ and $\kappa$ to the target values instead of being randomly
generated. The initial grid scan gives a rough estimate of $(\scd,\kappa)$
region populated by sequences. Next, we iteratively run the GA for unvisited
target bins adjacent to previously visited bins to map out the boundary of the
region. Fig.~S1a shows in the left panel the result of the GA scan, where bins
for which a sequence was found are shown in black. White bins indicate target
bins for which no sequence was found by the GA. For a few target bins, the GA
was not able to find a sequence but the subsequent WL scan (described below)
revealed sequences populating these bins. These mislabelled bins are shown in
red in in the left panel of Fig.~S1a.

The goal of the WL algorithm is to calculate the quantity $g_{i,j}$
representing the number of sequences with $\scd$ and $\kappa$ in bin $(i,j)$.
To avoid numerical errors associated with large numbers, we (as is customary)
formulate the algorithm in terms of $\ln(g_{i,j})$ rather than $g_{i,j}$.
Another central quantity in the WL algorithm is the histogram $h_{i,j}$ over
the number of visits in bin $(i,j)$. The WL algorithm operates as follows:
\begin{enumerate}
\item Initialize $h_{i,j }= \ln(g_{i,j})=0$ for all $(i,j)$ and set the 
update factor $f=1$.
\item Generate $n=100$ random electrically neutral 50mer K/E sequences, 
$\{\sigma_{\alpha}^{(a)}\}$, $a=1,\dots,n$, $\alpha=1,2,\dots,N$, 
and compute their associated $\scd$ and $\kappa$ values.
\item Perform the updates
\begin{equation}
\begin{aligned}
h_{i,j} &\rightarrow h_{i,j} + 1 , \\
\ln(g_{i,j}) &\rightarrow \ln(g_{i,j}) + f ,
\end{aligned}
\end{equation}
for the bin $(i,j)$ associated with each sequence.
\item Check if $h_{i,j}$ is sufficiently flat. In this work, $h_{i,j}$ is 
deemed flat if
\begin{equation}
 \frac{\langle h^2 \rangle - \langle h \rangle^2 }{\langle h \rangle^2}<0.5 \; ,
\end{equation}
where the averages are taken only over bins known to be populated by sequences.
If the flatness condition is satisfied we reset $h_{i,j} = 0$ for all $i,j$ and
update $f$ as $f \rightarrow f/2$.

\item Propose an update for every sequence $\{\sigma_{\alpha}^{(a)}\}$. 
The updates and their associated probabilities used in this work are
\begin{itemize}
\item \textit{Single flip {\rm (}80\%{\rm )}}: Flip the charges of a randomly selected pair of oppositely charged residues.
\item \textit{Multiple flips {\rm (}20\%{\rm )}}: 
Perform $n_{\rm flip}$ single flips, where $n_{\rm flip}$ is a uniformly distributed random integer between 2 and 10.
\end{itemize}
Compute $\scd$ and $\kappa$ for all proposed sequences 
$\{\sigma_{\alpha}^{(a')}\}$ and accept the updates with probabilities
\begin{equation}
P_{\rm acc}\left( \{\sigma_{\alpha}^{(a)}\} \rightarrow 
\{\sigma_{\alpha}^{(a')}\} \right) = 
\max \left( 1, \frac{g_{i,j}}{g_{i',j'}} \right) ,
\end{equation}
where $(i,j)$ and $(i',j')$ refer to the bins associated with 
$\{\sigma_{\alpha}^{(a)}\}$ and $\{\sigma_{\alpha}^{(a')}\}$, 
respectively. If a $(\scd,\kappa)$ combination is encountered for which 
the associated bin $(i,j)$ was estimated by the GA as empty, the bin 
is re-labelled as non-empty and kept in all subsequent WL iterations.
\item Repeat steps 3--5 until 25 WL ``levels'' have been completed, 
i.e.,~until $h_{i,j}$ is flat with $f=2^{-24}$.
\item Re-weight $g_{i,j} \rightarrow g_{i,j} / g_{\rm sv30}$ where $g_{\rm
sv30}$ is the value of $g_{i,j}$ for the bin associated with the di-block
sequence sv30. This step accounts for the double counting associated with
reversed sequences but relies on the discretization being sufficiently fine
such that $g_{\rm sv30}$ only received contributions from sv30 and its reverse.
\end{enumerate}

Our version of the WL algorithm differs from more standard 
implementations\cite{Irback2013,IrbackWessen2015} 
since $n=100$ instantiations, rather than
$n=1$, of the system are evolved while modifying the same $g_{i,j}$ and
$h_{i,j}$. As such, our setup shares characteristics with the
replica-exchange Wang-Landau method,\cite{Vogel2013} although we do not
e.g.,~constrain the individual sequences to subregions of the
$(\scd,\kappa)$-plane. We observe a dramatic decrease in computation time when
$n=1 \rightarrow 100$ and expect that the implementation can be made even more
efficient by parallel computing since the $n$ sequence updates in Step 5 can be
made independently in-between the $h_{i,j}$ and $g_{i,j}$ updates (see, e.g.,
refs.~\citen{Khan2005,Zhan2008,Yin2012} for more sophisticated parallel WL
algorithm implementations). The sought-after distribution, $P(\scd,\kappa)
\approx g_{i,j} / \Delta \scd \Delta \kappa$ up to discretization errors, is
now shown in the right panel of Fig.~S1a as well as maintext Fig.~1a.

Given this $P(\scd, \kappa)$, the moderate degree to which
$\scd$ and $\kappa$ are correlated may be quantified
by the Pearson correlation coefficient
\begin{equation}
r = \frac{\left\langle \left( \scd - \langle \scd \rangle \right) \left(
\kappa-\langle\kappa\rangle\right)\right\rangle}{\sqrt{\langle\scd^2\rangle 
- \langle \scd \rangle^2} \sqrt{ \langle \kappa^2 \rangle - 
\langle \kappa \rangle^2 }} \approx -0.684 \; ,
\end{equation}
where $\langle\dots\rangle$ here denotes average over the 
GA/WL-sampled $P(\scd, \kappa)$ distribution (thus the $r$ value between
$-$SCD and $\kappa$ is $0.684$). 
The 26 overall-neutral polyampholyte sequences studied in this work
are given in maintext Fig.~1b and in Table~\ref{tab:all_sequences}, their
$\scd$ and $\kappa$ values are listed in Fig.~S1 and plotted on the
$-$SCD versus $\kappa$ plane in maintext Fig.~1a.
\\

\noindent
{\large\bf Field-theoretic formulation of polyampholyte conformations
and phase separation} 

We study these model polyampholyte chain sequences using a field-theoretic
formulation---using field-theoretic simulation (FTS) and random phase
approximation (RPA)---as well as coarse-grained molecular dynamics (MD). 
As in our previous works (reviewed in ref.~\citen{MiMB2023}),
the present field-theoretic formulation is based on a Hamiltonian 
$\hat {H}$ that accounts for chain connectivity, short-spatial-range
excluded-volume repulsion and long-spatial-range electrostatic 
interactions for a system of $n_{\rm p}$ polymers (polyampholyte
chains) each consisting of $N_{\rm p}$ monomer (beads). $\hat{H}$ is
given by
\begin{equation}
\label{eq:particle_pic_hamil}
\beta \hat{H} = \frac{3}{2b^2} \sum_{i=1}^{n_{\rm p}} \sum_{\alpha=1}^{N_{\rm p}-1} \left( \bm{R}_{i,\alpha+1} -
\bm{R}_{i,\alpha} \right)^2 + \frac{v}{2}\int d\bm{r} \hat{\rho}_{\mathrm{tot}}(\bm{r})^2 + \frac{l_{\rm B}}{2} \int d\bm{r} \int d\bm{r}'
\frac{ \hat{c}(\bm{r}) \hat{c}(\bm{r}') }{\vert\bm{r}-\bm{r}'\vert} \; ,
\end{equation}
where $\beta\equiv 1/k_{\rm B}T$ ($k_{\rm B}$ is the Boltzmann constant and
$T$ is absolute temperature), $b$ is the reference root-mean-square bond 
length between two adjacent monomers along the chain sequence when 
non-bonded interactions are absent, ${\bm R}_{i,\alpha}$ is 
the position vector of the $\alpha$th bead of the $i$th chain,
$\hat{\rho}_{\mathrm{tot}}(\bm{r})$ is bead number density (matter density),
$\hat{c}(\bm{r})$ is charge density,
$v$ is the excluded volume parameter and 
$l_{\rm B}=e^2/(4\pi\epsilon_0\epsilon_{\rm r}\kB T)$ is 
the Bjerrum length that we use to quantify electrostatic interaction
strength, with $e$ denoting the protonic charge, 
$\epsilon_0$ and $\epsilon_{\rm r}$ are vacuum and relative
permittivities, respectively; and temperature in our field-theoretical
models is quantified by the reduced temperature  $T^*\equiv b/\lb$. 
An electric charge
$\sigma_{\alpha}$ is associated with the $\alpha$th bead 
of each chain. In this work, we consider only
overall charge neutral polyampholyte sequences, thus 
$\sum_{\alpha=1}^{N_{\rm p}} \sigma_{\alpha}=0$.
To avoid potential singularities arising from modeling polymer beads
as point particles, we model each bead as a normalized Gaussian 
distribution given by
$\Gamma(\bm{r})=\exp(-r^2/2\bar{a}^2)/(2\pi\bar{a}^2)^{3/2}$ 
where $r^2\equiv |\bm{r}|^2$
(refs.~\citen{Wang2010,RigglemanRajeevFredrickson2012}).
Accordingly, the bead number density and charge density are given,
respectively, by
\begin{equation}
\label{eq:part_den_defns}
\begin{aligned}
\hat{\rho}_{\mathrm{tot}}(\bm{r}) &= \sum_{i=1}^{n_{\rm p}} 
\sum_{\alpha=1}^{N_{\rm p}} \Gamma ( \bm{r} - \bm{R}_{i,\alpha} ) \; ,\\
\hat{c}(\bm{r}) &= \sum_{i=1}^{n_{\rm p}} 
\sum_{\alpha=1}^{N_{\rm p}} \sigma_{\alpha} 
\Gamma ( \bm{r} - \bm{R}_{i,\alpha} ) \; .
\end{aligned}
\end{equation}
The field-theoretic model system defined by Eqs.~\ref{eq:particle_pic_hamil} 
and \ref{eq:part_den_defns} is analyzed using random phase approximation (RPA) 
and field-theoretic simulation (FTS).
\\

{\bf Random phase approximation (RPA).}
An approximate analytical theory, termed RPA, can be derived from the
partition function formulated by path integrals based upon the
Hamiltonian in Eq.~\ref{eq:particle_pic_hamil}.
In the present study, all RPA calculations are performed within 
the context of an implicit solvent polymer field theory\cite{Pal2021}
(no explicit
solvent, unlike, e.g., in ref.~\citen{Wessen2021}) with contact-excluded 
volume interactions and unscreened Coulomb electrostatic interactions as
specified by Eq.~\ref{eq:particle_pic_hamil} and, as mentioned,
UV (short-distance)-divergences are regulated
by Gaussian-smeared beads involving the function $\Gamma(\bm{r})$
as described above. In the present RPA calculations, we use a fixed 
excluded volume parameter $v=0.0068 b^3$ and a Gaussian smearing length 
of ${\bar a}=b/\sqrt{6}$. 
The detailed mathematical formulation and the
computer code employed for our RPA calculations are documented and
available through our recent review.\cite{MiMB2023} 
Examples of phase diagrams computed using FTS and RPA for 
the 26 polyampholyte sequences we study (maintext Fig.~1b, Fig.~S1b, Table~S1)
are provided in maintext Fig.~1d. A complete list of these phase
diagrams are documented in Fig.~S2. The critical temperatures of these
sequences' phase transitions in RPA and FTS are tabulated in Fig.~S1b, whereas
the necessary details of our FTS methodology are provided below.
\\


{\bf Radius of gyration and contact maps of chain conformations in the
field-theoretic formulation.}
As a novel extension of our previous field-theoretic formulation,
we now connect the radius of gyration ($R_{\rm g}$) of a single polymer 
chain to the polymer beads' pair-correlation function\cite{Pal2021} as
follows:
\begin{equation}
\label{eq:Rg_defn_expl}
\begin{aligned}
{R_{\rm g}}^2 &= \frac{1}{2N_{\rm p}^2} \left\langle\sum_{\alpha=1}^{N_{\rm p}}
\sum_{\beta=1}^{N_{\rm p}} (\bm{R}_{\alpha} - \bm{R}_{\beta})^2\right\rangle\\
&= \frac{1}{2N_{\rm p}^2} \int d\bm{r} \int d\bm{r}' \langle \hat{\rho}_{\rm
c}(\bm{r}) \hat{\rho}_{\rm c}(\bm{r}') \rangle (\bm{r}-\bm{r}')^2 \; ,
\end{aligned}
\end{equation}
where the bead center density is defined here by $\hat{\rho}_{\rm c}(\bm{r}) =
\sum_{\alpha=1}^{N_{\rm p}} \delta(\bm{r} - \bm{R}_{\alpha})$ and $\langle
\cdot\cdot\cdot \rangle$ denotes thermal averaging. Since the
pair-correlation function $G(|\bm{r}-\bm{r}'|)=\langle \hat{\rho}_{\rm
c}(\bm{r}) \hat{\rho}_{\rm c}(\bm{r}') \rangle$ depends only on the relative
distance, we may perform a
coordinate transformation such that $\bm{R}=(\bm{r}+\bm{r}')/2$ and
$\bm{\mathsf{r}}=\bm{r}-\bm{r}'$, with the associated Jacobian's determinant
$|J|=1$, and thus rewrite Eq.~\eqref{eq:Rg_defn_expl} as
\begin{equation}
\label{eq:Rg}
\begin{aligned}
{R_{\rm g}}^2 &= \frac{1}{2N_{\rm p}^2} \int d\bm{R} \int d\bm{\mathsf{r}}
G(|\mathsf{r}|) |\mathsf{r}|^2\\ &= \frac{V}{2N_{\rm p}^2} \int d\bm{r}
G(|\bm{r}|) |\bm{r}|^2 \; ,
\end{aligned}
\end{equation}
where $V=\int d\bm{R}$ 
is the volume of the system. Because we have elected to use smeared
Gaussian packets instead of point particles in our field-theoretic formulation
to regularize short-range divergences as noted above, we will approximate
the $\delta$-function defined pair-correlation function defined above by
smeared densities instead. 
Accordingly, in the formulation below for calculating of $R_{\rm g}$, 
we replace $G(\vert \bm{r} - \bm{r}' \vert)$ in
Eq.~\eqref{eq:Rg_defn_expl} by $\mathcal{C}(|\bm{r}-\bm{r}'|)=\langle
\hat{\rho}(\bm{r}) \hat{\rho}(\bm{r}') \rangle$ 
where $\hat{\rho}(\bm{r}) = \sum_{\alpha=1}^{N_{\rm p}} \Gamma(\bm{r} -
\bm{R}_{\alpha})$ as in Eq.~\eqref{eq:part_den_defns}. 
Mathematically, $\mathcal{C}(r)$ (where $r=|\bm{r}|$) 
is expected to be different from 
$G(r)$ only in these functions' variations over short distances 
$\lesssim\bar{a}$. Since $G(r)$ is integrated and weighted by
$|\bm{r}|^2=r^2$ in Eq.~\ref{eq:Rg} for ${R_{\rm g}}^2$, the replacement 
of $G(r)$ by $\mathcal{C}(r)$ 
is not numerically significant for
the accuracy of the computed value of $R_{\rm g}$. For our purpose, however,
the smearing that replace $G(r)$ by $\mathcal{C}(r)$ is 
important as it allows $G(r)$ and thus $R_{\rm g}$ to be computed 
approximately using FTS.

By construction, Eq.~\ref{eq:Rg} is applicable only for a single chain 
and needs to be extended for a multi-chain system. To do so, consider 
$n_{\rm p}$ identical chains in a polyampholyte solution. 
We now choose one chain at random and recognize it as
``tagged''. We denote the rest of the $n_{\rm p}-1$ chains as 
``rest''.  The total bead density can now be rewritten as
$\hat{\rho}_{\mathrm{tot}}(\bm{r}) = 
\hat{\rho}^{(\mathrm{t})}(\bm{r}) + \hat{\rho}^{(\mathrm{r})}(\bm{r})$ where
the superscripts ``(t)'' and ``(r)'' refer to ``tagged'' and ``rest'', 
respectively. With this setup, we can now substitute the
smeared self-correlation function of the tagged chain 
$\mathcal{C}^{(\mathrm{t})}(r)$ for $G(r)$ in Eq.~\ref{eq:Rg} 
to find the tagged chain's ${R_{\rm g}}^2$ in the multi-chain system, and
can simply set $n_{\rm p}=1$ 
in this general formulation to recover the formula for the ${R_{\rm g}}^2$
of a single isolated chain.
For clarity, we will denote this latter isolated-chain smeared 
self-correlation function by $\mathcal{C}^{(\mathrm{isol})}(r)$. 

Correlation functions can also be utilized to 
compute residue-residue (bead-bead, or monomer-monomer) contact maps. 
To do so we only have to identify the monomers' sequence positions along the
polymer chains by rewriting the total bead center density for a 
multi-chain system  as
$\hat{\rho}_{\mathrm{c, tot}}(\bm{r}) 
= \sum_{\alpha=1}^{N_{\rm p}} 
\hat{\rho}_{\mathrm{c},\alpha}^{\mathrm{(t)}}(\bm{r})
+ \sum_{\alpha=1}^{N_{\rm p}} 
\hat{\rho}_{\mathrm{c},\alpha}^{\mathrm{(r)}}(\bm{r})$, where
$\hat{\rho}_{\mathrm{c},\alpha}^{\mathrm{(t/r)}}(\bm{r})$ is 
sequence-position-specific density of the center of the $\alpha$th bead 
in the ``tagged'' chain/``rest'' chains. As specified above for
Eq.~\ref{eq:Rg_defn_expl}, unlike the smeared
$\hat{\rho}(\bm{r})$ defined above, the bead center density 
$\hat{\rho}_{\mathrm{c}}(\bm{r})$ (with subscript ``c'') is defined
by $\delta$-function of position without $\Gamma$ smearing.
We use unsmeared bead center densities for the computation of contacts 
because contacts are defined by spatial separations between bead centers.
Now, the corresponding bead-specific unsmeared correlation 
functions can be defined as
$G^{(x),(y)}_{\alpha,\beta}(|\bm{R}|) = \left\langle \hat{\rho}_{\mathrm{c},\alpha}^{(x)}(\bm{r})
\hat{\rho}_{\mathrm{c},\alpha}^{(y)}(\bm{r}+\bm{R}) 
\right\rangle$ where $\{x,y\} \in \{\mathrm{t,r}\} $. With this definition, we
can compute the average contact frequency between the 
$(x,\alpha)$ and $(y,\beta)$ monomers (beads) by spatially integrating 
the corresponding $G^{(x),(y)}_{\alpha,\beta}(|\bm{R}|)$ 
from radial distance $|\bm{R}|=0$ up to a suitably 
chosen cutoff in $|\bm{R}|$ for defining a contact, with
the $x=y$ and $x\ne y$ cases accounting, respectively, for intrachain 
and interchain contacts. For the present work, we adopt
the definitions
\begin{subequations}
\label{eq:contact_freq_defns}
\begin{align}
\omega_{\alpha,\beta}^{\mathrm{(t),(t)}} &= V \int_{0}^{2b} d\bm{r} \; 
G^{(\mathrm{t}),(\mathrm{t})}_{\alpha,\beta}(|\bm{r}|) \; , 
\label{eq:contact_freq_defns_intra}\\
\omega_{\alpha,\beta}^{\mathrm{(t),(r)}} &= V \int_{0}^{2b} d\bm{r} \; 
G^{\mathrm{(t),(r)}}_{\alpha,\beta}(|\bm{r}|) \; , 
\label{eq:contact_freq_defns_inter}
\end{align}
\end{subequations}
as intrachain (Eq.~\ref{eq:contact_freq_defns_intra}) and
interchain (Eq.~\ref{eq:contact_freq_defns_inter}) 
contact frequencies, wherein 
a radial cutoff distance of $2b$ is used
for defining a bead-bead spatial contact.  
\\


{\bf Key steps in the field-theoretic simulation (FTS).}
Using the expression for the Hamiltonian in Eq.~\ref{eq:particle_pic_hamil},
the canonical partition function of our system of interest is given by
\begin{equation}
\label{eq:particle_pic_part_funcn}
\mathcal{Z} = \frac{1}{n_{\rm p}!} \left( \prod_{i=1}^{n_{\rm p}}
\prod_{\alpha=1}^{N_{\rm p}} \int d\bm{R}_{i,\alpha} \right) e^{-\beta
\hat{H}} \; .
\end{equation}
Following standard
procedures,\cite{FredricksonGanesanDrolet2002,Fredrickson2006,
Pal2021, MiMB2023, WessenDasPalChan2022} we derive a field theory
described by a field Hamiltonian $H$ (different
from $\hat{H}$) such that the partition function itself remains 
same as Eq.~\ref{eq:particle_pic_part_funcn} up to an inconsequential
overall multiplicative constant, viz.,
\begin{equation}
\label{eq:field_pic_part_funcn}
\mathcal{Z} = \frac{V^{n_{\rm p}}}{n_{\rm p}!} \int \mathcal{D}w \int \mathcal{D}\psi e^{-H[w, \psi]},
\end{equation}
in which two fluctuating auxiliary fields $w$ and $\psi$ are introduced, with
\begin{equation}
\label{eq:field_pic_hamil}
H[w,\psi] = \int d\bm{r} \left[ \frac{w(\bm{r})^2}{2v} + \frac{(\bm{\nabla}\psi(\bm{r}))^2}{8\pi l_{\rm B}} \right]
-n_{\rm p} \ln Q_{\rm p}[\breve{w},\breve{\psi}] \; ,
\end{equation}
where $\breve{\phi}(\bm{r}) = \Gamma \star \phi \equiv 
\int d\bm{r}' \Gamma (\bm{r} - \bm{r}') \phi(\bm{r}')$, $\phi = w,\psi$,
with ``$\star$'' denoting 
this spatial convolution henceforth.
In Eq.~\ref{eq:field_pic_hamil}, the single-chain partition function 
$Q_{\rm p}$ is given by
\begin{equation}
\label{eq:single_chain_part_func}
Q_{\rm p}[\breve{w}, \breve{\psi}] = \frac{1}{V} \int d\bm{R}_{N_{\rm p}} \left( \prod_{\alpha=1}^{N_{\rm p}-1} \int d\bm{R}_{\alpha}
G^0(\bm{R}_{\alpha+1} - \bm{R}_{\alpha} \vert b) \right) e^{-\mathrm{i} \sum_{\alpha=1}^{N_{\rm p}}
\left[ \breve{w}(\bm{R}_{\alpha}) + \sigma_{\alpha} \breve{\psi} (\bm{R}_{\alpha}) \right]} \; ,
\end{equation}
where $\mathrm{i}^2=-1$ is the imaginery unit and
\begin{equation}
\label{eq:chain_propagator}
G^0(\bm{r} \vert b) = \left( \frac{3}{2\pi b^2} \right)^{3/2} 
e^{-3|\bm{r}|^2/2b^2}.
\end{equation}

Based on this particle-to-field reformulation, the equilibrium properties 
of the system can now be studied through the dynamics of the fields. 
Since the field Hamiltonian $H$ involves complex variables, we 
adopt\cite{Fredrickson2006} a ``Complex Langevin'' (CL) 
prescription\cite{Parisi1983,Klauder1983} 
inspired by stochastic quantization\cite{ParisiWu1981,ChanHalpern1986} 
for computing averages for
the system defined by Eq.~\ref{eq:field_pic_hamil}
by introducing dependence on a fictitious time to
the fields, which then evolve dynamically in accordance to
the Langevin equations
\begin{equation}
\label{eq:CL_eqns}
\begin{aligned}
\frac{\partial w(\bm{r}, t)}{\partial t} &= - \left[ \mathrm{i} \tilde{\rho}_{\mathrm{tot}}(\bm{r}, t) + \frac{w(\bm{r}, t)}{v}\right]
+ \eta_{w}(\bm{r}, t) \; ,\\
\frac{\partial \psi(\bm{r}, t)}{\partial t} &= - \left[ \mathrm{i} \tilde{c}(\bm{r}, t)
- \frac{\bm{\nabla}^2\psi(\bm{r}, t)}{4\pi l_{\rm B}}\right] + \eta_{\psi}(\bm{r}, t) \; ,
\end{aligned}
\end{equation}
where $t$ is a fictitious time, 
$\eta_w$ and $\eta_{\psi}$ are fields of real-valued random numbers 
drawn from a normal distribution of zero mean and standard deviation 
that is nonzero only when $t=t'$ and $\bm{r}=\bm{r}'$
[$\propto 2\delta(t-t')\delta(\bm{r}-\bm{r}')$, specifically,
$\langle \eta_{\phi}(\bm{r},t) \eta_{\phi'}(\bm{r}',t') \rangle =
2 \delta_{\phi,\phi'} \delta(\bm{r} - \bm{r}') \delta(t-t')$]. This is the basic
Langevin formulation of FTS. In Eq.~\ref{eq:CL_eqns},
$\tilde{\rho}_{\mathrm{tot}}$ and $\tilde{c}$ are field operators 
for total bead density and charge density,
respectively, which may be expressed as
\begin{equation}
\label{eq:field_operators_total_bead_charge}
\begin{aligned}
\tilde{\rho}_{\mathrm{tot}}(\bm{r}) &= \Gamma \star \frac{n_{\rm p}}{Q_{\rm p}V}\sum_{\alpha=1}^{N_{\rm p}}
q_{B}(\bm{r}, \alpha) q_{F}(\bm{r}, \alpha) e^{ \mathrm{i} \breve{w}(\bm{r}) + \mathrm{i} \sigma_{\alpha} \breve{\psi}(\bm{r}) } \; ,\\
\tilde{c}(\bm{r}) &= \Gamma \star \frac{n_{\rm p}}{Q_{\rm p}V}\sum_{\alpha=1}^{N_{\rm p}} \sigma_{\alpha}
q_{B}(\bm{r}, \alpha) q_{F}(\bm{r}, \alpha) e^{ \mathrm{i} \breve{w}(\bm{r}) + \mathrm{i} \sigma_{\alpha} \breve{\psi}(\bm{r}) } \; , 
\end{aligned}
\end{equation}
where the forward ($F$) and backward ($B$) chain propagators
$q_F$ and $q_B$ are constructed iteratively:
\begin{equation}
\begin{aligned}
q_{F}(\bm{r},\alpha+1) &= 
e^{-\mathrm{i} \breve{w}(\bm{r}) - 
\mathrm{i} \sigma_{\alpha+1} \breve{\psi}(\bm{r}) }
\int d\bm{r}' G^0(\bm{r}-\bm{r}' \vert b) q_{F} (\bm{r}', \alpha) \; ,\\
q_{B}(\bm{r},\alpha-1) &= e^{-\mathrm{i} \breve{w}(\bm{r}) - \mathrm{i} \sigma_{\alpha-1} \breve{\psi}(\bm{r}) }
\int d\bm{r}' G^0(\bm{r}-\bm{r}' \vert b) q_{B} (\bm{r}', \alpha) \; ,
\end{aligned}
\end{equation}
with $q_{F}(\bm{r}, 1) = e^{-\mathrm{i} \breve{w}(\bm{r}) - \mathrm{i} \sigma_{1} \breve{\psi}(\bm{r}) } $ and
$q_{B}(\bm{r}, N_{\rm p}) = e^{-\mathrm{i} \breve{w}(\bm{r}) - \mathrm{i} \sigma_{N_{\rm p}} \breve{\psi}(\bm{r}) } $.

In practice, the differential field evolution equations Eqs.~\ref{eq:CL_eqns} 
have to be numerically solved in discretized space and discretized CL time.
For the present work, every FTS is conducted in a $32\times 32\times 32$ 
cubic grid with a side length $32\bar{a}$ and we set
$\bar{a} = b/\sqrt{6}$ following previous works. 
A CL time step $dt=0.005$ and a semi-implicit CL time integration 
scheme\cite{sii} are employed. 
To compute the thermal average of any thermodynamic observable by FTS,
the particle-based thermal average has to be replaced by an average 
over the field configurations of the corresponding field operator. 
The field operators could be derived by introducing appropriate ``source'' 
terms (fields) in the particle picture Hamiltonian (``$J$'' 
fields, see below), as is commonly practiced in theoretical particle 
physics.\cite{IZ} If an observable $\hat{\mathcal{O}}(\bm{r})$ has a
field operator $\tilde{\mathcal{O}}(w,\psi)$, then in practice the equilibrium field average of that specific operator is
evaluated as an asymptotic CL time average through
\begin{equation}
\left \langle \hat{\mathcal{O}}
(\{\bm{R}_{i,\alpha}\}) \right \rangle = 
\left\langle \tilde{\mathcal{O}}[w,\psi] \right\rangle_{\rm F} =
\frac{ \int \mathcal{D}w \int \mathcal{D} \psi \; \tilde{\mathcal{O}}(w,\psi)
e^{-H[w,\psi] } }{ \int \mathcal{D}w \int \mathcal{D} \psi \; e^{-H[w,\psi] } }
= \frac{1}{M} \sum_{j=1}^{M}
\tilde{\mathcal{O}}[w(\bm{r},t_j),\psi(\bm{r},t_j)] \; , 
\end{equation}
where $M$ is the maximum number of field configurations considered
(configurations are labeled by $j$ in the above equation 
for selected values $t_j$ of fictitious time $t$) and
$\langle \cdot\cdot\cdot \rangle_{\rm F}$ denotes field averaging. 
The configurations at $t_j$s and the total number of configurations
$M$ used in the averaging are chosen such that
the real part of the field averaged quantity has a reasonably small fluctuation over independent simulation runs and
the corresponding imaginary part is approximately zero. For all the FTS
computation, we set the excluded volume parameter $v=0.0068b^3$.
Further details of our approach can be found in ref.~\citen{MiMB2023}.
\\


{\bf FTS operators for correlation functions in the computation of
polymer radius of gyration.} 
As outlined above in the discussion preceding Eq.~\ref{eq:contact_freq_defns},
to compute equilibrium properties of individual polymer chains in 
a multiple-chain system, we consider one chain,
chosen at random, as tagged. The total bead density in 
Eq.~\ref{eq:part_den_defns} can then be rewritten as
\begin{equation}
\label{eq:defns_t_r}
\hat{\rho}_{\mathrm{tot}}(\bm{r}) = 
\hat{\rho}^{\mathrm{(t)}}(\bm{r}) + \hat{\rho}^{\mathrm{(r)}}(\bm{r}) \; ,
\end{equation}
where the superscripts ``(t)'' and ``(r)'' 
denote `tagged' and `rest', respectively, with
$\hat{\rho}^{\mathrm{(t)}}(\bm{r})$ for the tagged chain and 
$\hat{\rho}^{\mathrm{(r)}}(\bm{r})$ for the rest of the chains (untagged
chains) given by
\begin{equation}
\label{eq:defns_t_r_ex}
\begin{aligned}
\hat{\rho}^{\mathrm{(t)}}(\bm{r}) & = \sum_{\alpha=1}^{N_{\rm p}} \Gamma
\left(\bm{r} -\bm{R}^{\mathrm{(t)}}_{\alpha} \right) \; ,\\
\hat{\rho}^{\mathrm{(r)}}(\bm{r}) &= \sum_{i=1}^{n_{\rm
p}-1}\sum_{\alpha=1}^{N_{\rm p}} \Gamma \left(\bm{r}
-\bm{R}^{\mathrm{(r)}}_{i,\alpha} \right) \; .
\end{aligned}
\end{equation}

Field operators for total bead densities of ``tagged'' and ``rest'' 
chains are identified by introducing the source
terms $J^{\mathrm{(t)}}$ and $J^{\mathrm{(r)}}$ in the original
particle-picture Hamiltonian in Eq.~\ref{eq:particle_pic_hamil}
in the following manner:
\begin{equation}
\beta\hat{{H}} \rightarrow
\beta\hat{{H}} [ J^{\mathrm{(t)}}, J^{\mathrm{(r)}} ] 
= \beta\hat{H} - \int d\bm{r}
J^{\mathrm{(t)}}(\bm{r})\hat{\rho}^{\mathrm{(t)}}(\bm{r}) - \int d\bm{r}
J^{\mathrm{(r)}}(\bm{r})\hat{\rho}^{\mathrm{(r)}}(\bm{r}) \; .  
\end{equation}
The resulting source-dependent partition function is now given by
\begin{equation}
\label{eq:part_func_tot_source}
\mathcal{Z}[ J^{\mathrm{(t)}}, J^{\mathrm{(r)}} ] \sim \left(
\prod_{\alpha=1}^{N_{\rm p}} \int d\bm{R}^{\mathrm{(t)}}_{\alpha} \right)
\left( \prod_{i=1}^{n_{\rm p}-1} \prod_{\alpha=1}^{N_{\rm p}} \int
d\bm{R}^{\mathrm{(r)}}_{i,\alpha} \right) e^{-\beta \hat{{H}} [
J^{\mathrm{(t)}}, J^{\mathrm{(r)}} ]} \; .  
\end{equation}
From this last expression (Eq.~\ref{eq:part_func_tot_source}), we 
obtain the formal relations
\begin{equation}
\label{eq:tot_den_func_derivatives}
\begin{aligned}
\left\langle \hat{\rho}^{\mathrm{(t)}}(\bm{r}) \right \rangle &= \left( \frac{1}{\mathcal{Z}[ J^{\mathrm{(t)}}, J^{\mathrm{(r)}} ]}
\frac{\delta \mathcal{Z}[ J^{\mathrm{(t)}}, J^{\mathrm{(r)}} ] } {\delta
J^{\mathrm{(t)}}(\bm{r})} \right)_{J^{\mathrm{(t)}}=J^{\mathrm{(r)}}=0} \; ,\\
\left\langle \hat{\rho}^{\mathrm{(r)}}(\bm{r}) \right \rangle &= \left(
\frac{1}{\mathcal{Z}[ J^{\mathrm{(t)}}, J^{\mathrm{(r)}} ]}
\frac{\delta \mathcal{Z}[ J^{\mathrm{(t)}}, J^{\mathrm{(r)}} ] } {\delta
J^{\mathrm{(r)}}(\bm{r})} \right)_{J^{\mathrm{(t)}}=J^{\mathrm{(r)}}=0} \; ,
\end{aligned}
\end{equation}
where $\langle ... \rangle$ denotes thermal averaging. In the corresponding
field theory representation, we have
\begin{equation}
\label{eq:field_part_func_tot_source}
\mathcal{Z}[J^{\mathrm{(t)}}, J^{\mathrm{(r)}}] \sim \int \mathcal{D}w \int
\mathcal{D}\psi e^{-{H}[w, \psi, J^{\mathrm{(t)}}, J^{\mathrm{(r)}}]}
\; , 
\end{equation}
where
\begin{equation}
\label{eq:field_ham_tot_source}
\begin{aligned}
{H}[w, \psi, J^{\mathrm{(t)}}, J^{\mathrm{(r)}}] = & \int d\bm{r} \left[
\frac{w(\bm{r})^2}{2v} + \frac{(\bm{\nabla}\psi(\bm{r}))^2}{8\pi l_{\rm B}}
\right] \\ 
& \quad\quad\quad\quad - \ln Q_{\rm
p}[\breve{w}-\mathrm{i}\breve{J}^{\mathrm{(t)}},\breve{\psi}] - (n_{\rm p}-1)
\ln Q_{\rm p}[\breve{w}-\mathrm{i}\breve{J}^{\mathrm{(r)}},\breve{\psi}] \; .
\\
\end{aligned}
\end{equation}
In this equation (Eq.~\ref{eq:field_ham_tot_source}),  
$Q_{\rm p}$ has the same functional form as that in 
Eq.~\ref{eq:single_chain_part_func}. We can now apply
the definitions Eq.~\ref{eq:tot_den_func_derivatives} to
Eq.~\ref{eq:field_part_func_tot_source} to arrive at
$\langle \hat{\rho}^{\mathrm{(t)}}(\bm{r}) \rangle 
= \langle \tilde{\rho}^{\mathrm{(t)}}(\bm{r}) 
\rangle_{\mathrm{F}}$ and $\langle 
\hat{\rho}^{\mathrm{(r)}}(\bm{r}) \rangle = 
\langle \tilde{\rho}^{\mathrm{(r)}}(\bm{r}) \rangle_{\mathrm{F}}$ where
\begin{equation}
\begin{aligned}
\tilde{\rho}^{\mathrm{(t)}}(\bm{r}) &= \frac{1}{n_{\mathrm{p}}} \tilde{\rho}_{\mathrm{tot}}(\bm{r}) \; ,\\
\tilde{\rho}^{\mathrm{(r)}}(\bm{r}) &= \frac{n_{\mathrm{p}}-1}{n_{\mathrm{p}}} \tilde{\rho}_{\mathrm{tot}}(\bm{r}) \; ,
\end{aligned}
\end{equation}
and $\tilde{\rho}_{\mathrm{tot}}(\bm{r})$ is defined in 
Eq.~\ref{eq:field_operators_total_bead_charge} and, again,
$\langle ... \rangle_{\mathrm{F}}$ represents average over field configurations.

We can also compute the self-correlation function of the tagged chain 
from the partition function in Eq.~\ref{eq:part_func_tot_source} 
through the formal relation
\begin{equation}
\label{eq:tagged_self_corr_def}
\mathcal{C}^{\mathrm{(t)}}(\vert\bm{r}-\bm{r}'\vert) = \left\langle
\hat{\rho}^{\mathrm{(t)}}(\bm{r})
\hat{\rho}^{\mathrm{(t)}}(\bm{r}')\right\rangle = \left( \frac{1}{\mathcal{Z}[
J^{\mathrm{(t)}}, J^{\mathrm{(r)}} ]} \frac{\delta^2 \mathcal{Z}[
J^{\mathrm{(t)}}, J^{\mathrm{(r)}} ] } {\delta J^{\mathrm{(t)}}(\bm{r}) \delta
J^{\mathrm{(t)}}(\bm{r}')}  \right)_{J^{\mathrm{(t)}}=J^{\mathrm{(r)}}=0}.
\end{equation}
Applying this relation (Eq.~\ref{eq:tagged_self_corr_def}) to
Eq.~\ref{eq:part_func_tot_source} results in
\begin{equation}
\mathcal{C}^{\mathrm{(t)}}(\vert\bm{r}-\bm{r}'\vert) = \frac{\mathrm{i}}{v}
\left\langle w(\bm{r}) \tilde{\rho}_{\mathrm{tot}}(\bm{r}')
\right\rangle_{\mathrm{F}} -\left \langle \tilde{\rho}^{\mathrm{(t)}}(\bm{r})
\tilde{\rho}^{\mathrm{(r)}}(\bm{r}') \right\rangle_{\mathrm{F}} \; , 
\end{equation}
where, following ref.~\citen{Pal2021},
we have avoided computing a functional double derivative of $Q_{\rm p}$. 

When there is only a single isolated chain in the system, the self-correlation 
function is obtained by setting
$n_{\rm p}=1$ and $\tilde{\rho}^{\mathrm{(r)}}(\bm{r}')=0$ 
in the above derivation, yielding
\begin{equation}
\mathcal{C}^{\mathrm{(isol)}}(\vert\bm{r}-\bm{r}'\vert) = \frac{\mathrm{i}}{v}
\left\langle w(\bm{r}) \tilde{\rho}_{\mathrm{tot}}(\bm{r}')
\right\rangle_{\mathrm{F}} \; .  
\end{equation}
Utilizing translational invariance of our systems, we can write
\begin{equation}
\begin{aligned}
\left\langle \hat{\rho}(\bm{r}) \hat{\rho}(\bm{r}')\right\rangle &= \frac{1}{V}
\int d\bm{r}^{\prime\prime}\left\langle \hat{\rho}(\bm{r}+\bm{r}^{\prime\prime})
\hat{\rho}(\bm{r}'+\bm{r}^{\prime\prime})\right\rangle\\ 
&= \frac{1}{V} \int
\frac{d\bm{k}}{(2\pi)^3} e^{-\mathrm{i}\bm{k}\cdot(\bm{r}-\bm{r}')}
\left\langle \check{\rho}(\bm{k}) \check{\rho}(-\bm{k})\right\rangle,
\end{aligned}
\end{equation}
where, when the density smearing described above is applied, 
$\check{\rho}(\bm{k})\equiv
\int d\bm{r} e^{ {\mathrm{i}}\bm{k} \cdot \bm{r}}\hat{\rho}(\bm{r})$ 
is the Fourier 
transform of the smeared density $\hat{\rho}(\bm{r})$. In that case,
it follows that $\check{\rho}(\bm{k}) = 
\check{\Gamma}(\bm{k}) \check{\rho}_{\rm c}(\bm{k})$ where
$\check{\Gamma}(\bm{k})=e^{-\bar{a}^2 |\bm{k}|^2/2}$ 
is the Fourier transform of the smearing function $\Gamma(\bm{r})$
and $\check{\rho}_{\rm c}(\bm{k})$ is the Fourier transform of the bead center
density $\hat{\rho}_{\rm c}(\bm{r})$.  This implies that when
Gaussian smearing is utilized in our formulation, the bead center-bead
center pair correlation function [$G(r)\rightarrow \mathcal{C}(r)$] 
can now be expressed as
\begin{equation}
\left\langle \hat{\rho}_{\rm c}(\bm{r}) \hat{\rho}_{\rm c}(\bm{r}')\right\rangle =
\frac{1}{V} \int \frac{d\bm{k}}{(2\pi)^3} e^{-\mathrm{i}\bm{k}\cdot(\bm{r}-\bm{r}')} \frac{ \left\langle \check{\rho}(\bm{k}) \check{\rho}(-\bm{k})\right\rangle } {\check{\Gamma}(\bm{k})^2} \; .
\end{equation}
\\

{\bf FTS operators for correlation functions in the computation of
contact maps of polymers.} 
As noted above, the starting point for contact map computations is 
the (unsmeared) bead center density correlation functions.
As noted above in the discussion preceding Eq.~\ref{eq:contact_freq_defns}, 
we first rewrite the total bead center density as
\begin{equation}
\label{eq:defns_t_r_cont}
\hat{\rho}_{\mathrm{c,tot}}(\bm{r}) = \hat{\rho}_{\rm c}^{\mathrm{(t)}}(\bm{r}) + \hat{\rho}_{\rm c}^{\mathrm{(r)}}(\bm{r}),
\end{equation}
where $\hat{\rho}_{\rm c}^{\mathrm{(t)}}(\bm{r})$ and $\hat{\rho}_{\rm
c}^{\mathrm{(r)}}(\bm{r})$ could be expressed in terms of the 
bead-position-specific densities along the chains through
\begin{equation}
\label{eq:defns_t_r_ex_cont}
\begin{aligned}
\hat{\rho}_{\rm c}^{\mathrm{(t)}}(\bm{r}) &= \sum_{\alpha=1}^{N_{\rm p}}
\hat{\rho}_{\mathrm{c},\alpha}^{\mathrm{(t)}}(\bm{r}) = \sum_{\alpha=1}^{N_{\rm
p}} \delta \left(\bm{r} -\bm{R}^{\mathrm{(t)}}_{\alpha} \right) \; ,\\
\hat{\rho}_{\rm c}^{\mathrm{(r)}}(\bm{r}) &= \sum_{\alpha=1}^{N_{\rm p}}
\hat{\rho}_{\mathrm{c},\alpha}^{\mathrm{(r)}}(\bm{r}) = \sum_{i=1}^{n_{\rm
p}-1}\sum_{\alpha=1}^{N_{\rm p}} \delta \left(\bm{r}
-\bm{R}^{\mathrm{(r)}}_{i,\alpha} \right) \; .
\end{aligned}
\end{equation}
We now introduce monomer (residue)-specific source terms in the interaction
Hamiltonian Eq.~\ref{eq:particle_pic_hamil}:
\begin{equation}
\beta\hat{{H}} \rightarrow
\beta\hat{{H}}[\{J^{\mathrm{(t)}}_{\alpha}\},
\{J^{\mathrm{(r)}}_{\alpha}\}] = \beta\hat{H} - \int d\bm{r} \sum_{\alpha=1}^{N}
J^{\mathrm{(t)}}_{\alpha}(\bm{r})\hat{\rho}^{\mathrm{(t)}}_{\mathrm{c},\alpha}(\bm{r})
- \int d\bm{r} \sum_{\alpha=1}^{N}
J^{\mathrm{(r)}}_{\alpha}(\bm{r})\hat{\rho}^{\mathrm{(r)}}_{\mathrm{c},\alpha}(\bm{r}) \; .
\end{equation}
The corresponding partition function is 
\begin{equation}
\label{eq:partition_function_part_res_spec}
\mathcal{Z}[\{J^{\mathrm{(t)}}_{\alpha}\}, \{J^{\mathrm{(r)}}_{\alpha}\}] \sim
\left(  \prod_{\alpha=1}^{N_{\rm p}} \int d\bm{R}^{\mathrm{(t)}}_{\alpha}
\right) \left( \prod_{i=1}^{n_{\rm p}-1} \prod_{\alpha=1}^{N_{\rm p}} \int
d\bm{R}^{\mathrm{(r)}}_{i,\alpha} \right) e^{-\beta \hat{{H}} [
\{J^{\mathrm{(t)}}_{\alpha}\}, \{J^{\mathrm{(r)}}_{\alpha}\} ]} \; .
\end{equation}
From this expression (Eq.~\ref{eq:partition_function_part_res_spec}), 
the average density of the $\alpha$th bead center is formally 
\begin{equation}
\label{eq:defn_res_spec_den}
\left\langle \hat{\rho}^{(x)}_{\mathrm{c},\alpha}(\bm{r}) \right\rangle =
\left( \frac{1}{\mathcal{Z}[\{J^{\mathrm{(t)}}_{\alpha}\},
\{J^{\mathrm{(r)}}_{\alpha}\}]}\frac{\delta
\mathcal{Z}[\{J^{\mathrm{(t)}}_{\alpha}\},
\{J^{\mathrm{(r)}}_{\alpha}\}]}{\delta J^{(x)}_{\alpha}(\bm{r})}
\right)_{\{J_{\alpha}^{\mathrm{(t)}} \}=\{J_{\alpha}^{\mathrm{(r)}}\}=0} \; ,
\end{equation}
and the correlation function between any two bead center is given by
\begin{equation}
\label{eq:defn_res_spec_corr}
G^{(x),(y)}_{\alpha, \beta}(\vert \bm{r}-\bm{r}'\vert) = \left\langle
\hat{\rho}^{(x)}_{\mathrm{c},\alpha}(\bm{r})
\hat{\rho}^{(y)}_{\mathrm{c},\beta}(\bm{r}')  \right\rangle = \left(
\frac{1}{\mathcal{Z}[\{J^{\mathrm{(t)}}_{\alpha}\},
\{J^{\mathrm{(r)}}_{\alpha}\}]}\frac{\delta
\mathcal{Z}[\{J^{\mathrm{(t)}}_{\alpha}\},
\{J^{\mathrm{(r)}}_{\alpha}\}]}{\delta J^{(x)}_{\alpha}(\bm{r}) \delta
J^{(y)}_{\beta}(\bm{r}) } \right)_{\{J_{\alpha}^{\mathrm{(t)}} \}=
\{J_{\alpha}^{\mathrm{(r)}}\}=0}
\end{equation}
where $\{x,y\} \in \{\mathrm{t,r}\}$.
The corresponding partition function in the field picture is then given by
\begin{equation}
\label{eq:field_part_func_tot_source_residue_sp}
\mathcal{Z}[\{J_{\alpha}^{\mathrm{(t)}} \}, \{J_{\alpha}^{\mathrm{(r)}} \}] \sim \int \mathcal{D}w \int \mathcal{D}\psi e^{-{H}[w, \psi, \{J_{\alpha}^{\mathrm{(t)}} \}, \{ J_{\alpha}^{\mathrm{(r)}} \}]}
\end{equation}
where
\begin{equation}
\label{eq:field_ham_tot_source_residue_sp}
\begin{aligned}
{H}[w, \psi, \{J_{\alpha}^{\mathrm{(t)}} \},
\{J_{\alpha}^{\mathrm{(r)}}\}] =& \int d\bm{r} \left[ \frac{w(\bm{r})^2}{2v} +
\frac{(\bm{\nabla}\psi(\bm{r}))^2}{8\pi l_{\rm B}} \right] - \ln Q_{\rm
p}[\breve{w}-\mathrm{i} \{ J_{\alpha}^{\mathrm{(t)}}\},\breve{\psi}] \\ &-
(n_{\rm p}-1) \ln Q_{\rm p}[\breve{w}-\mathrm{i} \{
J_{\alpha}^{\mathrm{(r)}}\},\breve{\psi}] \; ,
\end{aligned}
\end{equation}
and the absence of breve in the source terms on the right hand side 
in the above equation 
means that the source terms are not smeared.
In Eq.~\ref{eq:field_ham_tot_source_residue_sp}, $Q_{\rm p}$ has the 
same functional form as that in Eq.~\ref{eq:single_chain_part_func},
which can be written explicitly as
\begin{equation}
\label{eq:Qp_res_spec_source}
\begin{aligned}
Q_{\rm p}[\breve{w}-\mathrm{i} \{ \breve{J}_{\alpha}\},\breve{\psi}] &=
\frac{1}{V} \int d\bm{R}_{N_{\rm p}} e^{- \mathrm{i} \Psi_{N_{\rm p}} } \int
d\bm{R}_{N_{\rm p}-1} G^0_{N_{\rm p},N_{\rm p}-1}  e^{- \mathrm{i} \Psi_{N_{\rm
p}-1}} \cdot\cdot\cdot \\ & 
\quad\quad\quad \cdot\cdot\cdot 
\int d\bm{R}_{\alpha} G^0_{\alpha+1, \alpha} e^{- \mathrm{i} \Psi_{\alpha}}
\cdot\cdot\cdot
\int d\bm{R}_{1} G^0_{2,1} e^{- \mathrm{i} \Psi_{1} } \; ,
\end{aligned}
\end{equation}
where $\Psi_{\alpha} = \breve{w}(\bm{R}_{\alpha}) + 
\sigma_{\alpha}\breve{\psi}(\bm{R}_{\alpha}) +
\mathrm{i} J_{\alpha}(\bm{R}_{\alpha})$, and
\begin{equation}
G^0_{\alpha+1,\alpha} 
\equiv G^0(\bm{R}_{\alpha+1}-\bm{R}_\alpha\vert b)
\end{equation}
with $G^0$ defined by Eq.~\ref{eq:chain_propagator}.
Substituting Eq.~\ref{eq:defn_res_spec_den} into
Eq.~\ref{eq:field_part_func_tot_source_residue_sp} then results in $\langle
\hat{\rho}^{\mathrm{(t)}}_{\mathrm{c},\alpha}(\bm{r}) \rangle = \langle
\tilde{\rho}^{\mathrm{(t)}}_{\mathrm{c},\alpha}(\bm{r}) \rangle_{\rm F}$ and
$\langle \hat{\rho}^{\mathrm{(r)}}_{\mathrm{c},\alpha}(\bm{r}) \rangle =
\langle \tilde{\rho}^{\mathrm{(r)}}_{\mathrm{c},\alpha}(\bm{r}) \rangle_{\rm
F}$, with 
\begin{equation}
\begin{aligned}
\tilde{\rho}^{\mathrm{(t)}}_{\mathrm{c},\alpha}(\bm{r}) &= \frac{1}{Q_{\rm
p}[\breve{w},\breve{\psi}]V} q_{B}(\bm{r}, \alpha) q_{F}(\bm{r},
\alpha) e^{ \mathrm{i}
\breve{w}(\bm{r}) + \mathrm{i} \sigma_{\alpha} \breve{\psi}(\bm{r}) } \; ,\\
\tilde{\rho}^{\mathrm{(r)}}_{\mathrm{c},\alpha}(\bm{r}) &= \frac{n_{\rm
p}-1}{Q_{\rm p}[\breve{w},\breve{\psi}]V}
q_{B}(\bm{r}, \alpha) q_{F}(\bm{r}, \alpha) e^{ \mathrm{i}
\breve{w}(\bm{r}) + \mathrm{i} \sigma_{\alpha} \breve{\psi}(\bm{r}) } \; .
\end{aligned}
\end{equation}
As discussed above in conjunction with eq.~\ref{eq:contact_freq_defns_inter}, 
the numbers of interchain contacts and thus interchain contact maps are 
determined in our formulation via the cross correlation function 
between the $\alpha$th bead center of the 
tagged chain and the $\beta$th bead center of the rest of the chains.
This cross correlation is readily obtained by taking double functional
derivatives of the partition function in 
Eq.~\ref{eq:field_part_func_tot_source_residue_sp} 
in accordance with Eq.~\ref{eq:defn_res_spec_corr} to yield
\begin{equation}
G^{\mathrm{(t)},\mathrm{(r)}}_{\alpha,
\beta}(\vert\bm{r}-\bm{r}'\vert) = \langle
\tilde{\rho}^{\mathrm{(t)}}_{\mathrm{c},\alpha}(\bm{r})
\tilde{\rho}^{\mathrm{(r)}}_{\mathrm{c},\beta}(\bm{r}') \rangle_{\rm F} \; 
\end{equation}
for the calculation of interchain numbers of contacts in
Eq.~\ref{eq:contact_freq_defns_inter}.
Similarly, for the numbers of intrachain contacts and intrachain contact maps, 
we compute, in accordance with Eq.~\ref{eq:contact_freq_defns_intra}, 
the correlations between the beads of the tagged chain as follows:
\begin{equation}
\label{eq:defn_res_spec_corr_intra}
\begin{aligned}
G^{\mathrm{(t)},\mathrm{(t)}}_{\alpha, \beta}(|\bm{r}|) &=
\frac{1}{V}\int d\bm{R} \left\langle
\hat{\rho}^{\mathrm{(t)}}_{\mathrm{c},\alpha}(\bm{R})
\hat{\rho}^{\mathrm{(t)}}_{\mathrm{c},\beta}(\bm{r}+\bm{R}) \right\rangle \\
&= \frac{1}{V}\int d\bm{R}
\left( \frac{1}{\mathcal{Z}[\{J^{\mathrm{(t)}}_{\alpha}\},
\{J^{\mathrm{(r)}}_{\alpha}\}]}\frac{\delta
\mathcal{Z}[\{J^{\mathrm{(t)}}_{\alpha}\},
\{J^{\mathrm{(r)}}_{\alpha}\}]}{\delta J^{\mathrm{(t)}}_{\alpha}(\bm{R}) \delta
J^{\mathrm{(t)}}_{\beta}(\bm{r}+\bm{R}) } 
\right)_{\{J_{\alpha}^{\mathrm{(t)}}\}=\{J_{\alpha}^{\mathrm{(r)}}\}=0} \\ 
& = \frac{1}{V} \int d\bm{R} \left(
\frac{1}{Q_{\rm p}[\breve{w}-\mathrm{i} \{
J_{\alpha}^{\mathrm{(t)}}\},\breve{\psi}]} \frac{\delta^2 Q_{\rm
p}[\breve{w}-\mathrm{i} \{ J_{\alpha}^{\mathrm{(t)}}\},\breve{\psi}]}{\delta
J^{\mathrm{(t)}}_{\alpha}(\bm{R}) \delta
J^{\mathrm{(t)}}_{\beta}(\bm{R}+\bm{r})} \right)_{\{J_{\alpha}^{\mathrm{(t)}}
\}=0} \\
& = \left\langle \tilde{G}^{\mathrm{(t)},\mathrm{(t)}}_{\alpha,
\beta}(\bm{r})\right\rangle_{\rm F} \; , 
\end{aligned}
\end{equation}
where
\begin{equation}
\label{eq:intra_corr_func_operator}
\begin{aligned}
\tilde{G}^{\mathrm{(t)},\mathrm{(t)}}_{\alpha, \beta}(\bm{r}) & =
\frac{1}{V} \int d\bm{R} \left[ \frac{1}{VQ_{\rm p}[\breve{w}, \breve{\psi}]}
\int d\bm{R}^{\mathrm{(t)}}_{N_{\rm p}} e^{- \mathrm{i} \Psi_{N_{\rm
p}}(\bm{R}^{\mathrm{(t)}}_{N_{\rm p}})} \int d\bm{R}^{\mathrm{(t)}}_{N_{\rm
p}-1} G^0_{N_{\rm p},N_{\rm p}-1} e^{- \mathrm{i} \Psi_{N_{\rm
p}-1}(\bm{R}^{\mathrm{(t)}}_{N_{\rm p}-1})} \cdots \right.\\ 
&\left. \quad\quad \cdot\cdot\cdot
\int d\bm{R}^{\mathrm{(t)}}_{2} G^0_{3,2} e^{- \mathrm{i}
\Psi_{2}(\bm{R}^{\mathrm{(t)}}_{2})} \int d\bm{R}^{\mathrm{(t)}}_{1} G^0_{2,1}
e^{- \mathrm{i} \Psi_{1}(\bm{R}^{\mathrm{(t)}}_{1})} \right] \delta(\bm{R} -
\bm{R}^{\mathrm{(t)}}_{\alpha}) \delta(\bm{R} + \bm{r} -
\bm{R}^{\mathrm{(t)}}_{\beta}) \\
& = \frac{1}{VQ_{\rm p}[\breve{w}, \breve{\psi}]} \int \frac{d\bm{k}}{(2\pi)^3}
\left[ \frac{1}{V} \int d\bm{R}^{\mathrm{(t)}}_{N_{\rm p}} e^{- \mathrm{i}
\Psi_{N_{\rm p}}(\bm{R}^{\mathrm{(t)}}_{N_{\rm p}})} \int
d\bm{R}^{\mathrm{(t)}}_{N_{\rm p}-1} G^0_{N_{\rm p},N_{\rm p}-1} e^{- \mathrm{i}
\Psi_{N_{\rm p}-1}(\bm{R}^{\mathrm{(t)}}_{N_{\rm p}-1})}\cdots \right.\\ 
&\left.  \quad\quad
\cdot\cdot\cdot \int d\bm{R}^{\mathrm{(t)}}_{2} G^0_{3,2} e^{- \mathrm{i}
\Psi_{2}(\bm{R}^{\mathrm{(t)}}_{2})} \int d\bm{R}^{\mathrm{(t)}}_{1} G^0_{2,1}
e^{- \mathrm{i} \Psi_{1}(\bm{R}^{\mathrm{(t)}}_{1})} \right]
e^{\mathrm{i}\bm{k}\cdot(\bm{R}^{\mathrm{(t)}}_{\beta} -
\bm{R}^{\mathrm{(t)}}_{\alpha} - \bm{r})}\\
&= \frac{1}{VQ_{\rm p}} \int \frac{d\bm{k}}{(2\pi)^3} e^{-\mathrm{i}\bm{k}\cdot
\bm{r}}Q^{(\alpha,\beta)}_{\rm p}(\bm{k}) \; , 
\end{aligned}
\end{equation}
with $\Psi_{\alpha} = \breve{w}(\bm{R}^{\mathrm{(t)}}_{\alpha}) +
\sigma_{\alpha}\breve{\psi}(\bm{R}^{\mathrm{(t)}}_{\alpha})$. In
the last Eq.~\ref{eq:intra_corr_func_operator}, 
$Q^{(\alpha,\beta)}_{\rm p}(\bm{k})$
is a $\bm{k}$-dependent (Fourier-transformed) 
single chain restricted partition function, given by
\begin{equation}
\label{eq:single_chain_k_dep_restricted_part_funcn}
\begin{aligned}
Q^{(\alpha,\beta)}_{\rm p}(\bm{k}) & = \left[ \frac{1}{V} \int
d\bm{R}^{\mathrm{(t)}}_{N_{\rm p}} e^{- \mathrm{i} \Psi_{N_{\rm
p}}(\bm{R}^{\mathrm{(t)}}_{N_{\rm p}})} \int d\bm{R}^{\mathrm{(t)}}_{N_{\rm
p}-1} G^0_{N_{\rm p},N_{\rm p}-1} e^{- \mathrm{i} \Psi_{N_{\rm
p}-1}(\bm{R}^{\mathrm{(t)}}_{N_{\rm p}-1})} \right.\\ 
&\left. \quad\quad \cdot\cdot\cdot
\int d\bm{R}^{\mathrm{(t)}}_{2} G^0_{3,2} e^{- \mathrm{i}
\Psi_{2}(\bm{R}^{\mathrm{(t)}}_{2})} \int d\bm{R}^{\mathrm{(t)}}_{1} G^0_{2,1}
e^{- \mathrm{i} \Psi_{1}(\bm{R}^{\mathrm{(t)}}_{1})} \right]
e^{\mathrm{i}\bm{k}\cdot(\bm{R}^{\mathrm{(t)}}_{\beta} -
\bm{R}^{\mathrm{(t)}}_{\alpha})} \; .
\end{aligned}
\end{equation}
This implies that in order to evaluate
$\tilde{G}^{\mathrm{(t)},\mathrm{(t)}}_{\alpha, \beta}(\bm{r})$
in Eq.~\ref{eq:intra_corr_func_operator}, we need to first evaluate
$Q^{(\alpha,\beta)}_{\rm p}(\bm{k})$ at each $\bm{k}$ and then perform
an inverse Fourier transform according to the last equality in
Eq.~\ref{eq:intra_corr_func_operator}. Therefore,
for a $N_{\rm L}\times N_{\rm L}\times N_{\rm L}$ spatial $(x,y,z)$ lattice 
used for FTS, ideally we should calculate the values of
$Q^{(\alpha,\beta)}_{\rm p}({\bm k})$ on a (reciprocal) ${\bm k}$-space lattice
of the same size, i.e., determine $Q^{(\alpha,\beta)}_{\rm p}({\bm k})$ 
$(N_{\rm L})^3$ times for each $(\alpha,\beta)$ pair for each field 
configuration (Eq.~\ref{eq:single_chain_k_dep_restricted_part_funcn}). 
This would be exceedingly computationally intensive. 
For computational efficiency, we consider instead a slightly more
coarse-grained $\bm{k}$-space
lattice of dimensions $N_{\bm k}\times N_{\bm k}\times N_{\bm k}$ 
with $N_{\bm k} = N_{\rm L}/2$, such that the resolution of its
reciprocal space, i.e., the original $(x,y,z)$ space, becomes 
$(\Delta k_x, \Delta k_y, \Delta k_z) 
\equiv\left( 4\pi/L,4\pi/L,4\pi/L \right)$
with $L = N_{\rm L} \bar{a}$. 
Nonetheless, we still compute the 
$\left(\prod_{\alpha} \int d\bm{R}_{\alpha}\right)$ volume integrals
in Eq.~\eqref{eq:single_chain_k_dep_restricted_part_funcn} in the original
$N_{\rm L}\times N_{\rm L}\times N_{\rm L}$ spatial lattice with 
resolution $(\Delta x, \Delta y, \Delta z)=(\bar{a}, \bar{a}, \bar{a})$.
In this way, we reduce the computational cost by $\sim 8$ times.
Thus, for all the intrachain FTS contact maps presented here, $N_{\rm L}=32$, 
$N_{\bm k}=16$, and the final equilibrium field configurations of $32$ 
independent runs are used for field averaging. As a check on the accuracy
of this coarse-graining, we have also computed contact maps with 
$N_{\bm k}=8$ reciprocal lattice and saw no visible difference from
the $N_{\bm k}=16$ results.
\\


{\bf Volume integration in simulation boxes with periodic boundary conditions.}
To perform volume integration of an isotropic function 
$f(r)$ in a $L\times L\times L$ cubic box ($r\equiv\vert \bm{r} \vert$)
with periodic boundary conditions, the integration measure $dV$ has to
be modified owing to the periodic boundary conditions, as follows:
\begin{eqnarray}
\label{eq:dV1}
dV(r) &=& \left\lbrace \begin{matrix}
4 \pi r^2 dr          \, , &   0\leq r \leq L/2 \, ,  \\
2 \pi r (3L - 4 r) dr   \, , & L/2 < r \leq \sqrt{2}L/2 \, ,  \\
2 r L [3 \pi - 12 g_1(r/L) + g_2(r/L) ] dr \, , 
                         & \sqrt{2}L/2 < r \leq \sqrt{3}L/2 \, ,\\
0, & r >  \sqrt{3}L/2 \, ,
\end{matrix} \right.            
\end{eqnarray}
where
\begin{eqnarray}
\label{eq:dV2}
g_1(x) &=& \tan^{-1} \sqrt{4 x^2 - 2} \, , \nonumber \\
g_2(x) &=& 8 x \left\{ \tan^{-1}
\left[ \frac{ 2 x (4 x^2-3) }{\sqrt{4 x^2 - 2} (4 x^2 + 1) } \right]\right\}\, ,
\end{eqnarray}
such that integration of any isotropic function $f(r)$ from $r=0$
to any upper integration limit $r=r_{\rm max}$ on the periodic
lattice is specified by
\begin{equation}
\int_{|\bm{r}|\le r_{\rm max}} d\bm{r} f(r) = 
\int_{0}^{r_{\rm max}} dV(r) f(r) \; 
\end{equation}
with $dV(r)$ defined in Eqs.~\ref{eq:dV1} and \ref{eq:dV2} above.
\\


{\bf Numerical estimation of $T^*_{\rm cr}$ in FTS}
Binodal phase boundaries in FTS are computed using the method described
in ref.~\citen{MiMB2023}. In the present study, to estimate the critical 
temperature $T^*_{\rm cr}$ from the FTS binodals, we adopt---as in recent 
coarse-grained MD studies of biomolecular 
condensates\cite{suman2,dignon18}---the scaling approach 
outlined in ref.~\citen{Panagiotopoulos_Scaling_2017}, which assumes
that low- and high-density phase concentrations, denoted by $\rho_{\rm L}$ and
$\rho_{\rm H}$, respectively, follow the relations
\begin{subequations}
\label{eq:fitting_a_la_panagiotopoulos}
\begin{align}
\frac{\rho_{\rm H} + \rho_{\rm L}}{2} & = \rho_{\rm cr}+A(T^*_{\rm cr}-T^*) \; ,
\label{eq:avg_func}\\
\rho_{\rm H} - \rho_{\rm L} & = \Delta \rho_0 \left(1-\frac{T^*}{T^*_{\rm cr}} 
\right)^{\nu} \; , \label{eq:diff_func}
\end{align}
\end{subequations}
where $\nu=0.325$, and $A$, $\Delta \rho_0$, critical density $\rho_{\rm cr}$,
and $T^*_{\rm cr}$ are free fitting parameters. Now, for
each sequence, we use the highest two simulated $T^*$ for fitting. 
First, we estimate $\Delta \rho_0$ and $T^*_{\rm cr}$ by fitting the 
numerical values of $(\rho_{\rm H} - \rho_{\rm L})$ to
Eq.~\ref{eq:diff_func}.
Next, we apply this fitted $T^*_{\rm cr}$ to Eq.~\eqref{eq:avg_func} 
to fit the numerical values of $(\rho_{\rm H} + \rho_{\rm L})/2$
to yield fitted values for $\rho_{\rm cr}$ and $A$. 
These fitted parameters are then applied to obtain
$\rho_{\rm H}$ and $\rho_{\rm L}$ as functions of $T^*$ 
from the two relations in 
Eq.~\ref{eq:fitting_a_la_panagiotopoulos}. In Fig.~S2 (see below)
and maintext Fig.~1d, these fitted functions are used to construct 
continuous curves through the fitted $(\rho_{\rm cr},T^*_{\rm cr})$
critical point and the four simulated binodal 
$(\rho_{\rm L},T^*)$ and $(\rho_{\rm H},T^*)$ 
datapoints with the two highest $T^*$, whereas
the numerical FTS $(\rho_{\rm L},T^*)$ and $(\rho_{\rm H},T^*)$
datapoints for the dilute and dense branches of the binodals
for the rest of the simulated $T^*$ are simply connected by lines
as guides for the eye.
\\

\noindent
{\large\bf Coarse-grained molecular dynamics (MD) model of polyampholyte
conformations and phase separation}

The coarse-grained MD model here is essentially identical to the ``hard-core 
repulsion'' model we used previously in ref.~\citen{suman2} for the 
simulation of ``sv''\cite{rohit2013} and ``as''\cite{suman2} polyampholyte 
sequences. The only difference is that the repulsive part of the 
Lennard-Jones (LJ) potential is set to zero for $r>2^{1/6}a$ in the 
present study---where $a$ is the reference (equilibrium) bond length 
between successive beads (monomers) along the chain sequence---instead 
of for $r>a$ in ref.~\citen{suman2}. MD simulation in the present work is
carried out using the protocol described in 
refs.~\citen{suman2,WessenDasPalChan2022}.
A contact is defined to exist between two monomers when their center-of-mass
spatial separation is within $2a$.

\vfill\eject
\setcounter{table}{0}
\renewcommand{\tablename}{{\bf Table}}
\renewcommand{\thetable}{{\bf S}{\bf \arabic{table}}}

\begin{table*}
 \caption{All 26 polyampholyte sequences studied in this work. Following
previous notation,\cite{rohit2013,kings2015,lin2017,joanJPCL2019,suman2}
positively and negatively charged beads along the
sequences are symbolized here, respectively, by ``K'' (lysine, charge $+1$) 
and ``E'' (glutamic acid, charge $-1$).}\label{tab:all_sequences} 
\vskip 0.5cm
 \begin{tabular}{ll}
 {\bf Name} & $\null\hskip 4.3cm${\bf Sequence} \\  \hline \hline
sv1 & \texttt{EKEKEKEKEKEKEKEKEKEKEKEKEKEKEKEKEKEKEKEKEKEKEKEKEK}\\ 
sv2 & \texttt{EEEKKKEEEKKKEEEKKKEEEKKKEEEKKKEEEKKKEEEKKKEEEKKKEK}\\
sv10    & \texttt{EKKKKKKEEKKKEEEEEKKKEEEKKKEKKEEKEKEEKEKKEKKEEKEEEE}\\
sv12    & \texttt{EKKEEEEEEKEKKEEEEKEKEKKEKEEKEKKEKKKEKKEEEKEKKKKEKK}\\
sv15    & \texttt{KKEKKEKKKEKKEKKEEEKEKEKKEKKKKEKEKKEEEEEEEEKEEKKEEE}\\
sv17    & \texttt{EKEKKKKKKEKEKKKKEKEKKEKKEKEEEKEEKEKEKKEEKKEEEEEEEE}\\
sv20    & \texttt{EEKEEEEEEKEEEKEEKKEEEKEKKEKKEKEEKKEKKKKKKKKKKKKEEE}\\
sv23    & \texttt{EEEEEKEEEEEEEEEEEKEEKEKKKKKKEKKKKKKKEKEKKKKEKKEEKK}\\
sv24    & \texttt{EEEEKEEEEEKEEEEEEEEEEEEKKKEEKKKKKEKKKKKKKEKKKKKKKK}\\
sv28    & \texttt{EKKKKKKKKKKKKKKKKKKKKKEEEEEEEEEEEEEEEEEEKKEEEEEKEK}\\
sv29    & \texttt{KEEEEKEEEEEEEEEEEEEEEEEEEEEKKKKKKKKKKKKKKKKKKKKKKK}\\
sv30    & \texttt{EEEEEEEEEEEEEEEEEEEEEEEEEKKKKKKKKKKKKKKKKKKKKKKKKK}
\vspace{0.2cm} \\
as1     & \texttt{KKKKKEKKKKKEKKKKEKKKKEKKEKEEEKEEEEKEEEEKEEEEKEEEEE}\\
as2     & \texttt{EEEEEEEEKKEKKKKEEEEEEEEEEEKKKKEEEKEKEKKKKKKKKKKKKK}\\
as3     & \texttt{KKKKEEEEEEEEEEEEEEEEEEKKEKKKKKKKKKKKKKKEKEEEEEKKKK}\\
as4     & \texttt{KKKKKKEEEEEEEEEEEEEEEKKKKKKKKKKKKKKKKKKEEEEEEEEEEK}
\vspace{0.2cm} \\
c$\kappa$1& \texttt{KKKEEEEEEEEKKKKKKKKKKEEEEEEEEEEKKKKKKKKKEEKEEEKEKE}\\
c$\kappa$2 & \texttt{KKKEEEEEEEKKEEEEEKKKEKKKKKKKKEEEEEEEEEEEEKKKKKKKKK}\\
c$\kappa$3 & \texttt{KKKKKKKKKKKKKKKEKKKKKEKKEEEEKEEEEEKEEEEEKEEEEEEEEE}\\
c$\kappa$4 & \texttt{KKKKKKKKKKEKKKKKEKKKKEKKKKEEKEEEEEKEEEEEEEEEEEEEEE}
\vspace{0.2cm}  \\
cSCD1 & \texttt{KKKKEKKKKEEKKKKKKKEEKKKKKEEEEKKEEEEEEEKEEKEEEKEEEE}\\
cSCD2 & \texttt{KEKKKKKKKKEEKKKKKKEEKKKKKEEEEEKKEEEEEEEEKKKEEEEEEE}\\
cSCD3   & \texttt{EEEEEEEKKKKKKKKKKKKKKKEEEKKKKKKKKKKEEEEEEEEEEEEEEE}\\
cSCD4 & \texttt{EEEEEKKKKKKKKKKKKKKKKKKKKEEEEEEEEEEEEEEEEEEEEKKKKK}
\vspace{0.2cm} \\
obs1 &  \texttt{KKKKKKKKKKEEEEEEEEEEEEKKKKKEEEEEEEEEEEEEKKKKKKKKKK}\\
ebs1    & \texttt{EEEEEEEEEKKKKKKKKEEEEEEEEKKKKKKKKEEEEEEEEKKKKKKKKK}\\
 \end{tabular}
 \end{table*}

\vfill\eject
\setcounter{figure}{0}
\renewcommand{\figurename}{{\bf Fig.}}
\renewcommand{\thefigure}{{\bf S}{\bf \arabic{figure}}}


%
%



%



\begin{figure*}[ht]
\centerline{\large\bf Supporting Figures}
{\includegraphics[width=0.78\columnwidth,angle=0]{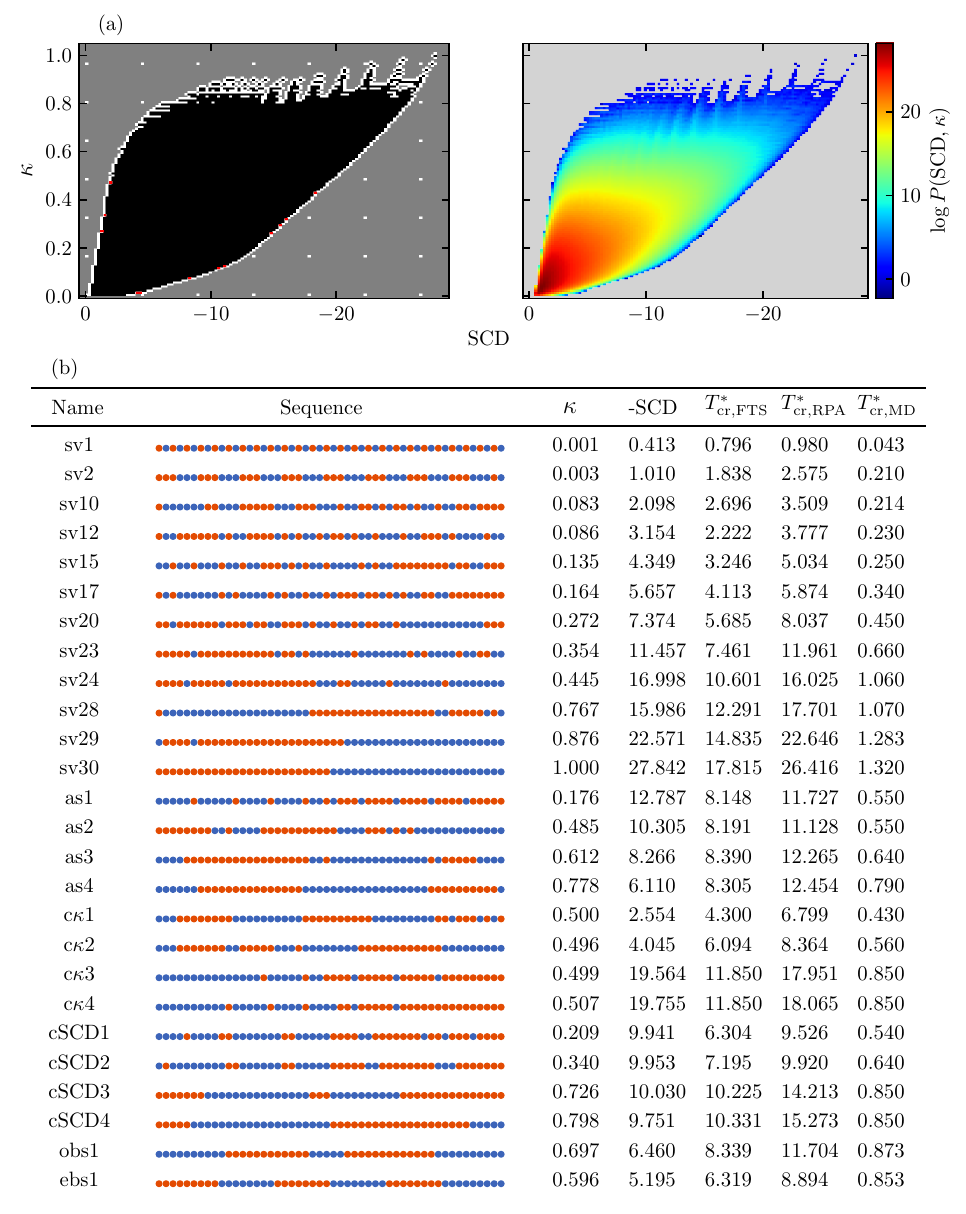}}
\caption{Genetic-algorithm (GA) and Wang-Landau (WL) sampling of 
the SCD--$\kappa$ sequence space of net-neutral 50mer polyampholytes. 
(a) {\it Left}-- Results of the diversity-enhanced 
GA scan. Black region represents the region found by the GA to be 
populated by sequences.  Bins containing target $({\rm SCD},\kappa)$ 
values for which no sequence was found by the GA are shown in white. Red bins 
indicate bins mislabelled by the GA as empty, i.e.,~bins for which the GA 
was not able to find a sequence but that the subsequent WL calculation 
revealed to be non-empty. 
Other regions where no sequence was found by either GA or WL are in gray.
{\it Right}-- Joint 
distribution $P({\rm SCD},\kappa)$ 
computed using the WL algorithm. (b) The 26 net-neutral polyampholyte 
sequences studied in this work (same as those depicted in maintext Fig.~1b), 
their $\kappa$, SCD values as well their critical temperatures in the FTS, 
RPA, and MD models. Positively and negatively charged monomers along 
the sequences are represented, respectively, by blue and red beads.
}
\label{figS1}
\end{figure*}

\vfill\eject

\begin{figure*}[ht]
{\includegraphics[width=0.90\columnwidth,angle=0]{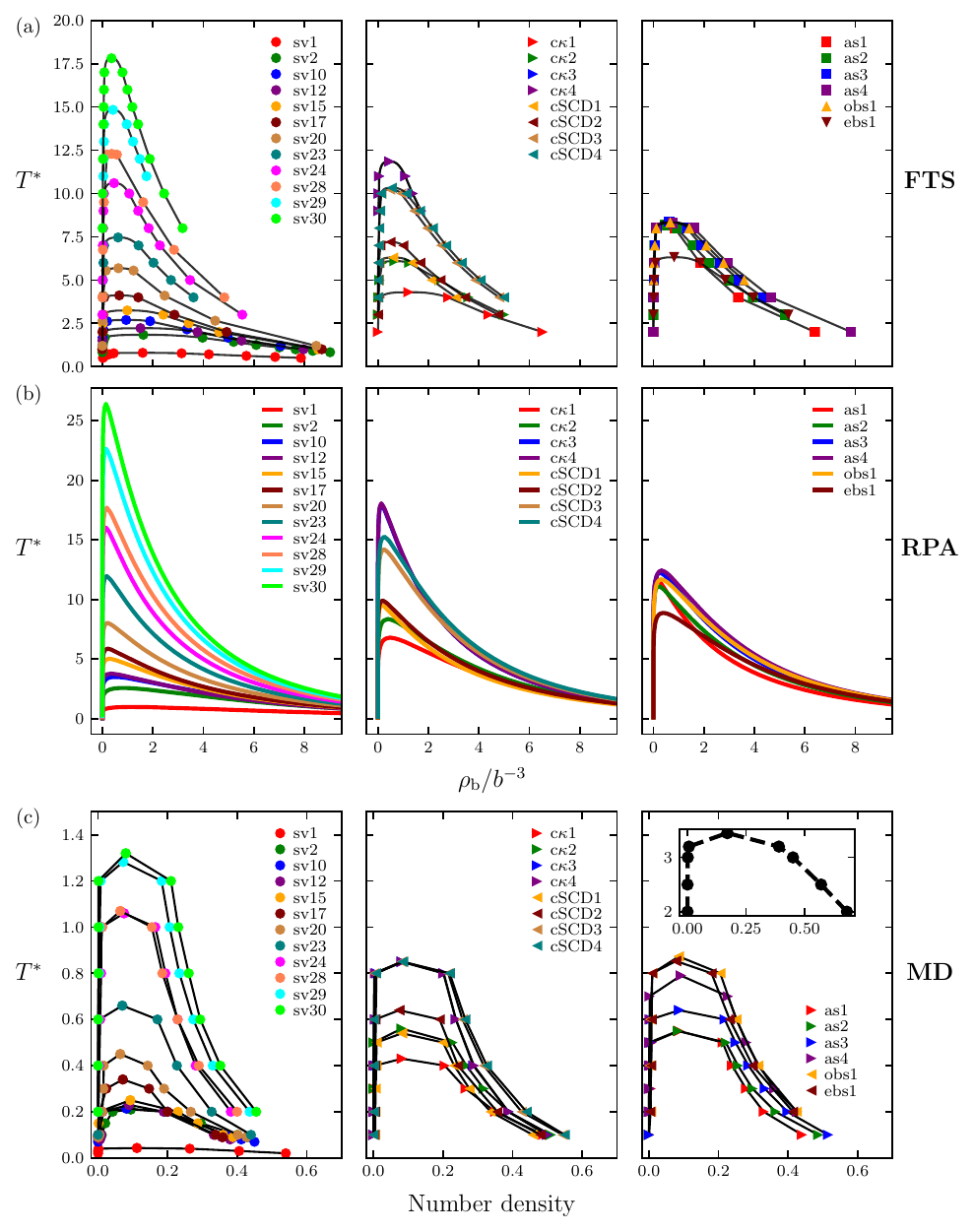}}
\caption{Phase diagrams of all 26 polyampholytes we consider in the FTS, 
RPA, and MD models. The phase diagram for the baseline homopolymer model 
in MD is shown as an inset in the bottom-right panel. 
Results in maintext Fig.~1d,e are included here as well for completeness
and to facilitate comparison.
}
\label{figS2}
\end{figure*}

\vfill\eject

\begin{figure*}[ht]
{\includegraphics[width=0.90\columnwidth,angle=0]{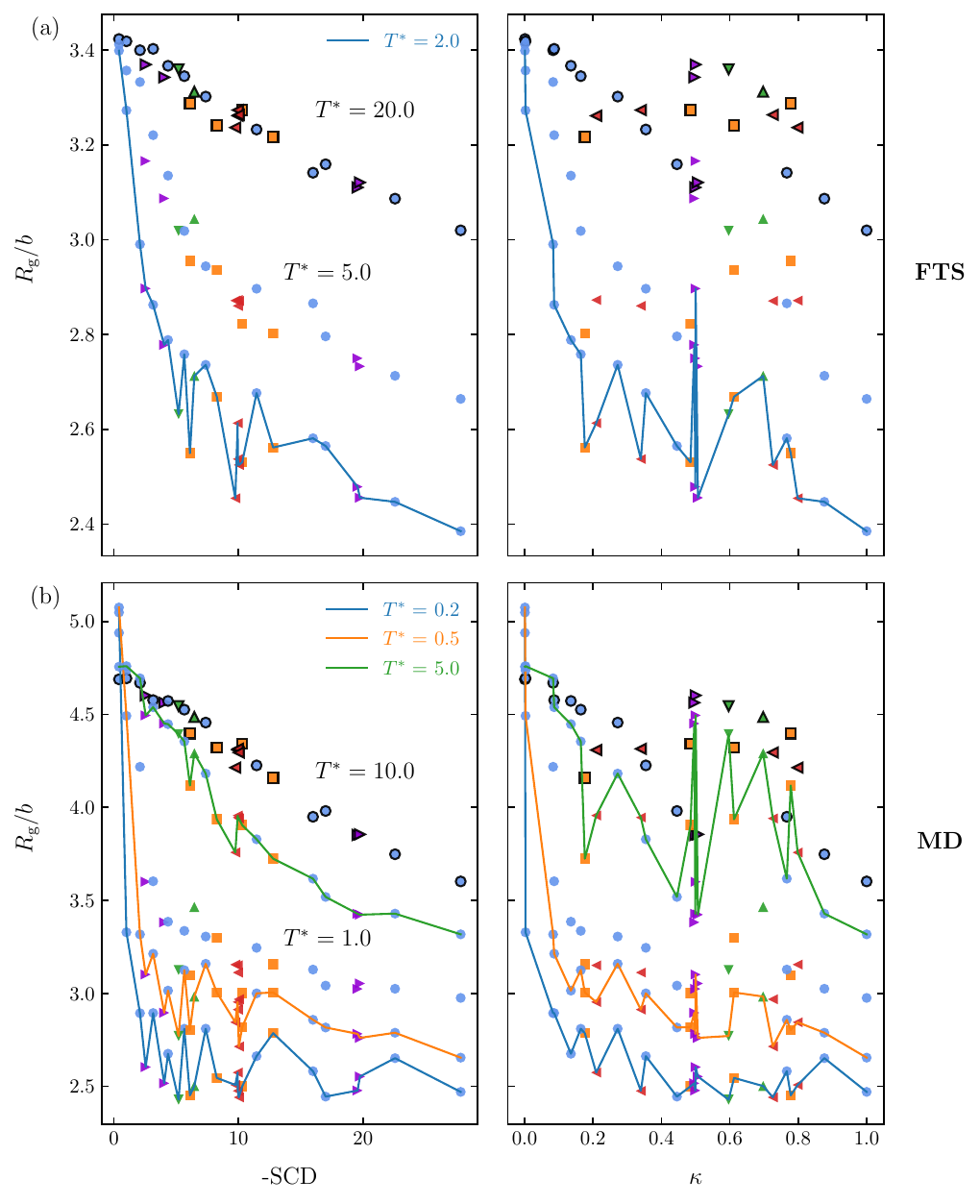}}
\caption{Variation of polyampholyte root-mean-square radius of gyration 
in the FTS (top) and MD (bottom) models
with SCD (left), $\kappa$ (right), and temperature (as indicated). 
Unless specified otherwise, root-mean-square radius of gyration, 
$\sqrt{{R_{\rm g}}^2}$, is denoted simply as $R_{\rm g}$ for notational
simplicity in the present study.
Results in maintext Fig.~2a,b,e,f are included here as well
to facilitate comparison.
}
\label{figS3}
\end{figure*}

\vfill\eject

\begin{figure*}[ht]
{\includegraphics[width=0.95\columnwidth,angle=0]{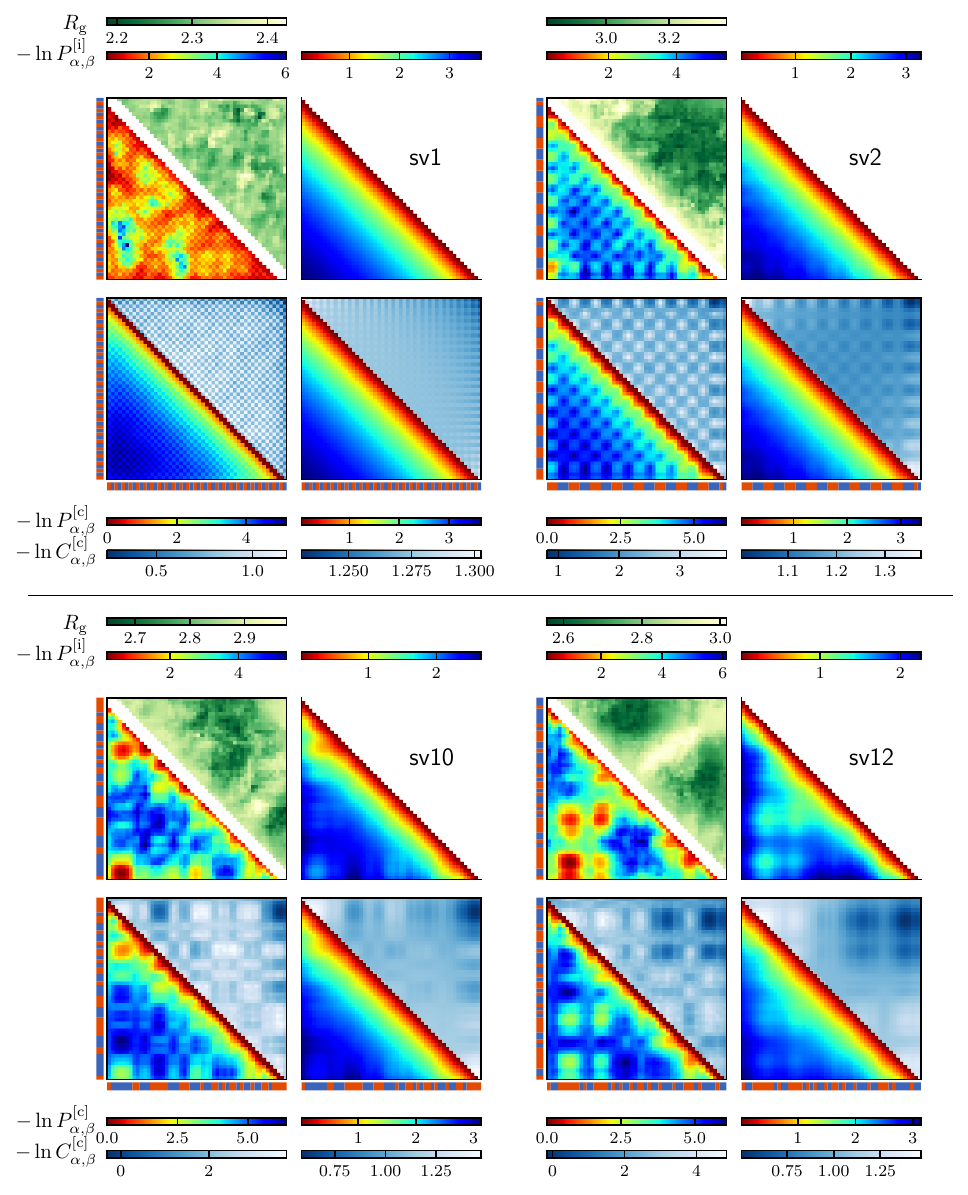}}
\caption{Contact patterns of polyampholytes as an isolated chain
versus in the multiple-chain condensed phase. Results are presented
in the same style as that in maintext Fig.~3b,c for the sequence obs1.
The corresponding results for all the other 25 polyampholyte sequences
are provided here. To facilitate comparison, 
the obs1 contact patterns in maintext Fig.~3b,c are
shown here again as the last item in this supporting figure
(1st page, {\it to be cont'd}).
}
\label{figS4a}
\end{figure*}

\vfill\eject

\setcounter{figure}{3}
\begin{figure*}[ht]
{\includegraphics[width=0.95\columnwidth,angle=0]{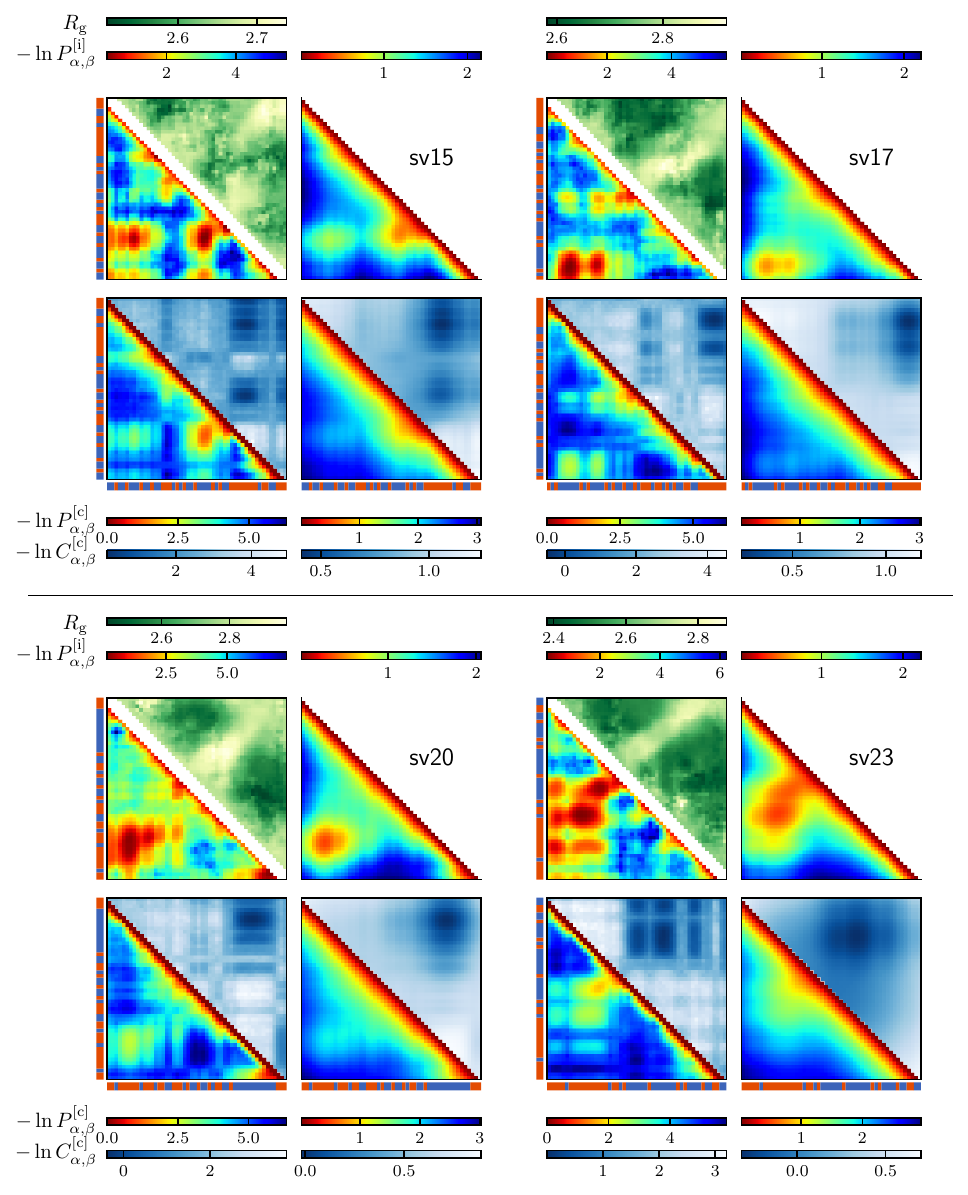}}
\caption{(2nd page, {\it cont'd}).}
\label{figS4b}
\end{figure*}

\vfill\eject

\setcounter{figure}{3}
\begin{figure*}[ht]
{\includegraphics[width=0.95\columnwidth,angle=0]{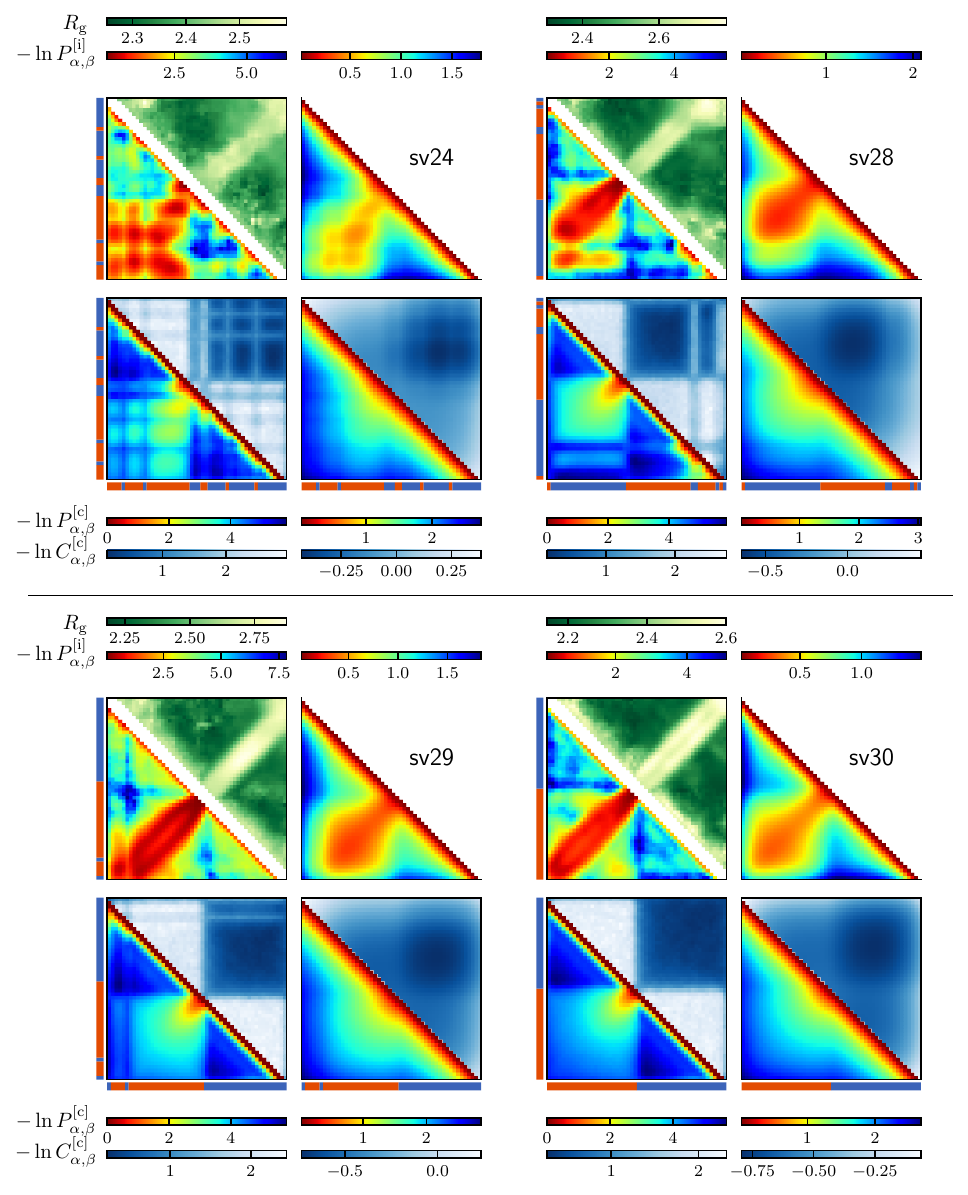}}
\caption{(3rd page, {\it cont'd}).}
\label{figS4c}
\end{figure*}

\vfill\eject

\setcounter{figure}{3}
\begin{figure*}[ht]
{\includegraphics[width=0.95\columnwidth,angle=0]{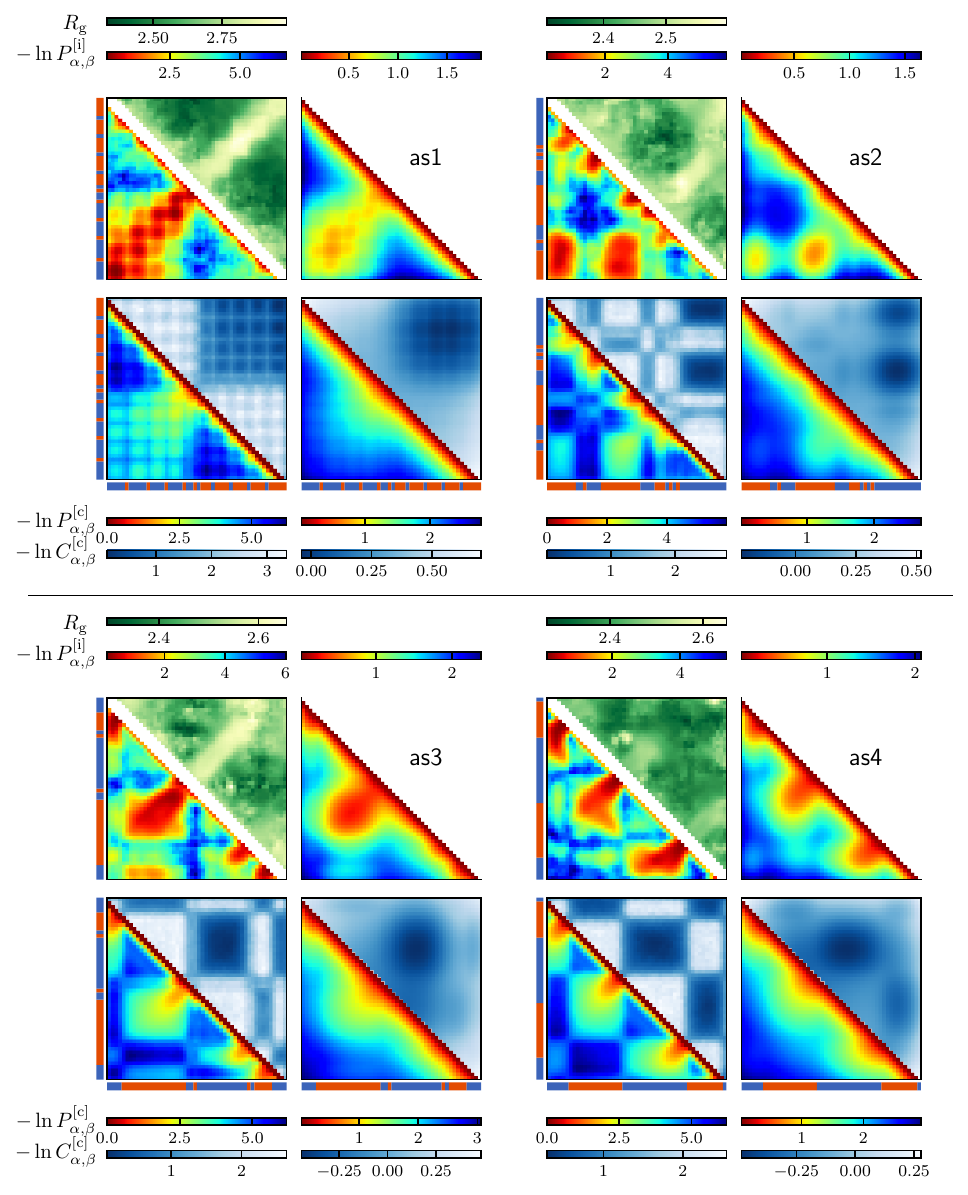}}
\caption{(4th page, {\it cont'd}).}
\label{figS4d}
\end{figure*}

\vfill\eject

\setcounter{figure}{3}
\begin{figure*}[ht]
{\includegraphics[width=0.95\columnwidth,angle=0]{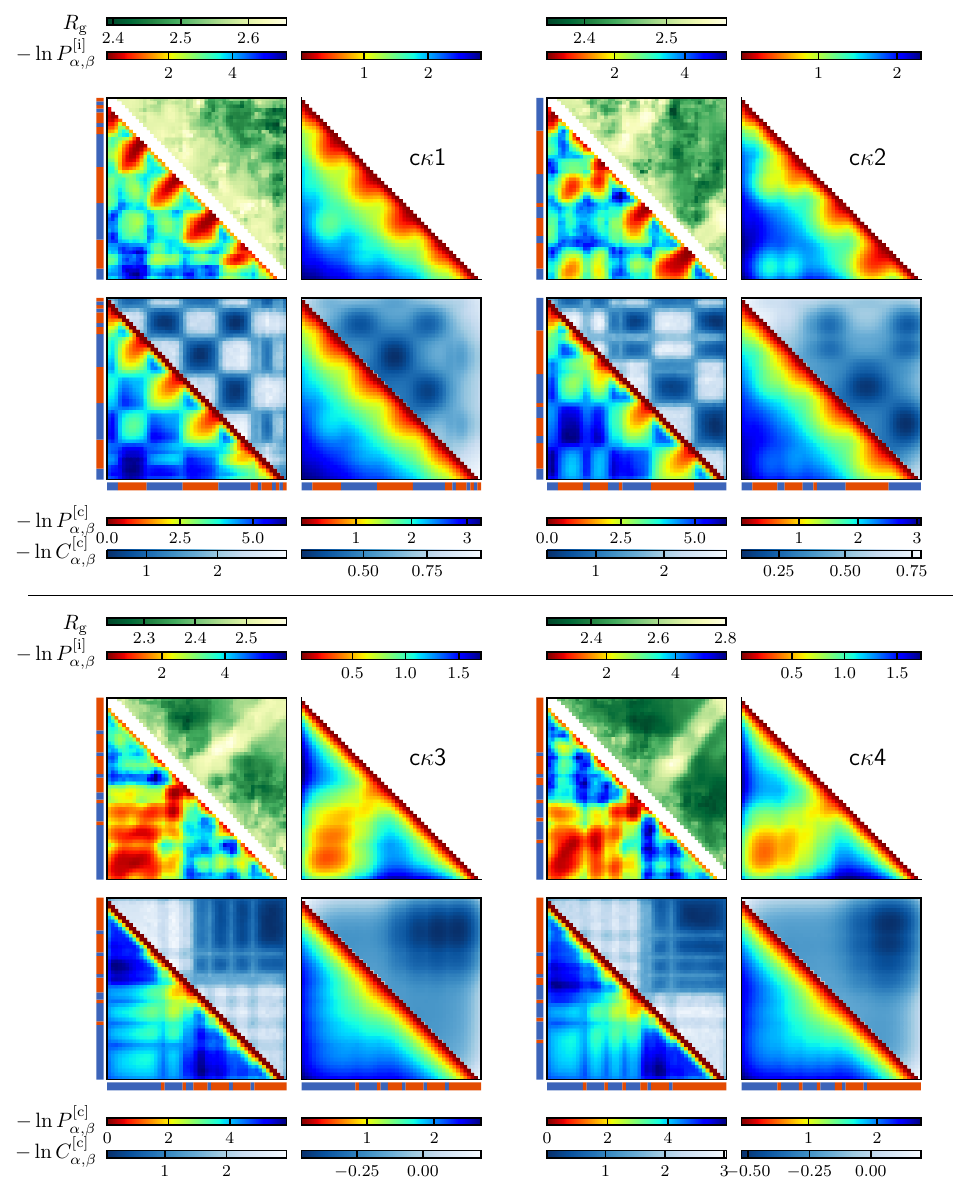}}
\caption{(5th page, {\it cont'd}).}
\label{figS4e}
\end{figure*}

\vfill\eject

\setcounter{figure}{3}
\begin{figure*}[ht]
{\includegraphics[width=0.95\columnwidth,angle=0]{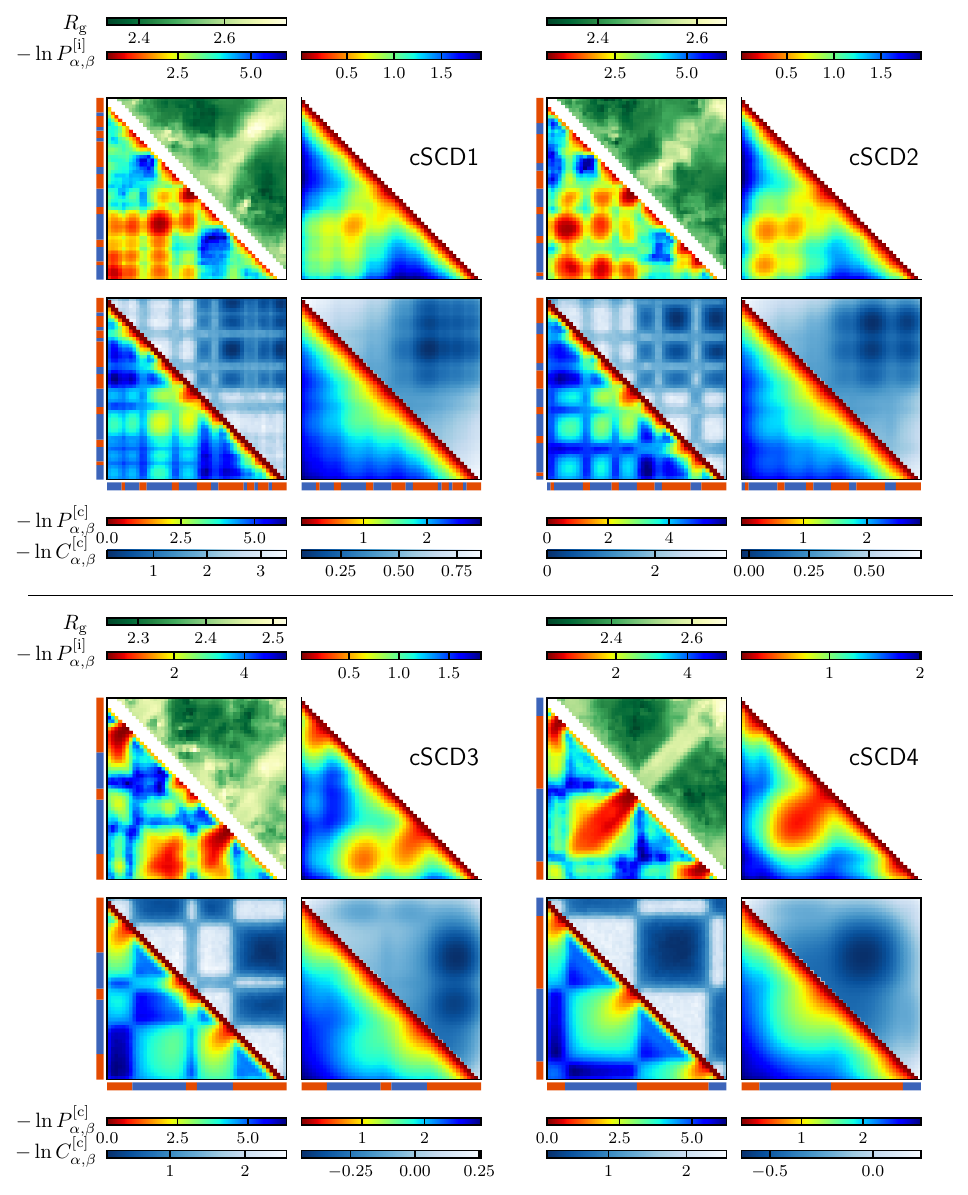}}
\caption{(6th page, {\it cont'd}).}
\label{figS4bf}
\end{figure*}

\vfill\eject

\setcounter{figure}{3}
\begin{figure*}[ht]
{\includegraphics[width=0.95\columnwidth,angle=0]{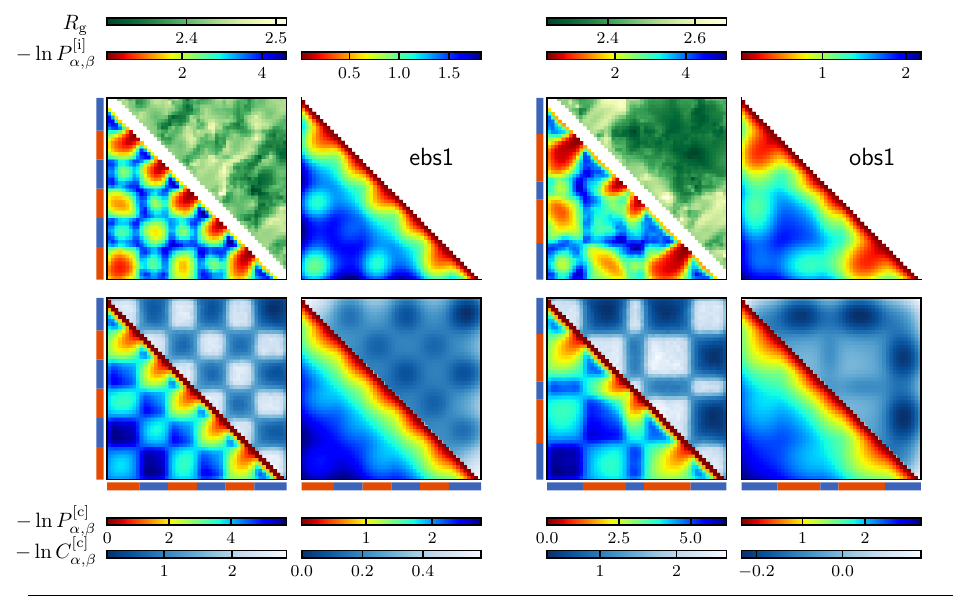}}
\caption{(7th page, {\it cont'd}). This is the last page of this supporting
figure.}
\label{figS4g}
\end{figure*}

\vfill\eject


\begin{figure*}[ht]
\vskip -1.0cm
{\includegraphics[width=0.90\columnwidth,angle=0]{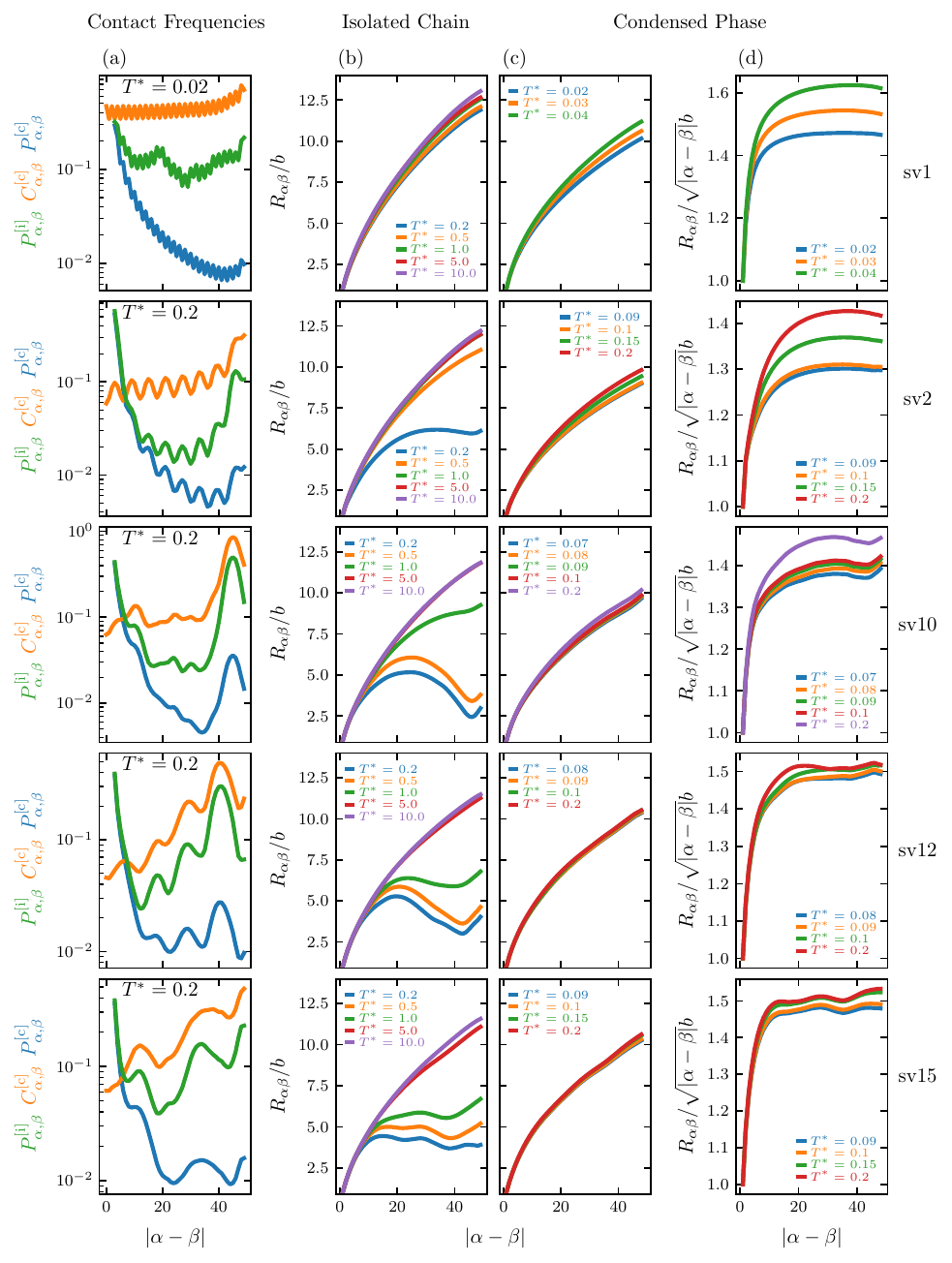}}
\vskip -0.5cm
\caption{Intrachain and interchain contact frequencies and intrachain
root-mean-square distance $R_{\alpha\beta}$ of polyampholytes, 
in essentially the same style as that in maintext Fig.~3e--g for the 
sequence obs1. The corresponding results for all the other 25 polyampholyte 
sequences are provided here. 
The data here in columns (a) and (b) for other sequences correspond, 
respectively, to the contact frequency data in maintext Fig.~3e and 3f. 
Similarly, the data here in columns (c) and (d) correspond, respectively to
the $R_{\alpha\beta}$ and $R_{\alpha\beta}/\sqrt{|\alpha-\beta|}$ 
data in maintext Fig.~3g,
with the dashed curves in Fig.~3g now replaced by solid curves in column (d).
In (d), positive slopes at large $|\alpha - \beta|$
observed for some sequences such as sv28, sv29, sv30, as1, c$\kappa$4, 
and cSCD1 indicate that their condensed-phase conformations are more 
expanded than Gaussian chains (1st page, {\it to be cont'd}).
}
\label{figS5a}
\end{figure*}

\vfill\eject

\setcounter{figure}{4}
\begin{figure*}[ht]
{\includegraphics[width=0.95\columnwidth,angle=0]{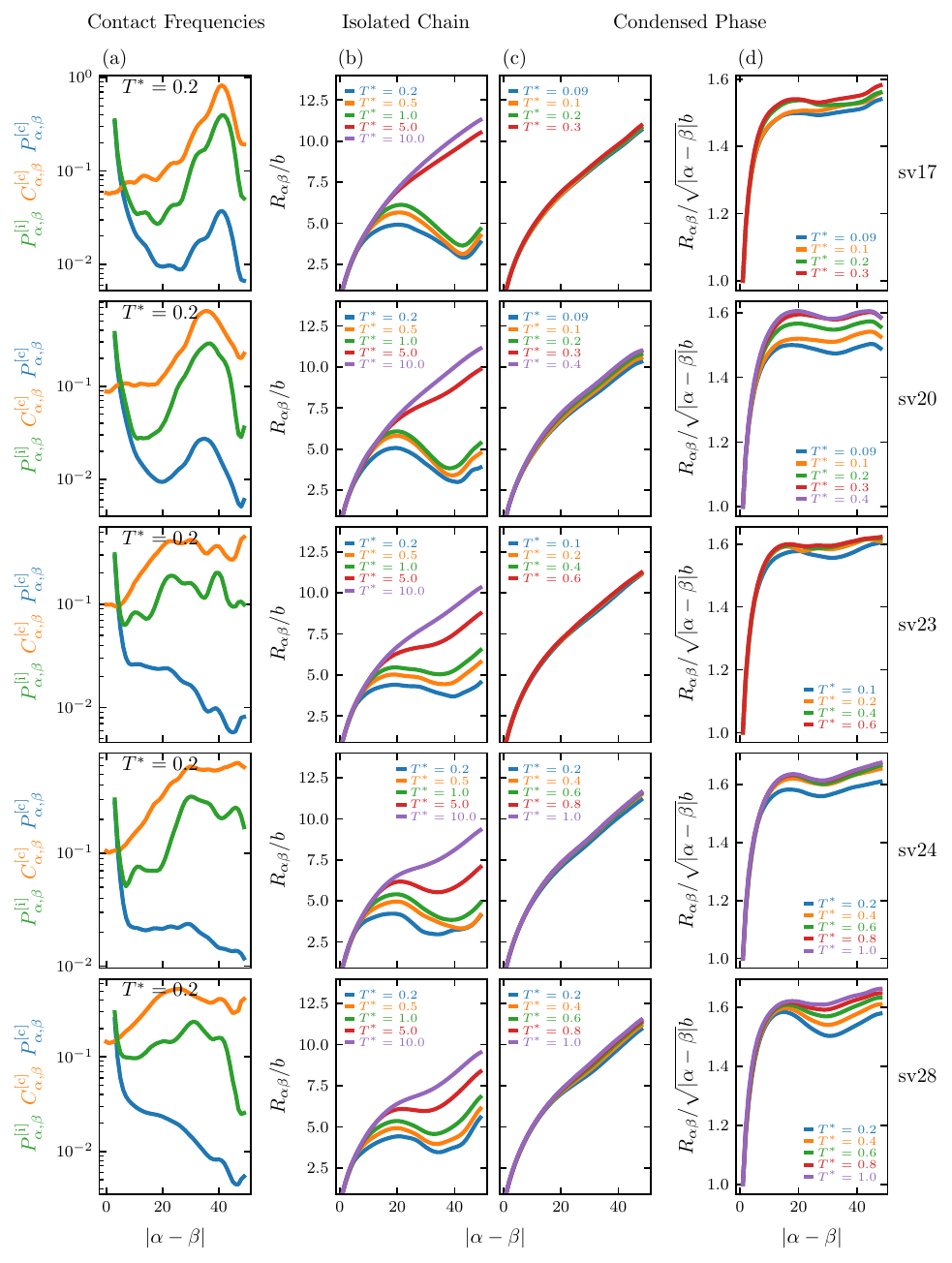}}
\caption{(2nd page, {\it cont'd}).}
\label{figS5b}
\end{figure*}

\vfill\eject

\setcounter{figure}{4}
\begin{figure*}[ht]
{\includegraphics[width=0.95\columnwidth,angle=0]{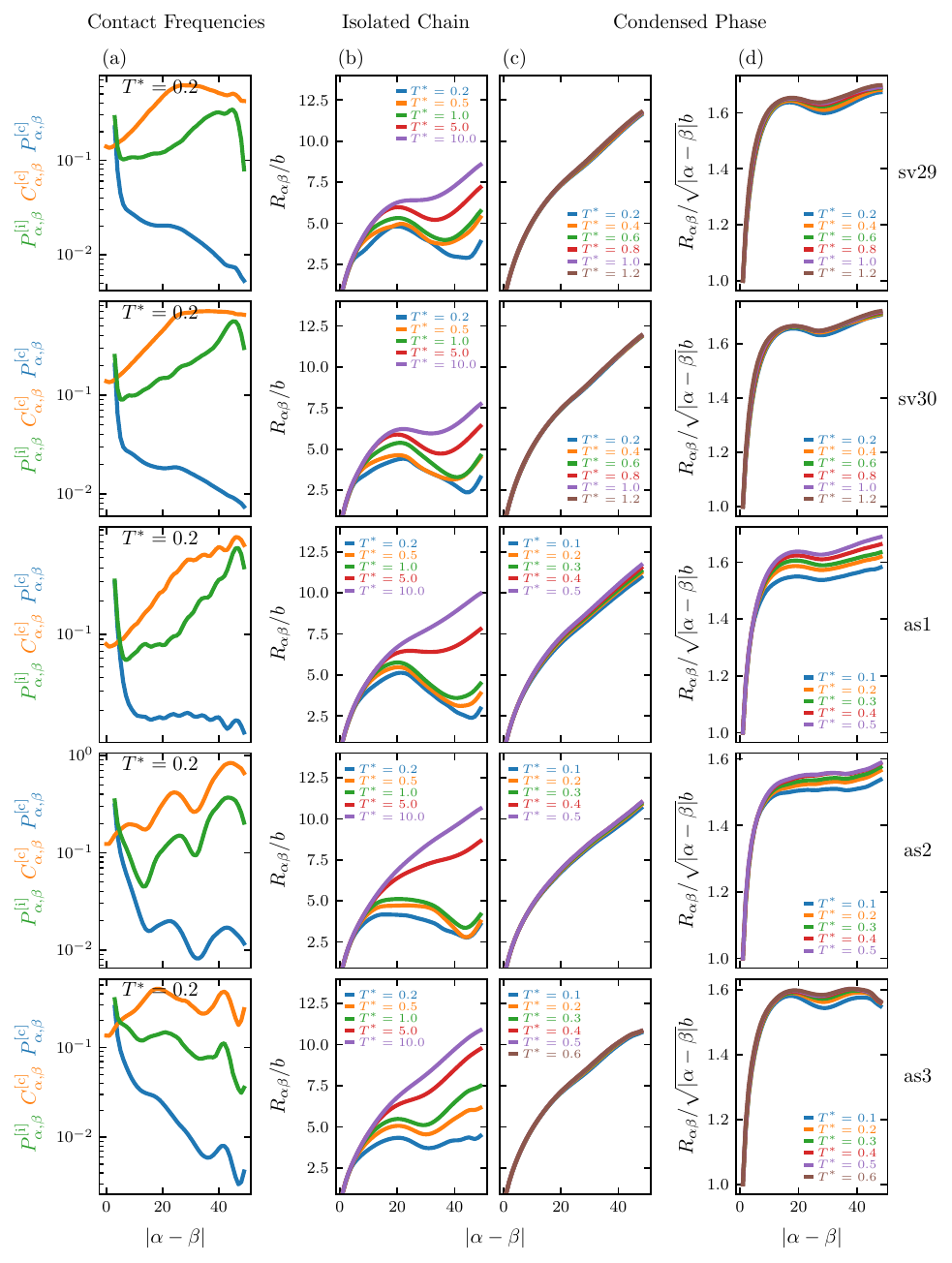}}
\caption{(3rd page, {\it cont'd}).}
\label{figS5c}
\end{figure*}

\vfill\eject

\setcounter{figure}{4}
\begin{figure*}[ht]
{\includegraphics[width=0.95\columnwidth,angle=0]{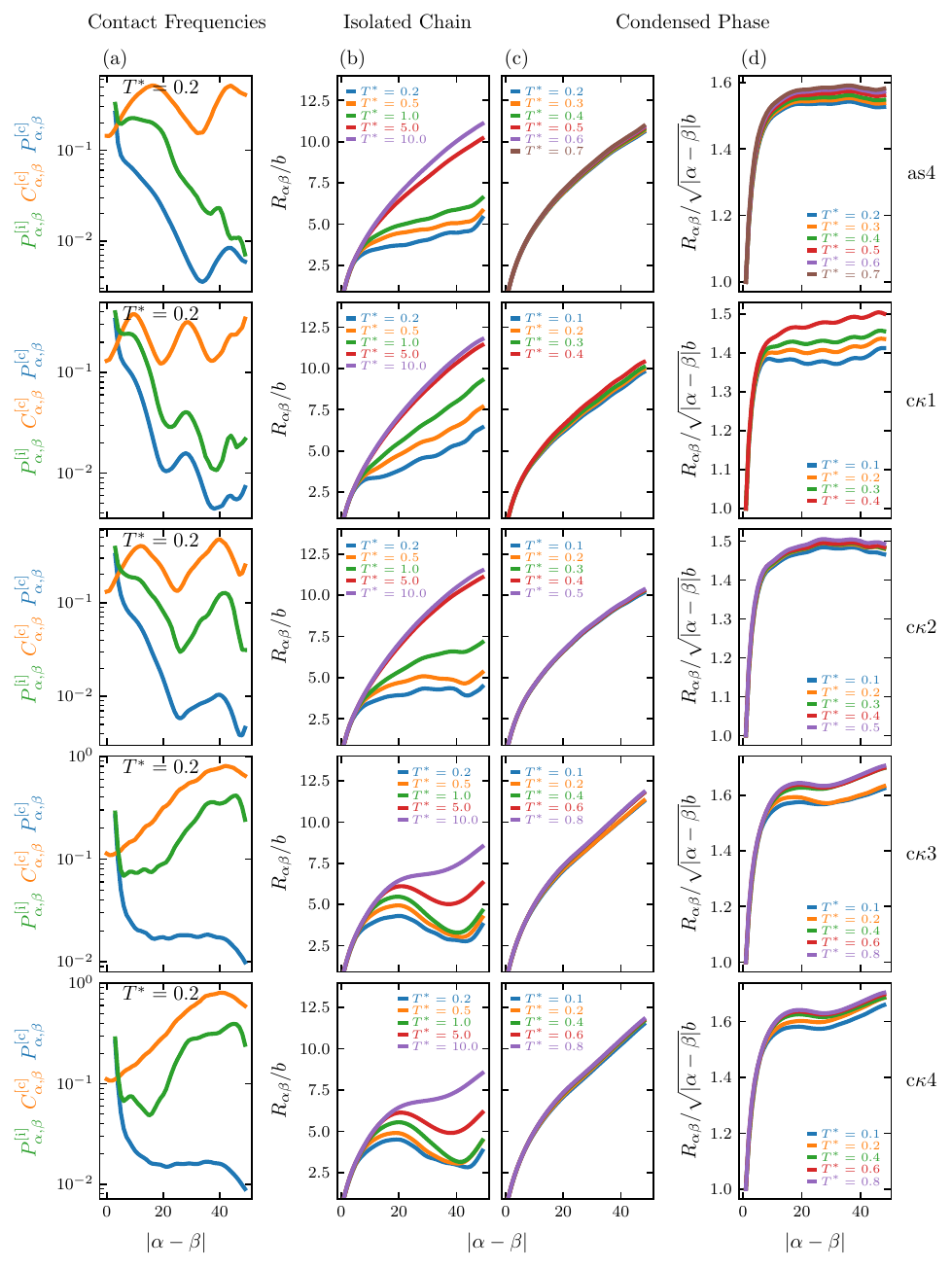}}
\caption{(4th page, {\it cont'd}).}
\label{figS5d}
\end{figure*}

\vfill\eject

\setcounter{figure}{4}
\begin{figure*}[ht]
{\includegraphics[width=0.95\columnwidth,angle=0]{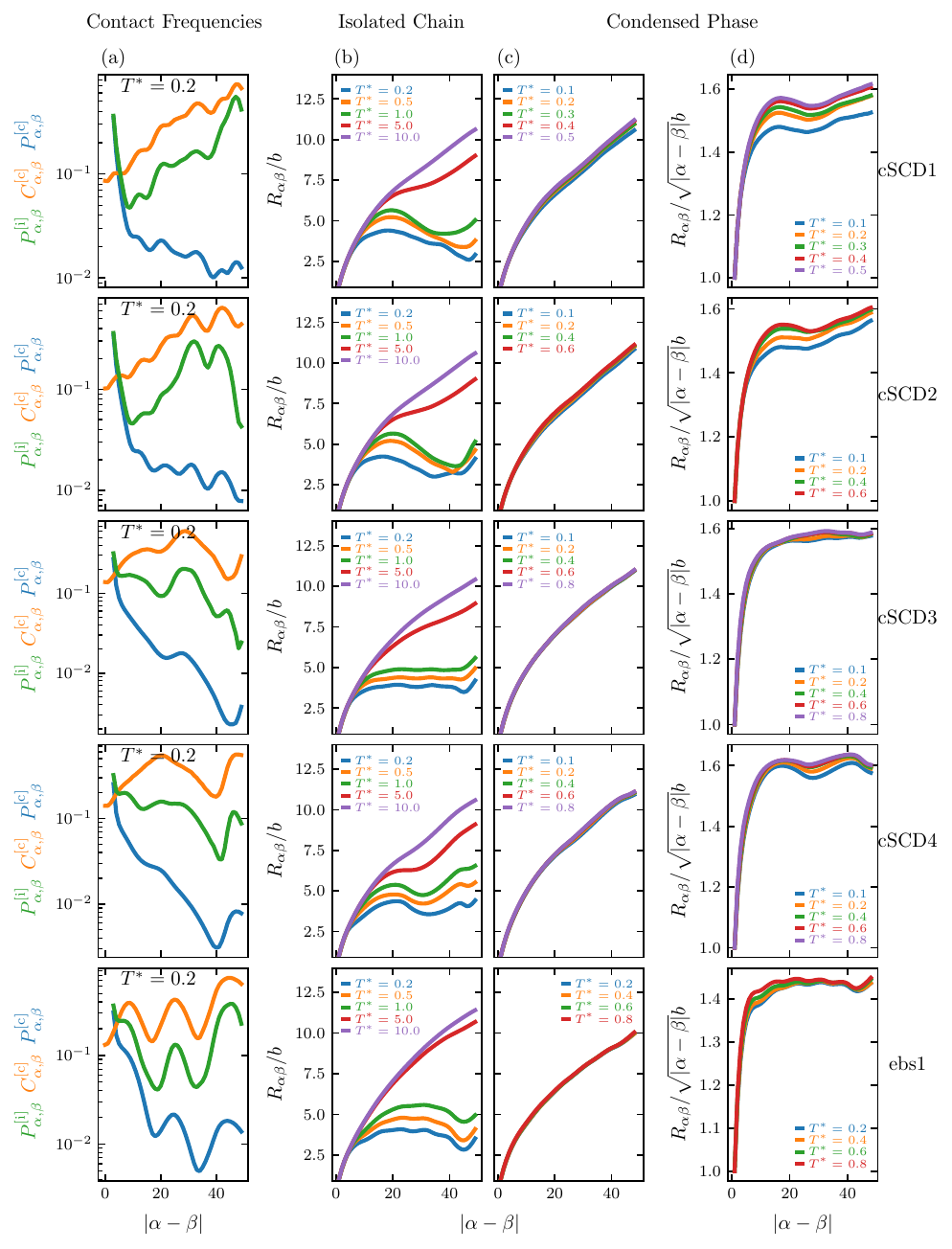}}
\caption{(5th page, {\it cont'd}). This is the last page of this supporting
figure.}
\label{figS5e}
\end{figure*}

\vfill\eject


\begin{figure*}[ht]
{\includegraphics[width=0.95\columnwidth,angle=0]{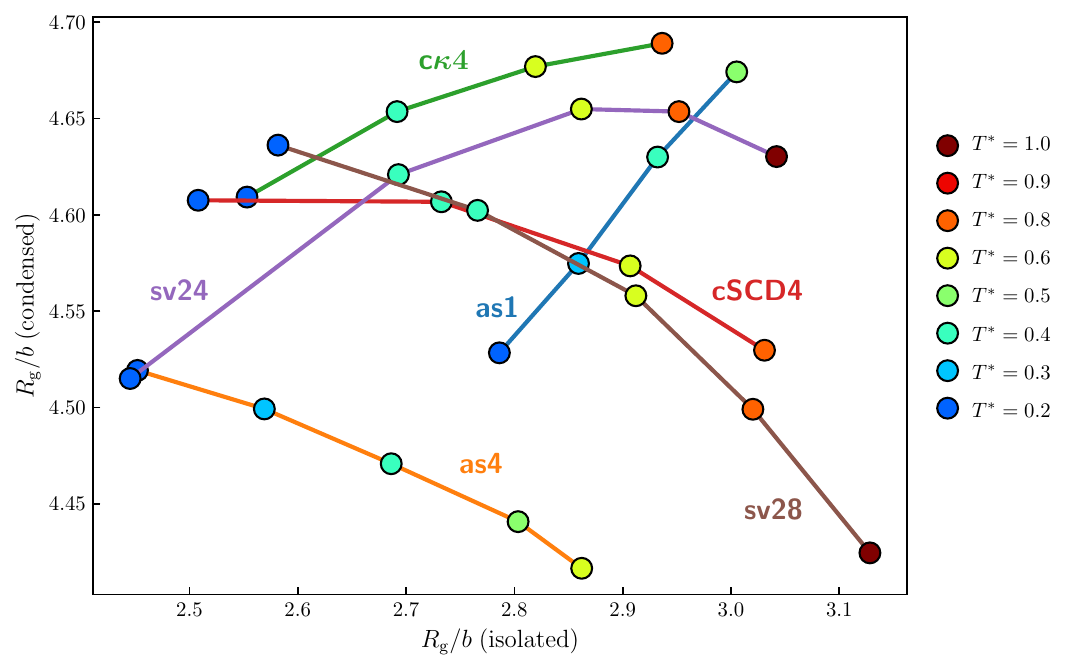}}
\caption{Temperature- and sequence-pattern-dependent relationship between 
root-mean-square radii of gyration of isolated versus condensed-phase 
polyampholyte chains in the MD model. Temperatures for
root-mean-square $R_{\rm g}$ values are color-coded as indicated on the right.
Datapoints for the same polyampholyte sequence (as labeled in different
colors) are connected by lines with matching color. Lines are otherwise
merely guides for the eye. Note that the range of $R_{\rm g}$ variation
in the condensed phase ($\sim 0.3b$, vertical axis) is small compared to
the corresponding variation for isolated chains ($\sim 0.7b$, horizontal axis).
Whereas isolated-chain $R_{\rm g}$s uniformly increase with increasing
$T^*$, condensed-phase $R_{\rm g}$s for different sequences can
increase or decrease (albeit only slightly) with increasing $T^*$. 
The physical origin of this sequence-specific feature and its possible
biophysical ramifications should be further investigated in future
studies. 
}
\label{figS6}
\end{figure*}

\vfill\eject

\begin{figure*}[ht]
{\includegraphics[width=0.80\columnwidth,angle=0]{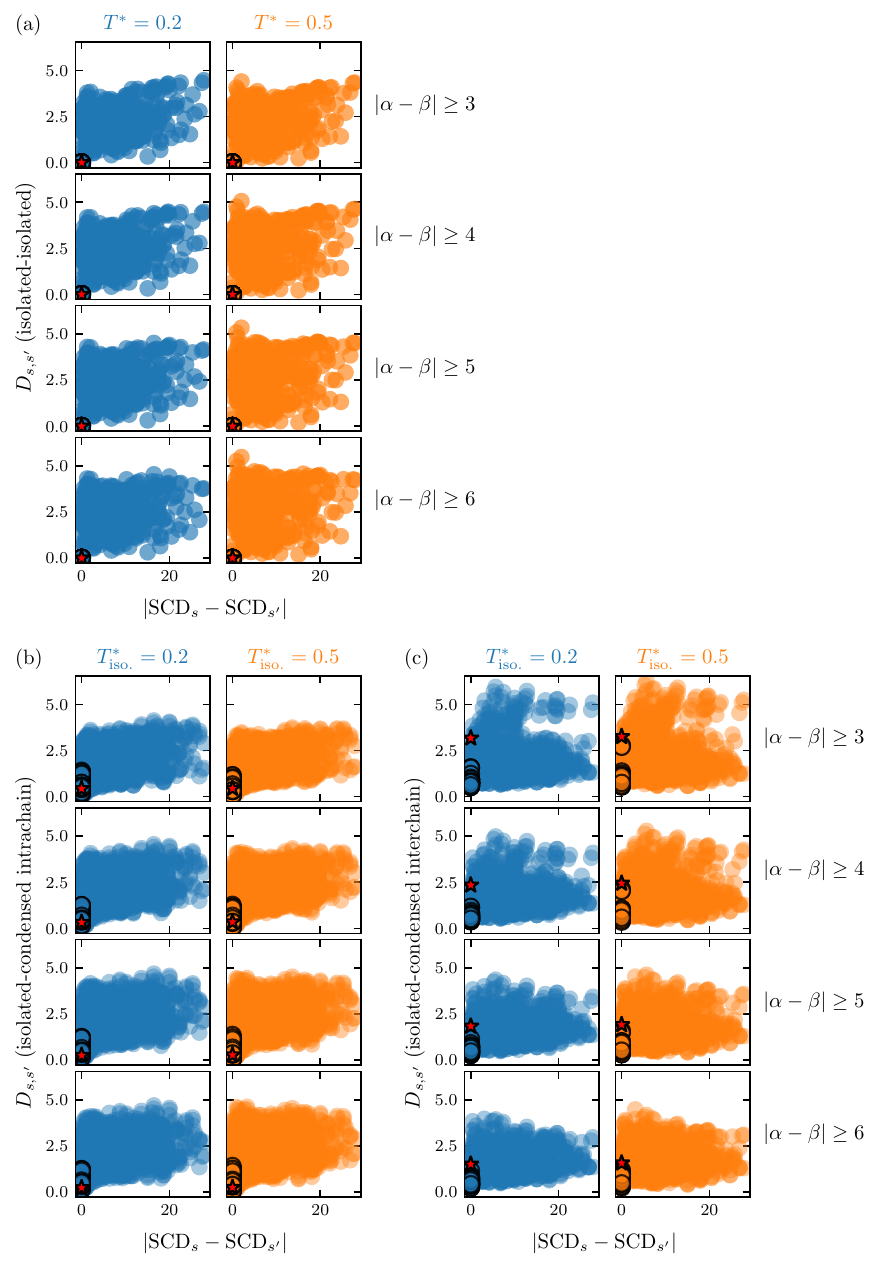}}
\vskip -0.4cm
\caption{Comparing contact patterns in the MD model using
symmetrized Kullback-Leibler divergences. 
(a) $D_{s,s^\prime}$ between the $|\alpha - \beta|\ge d_{\rm m}$ 
intrachain contacts 
of two separately isolated polyampholyte sequences $s$ and $s^\prime$ 
versus $s$--$s^\prime$ SCD difference ($s,s^\prime=$ all 26 sequences
in Fig.~S1). Results are shown for two temperatures $T^*=0.2$ and 
$0.5$ and different cutoffs $d_{\rm m}=3$--$6$ for local contacts, 
with $s=s^\prime$ datapoints identified by 
black circles and the value for baseline homopolymer at $T^*=2.0$ shown as 
a red star. (b,c) Same as maintext Fig.~4a and Fig.~4b 
(which are for $T^*_{\rm iso.}=0.2$ and $d_{\rm m}=4$),
respectively, 
now including data also for $T^*_{\rm iso.}=0.5$ and $d_{\rm m}=3,5$, and $6$.
}
\label{figS7}
\end{figure*}

\vfill\eject

\begin{figure*}[ht]
{\includegraphics[width=0.90\columnwidth,angle=0]{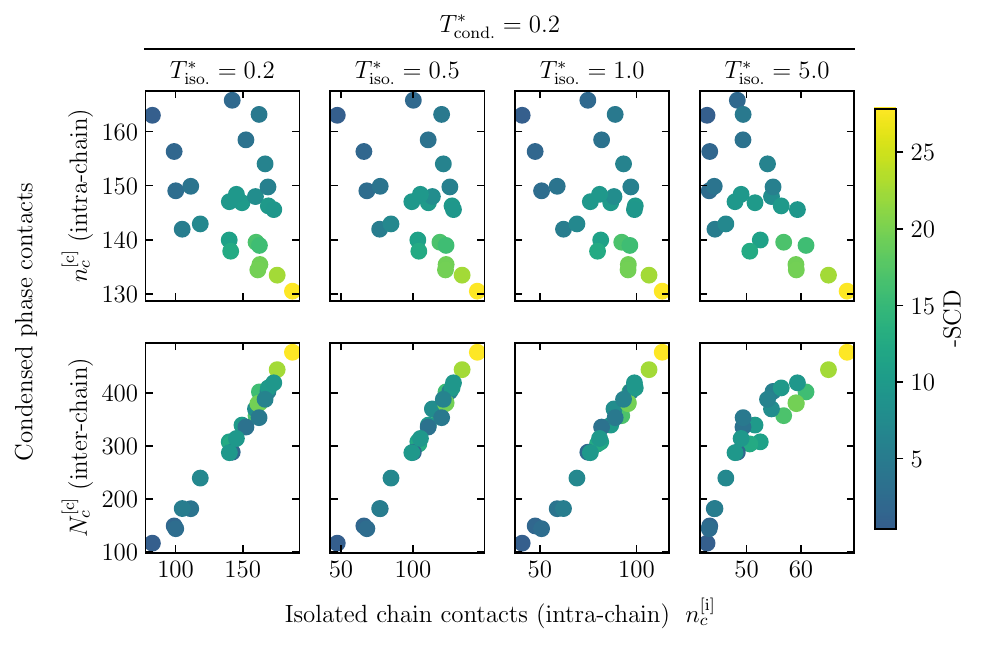}}
\caption{Variation in the number of isolated-chain intrachain contacts 
$n_c^{[{\rm i}]}$ (horizontal axes) with the number of  
condensed-phase intrachain contacts $n_c^{[{\rm c}]}$ (vertical axis, top
panels) and the number of condensed-phase interchain contacts $N_c^{[{\rm c}]}$
(vertical axis, bottom panels) in the MD model. 
Results are present in the same style as that
for maintext Fig.~4c,d. In addition to the $T^*_{\rm iso.}=0.2$
data in maintext Fig.~4c,d, corresponding data 
for $T^*_{\rm iso.}=0.5$, $1.0$, and $5.0$ are also provided here.
}
\label{figS8}
\end{figure*}

\vfill\eject

\begin{figure*}[ht]
{\includegraphics[width=0.90\columnwidth,angle=0]{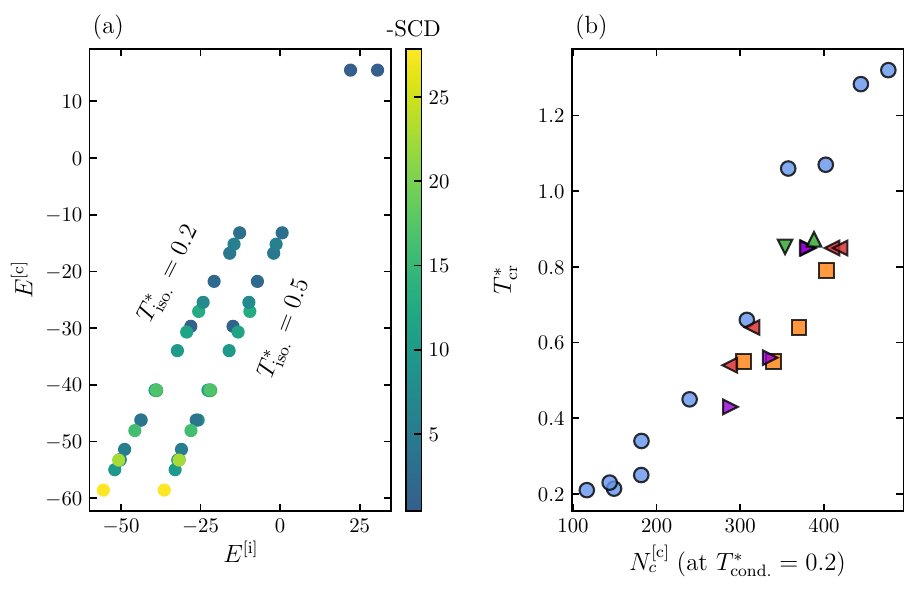}}
\caption{Relationship between condensed-phase contact frequencies
and interaction energies with phase-separation propensity in the MD model.
(a) Scatter plots of isolated-chain and condensed-phase potential 
energies $E^{[{\rm i}]}$ and energies $E^{[{\rm c}]}$ (both in units 
of $\epsilon$)
at two simulation temperatures
as indicated. The $T^*_{\rm iso.}=0.2$ datapoints are identical to those
provided in maintext Fig.~4e, and are included here to facilitate
comparison with the $T^*_{\rm iso.}=0.5$ datapoints.
(b) Critical temperature $T^*_{\rm cr}$ for all sequences except 
sv1 in maintext Fig.~1b and Fig.~S1. Symbols for sequences 
are the same as those in maintext Fig.~1a.
}
\label{figS9}
\end{figure*}

\vfill\eject

\vfill\eject

\clearpage

\vfill\eject
\clearpage

\noindent
{\Large\bf References}\\
 

\vfill\eject

\vfill\eject

$\null$
\begin{center}
   \includegraphics[height=44.5mm]{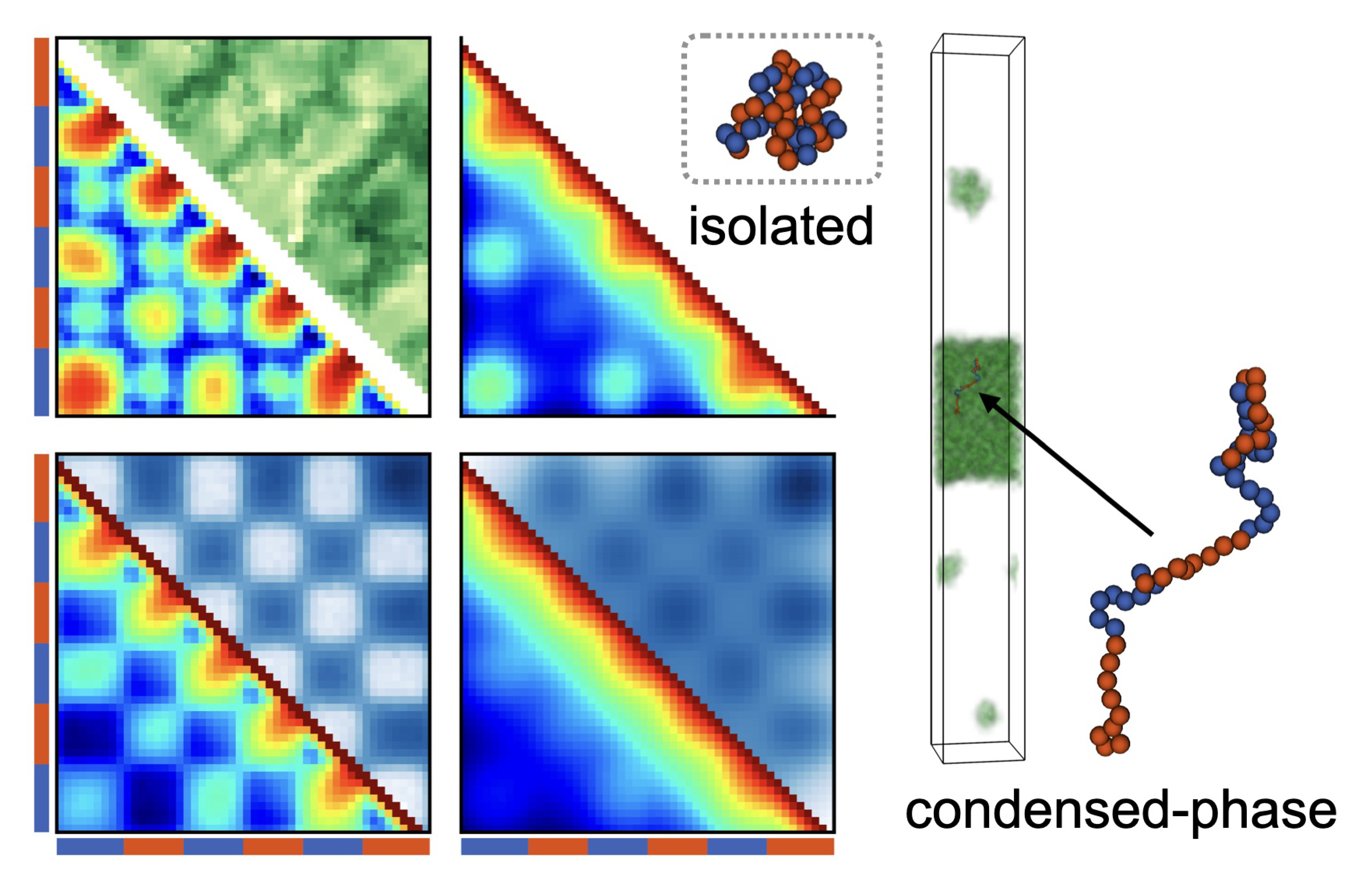}
\end{center}
\centerline{\bf TOC graphics}

\vfill\eject

\end{document}